\documentclass[aps,amssymb,amsmath,prd,twocolumn,
showpacs,preprintnumbers,superscriptaddress,nofootinbib,floatfix]{revtex4-1}
\usepackage{graphicx}
\usepackage{bm}
\usepackage{mathrsfs}
\usepackage{latexsym}
\usepackage{multirow}
\usepackage{float}
\usepackage{color}
\usepackage[normalem]{ulem} 
\usepackage{dcolumn}
\usepackage[colorlinks=true,citecolor=blue,urlcolor=blue]{hyperref}
\usepackage[usenames,dvipsnames]{xcolor}
\usepackage[caption=false]{subfig}
\usepackage{slashed}
\newcommand{\apjs}{The Astrophysical Journal Supplement}

\newcommand{\be}{\begin{equation}}
\newcommand{\ee}{\end{equation}}
\newcommand{\beq}{\begin{eqnarray}}
\newcommand{\eeq}{\end{eqnarray}}

\begin{document}
\title{Impact of the equation of state on $f$- and $p$- mode oscillations of neutron stars} 

\author{Athul Kunjipurayil}
\email{mywhatsappkp@gmail.com }
\affiliation{Department of Physics \& Astronomy, National Institute of Technology, Rourkela 769008, India}

\author{Tianqi Zhao}
\email{zhaot@ohio.edu}
\affiliation{Department of Physics and Astronomy, Ohio University,
Athens, OH~45701, USA}

\author{Bharat Kumar}
\email{\text{Correspondence:} kumarbh@nitrkl.ac.in }
\affiliation{Department of Physics \& Astronomy, National Institute of Technology, Rourkela 769008, India}

\author{Bijay K. Agrawal}
\email{bijay.agrawal@saha.ac.in}
\affiliation{Saha Institute of Nuclear Physics, 1/AF Bidhannagar, Kolkata 700064, India}
\affiliation{Homi Bhabha National Institute, Anushakti Nagar, Mumbai 400094, India}

\author{Madappa Prakash}
\email{prakash@ohio.edu}
\affiliation{Department of Physics and Astronomy, Ohio University,
Athens, OH~45701, USA}

\date{\today}

\begin{abstract}

We investigate the impact of the neutron-star matter equation of state  on the $f$- and $p_1$-mode oscillations of neutron stars obtained within the Cowling approximation and linearized general relativity. The $f$- and  $p_1$-mode oscillation frequencies, and their damping times are calculated using representative sets of Skyrme Hartree-Fock and relativistic mean-field models, all of which reproduce nuclear systematics and support $2M_\odot$ neutron stars. Our study shows strong correlations between the frequencies of $f$- and $p_1$-modes and their damping times with the pressure of $\beta$-equilibrated matter at densities equal to or slightly higher than the nuclear saturation density $\rho_0$. Such correlations are found to be  almost independent of the composition of the stars. The frequency of the $p_1$-mode of $1.4M_\odot$ star is strongly correlated with the slope of the symmetry energy $L_0$ and $\beta$-equilibrated pressure at density $\rho_0$. 
Compared to GR calculations, the error in the Cowling approximation for the $f$-mode is about 30\% for neutron stars of low mass, whereas it decreases with increasing mass. The accuracy of the $p_1$-mode  is  better than 15\% for neutron stars of maximum mass, and improves for lower masses and higher number of radial nodes. \

\end{abstract}

\maketitle


\section{Introduction}
\label{sec:intro}
Oscillation modes are generated when a neutron star (NS) is perturbed by an external or an internal disturbance. An oscillating NS can have different quasinormal modes (QNMs), categorized according to the restoring force that brings them back to equilibrium. QNMs can be fundamental mode ($f$-mode), pressure mode ($p$-mode), gravity mode ($g$-mode), rotational mode ($r$-mode),  space-time mode ($w$-mode), etc \cite{Kokkotas_1999}. The frequency of these oscillations depends upon the interior structure, and in some cases, the composition of the stars \cite{zhao2022quasi}. Moreover, these modes mainly contribute to a NS's gravitational wave emission. In order to study these oscillation frequencies, we need to solve the fluid perturbation equations along with the Tolman-Oppenheimer-Volkoff (TOV) equations in general relativity (GR) \cite{zhao2022quasi,zhao2022universal}. \\

Among the various QNMs studied theoretically, the $f$-mode is the most promising one to be observed first. The proto-neutron star (PNS) formed in a core-collapse supernova has long been considered as a  potential gravitational-wave source for LIGO and Virgo detectors \cite{ferrari2003gravitational}. Numerical simulation shows that about 10\% of the gravitational radiation is associated with the $\ell=2, m=1, 2$ $f$-mode oscillation \cite{shibagaki2020new}. The total gravitational radiation energy could reach $10^{44}-10^{47}$ ergs depending on the mass and rotation rate of the progenitor  \cite{radice2019characterizing}. Only galactic sources with distance  $D<20$ kpc are likely to be observed in  advanced LIGO observations, at a rate of a few per century \cite{li2011nearby}. The $f$-mode frequency of a proto-neutron star with high temperature  and high lepton fraction is lower than that for the NS which are considered here.  More likely to be observed are signals from the remnants of NS mergers, but  they will likely be complicated.
Numerical simulation of an equal-mass NS merger shows that the dominant fluid oscillation of a supermassive NS remnant coincides with the $m=2$ $f$-mode  \cite{stergioulas2011gravitational}, and has a strong correlation with the (zero temperature) isolated NS $f$-mode frequency  \cite{lioutas2021frequency}. The peak frequency in supermassive NSs is almost equal to that of the non-rotating $f$-mode frequency of isolated NSs with the same mass as each of the merging components \cite{ng2020gravitational}. With third generation gravitational wave telescopes such as the Cosmic Explorer \cite{reitze2019cosmic} and Einstein telescope \cite{punturo2010einstein}, the predicted event rate improves further to 0.06 yr$^{-1}$ to 4 yr$^{-1}$, which is
promising \cite{abbott2021population,zhao2022universal}. \\

Besides directly observing gravitational radiation from the fluid oscillations in a proto-neutron star and supramassive NS remnant, there's another direct method to measure QNMs of NSs by analysing NS merger waveform through dynamical tidal coupling \cite{hinderer2016effects,steinhoff2016dynamical,schmidt2019frequency}. During the inspiral of binary NSs, 
QNM oscillation can be excited by the dynamical tide 
when the orbital frequency approaches QNM frequency, which
leads to loss of orbital angular momentum. The significance of the signal is determined by the QNM frequency as well as its coupling to dynamical tidal field. Because the $g$-mode has a low frequency, tidal interactions could excite $g$-mode oscillations well before resonances with the $f$-mode is reached during the last part of the inspiral. However, the fluid perturbation of the $g$-mode peaks at the stellar center and thus has a weaker coupling to the tidal field. In contrast, the fluid perturbation of the $f$- and $p$-modes peak at the stellar surface with a stronger coupling to the tidal field. Indeed, the $f$- and $p$-mode frequencies of non-rotating NSs are too high to be excited in NS mergers. QNMs of rotating NS can be promoted by the NS spin $\omega_s$, because the resonant frequency could be reduced to $\omega_f-2\omega_s$, where $\omega_s$ is the spin frequency of the NS. For a millisecond pulsar, $f$-mode resonances 
could cause phase advances up to hundreds of cycles  \cite{wynn1999resonant,steinhoff2021spin}. A lower bound to the $f$-mode frequency has been estimated from the non-detection of a significant phase shift due to $f$-mode tidal resonance in GW170817, while an upper bound can be estimated from the $\Omega_f-\Lambda$ universal relation. The resulting 90\% credible interval of $f$-mode frequency for GW170817 is 1.43 kHz $<\nu_f<2.90$ kHz (1.48 kHz $<\nu_f<3.18$ kHz) for the more (less) massive NS in the binary. \\

The $f$-mode of neutron stars correlates with many NS properties including compactness~\cite{andersson1998towards}, moment of inertia~\cite{lau2010inferring} and static tidal polarizability~\cite{chan2014multipolar,sotani2021universal}. Such universal relations hold even for bare quark stars without crust or hybrid NSs with first order transitions \citep{zhao2022universal}. However, $p$-modes have very weak correlation with other NS properties \cite{andersson1998towards}. Because $p$-mode oscillation dominates in the stellar surface, it is sensitive to the equation of state (EoS) at much lower densities.  \\

One of the objectives in this work is calculate $f$-mode and lowest order $p$-mode frequencies and damping times by solving general relativistic non-radial oscillation equations  \cite{lindblom1983quadrupole,detweiler1985nonradial} 
in the
form an ordinary differential equation (ODE) eigenvalue problem.
These equations are for even-parity modes that include $f$-, $g$- $p$- and $w$-modes. We ignore the rotation of NSs which could slightly increase the $f$- and $p$-mode frequencies  \cite{kojima1993normal,kruger2020dynamics}. The results so obtained will then be compared with those obtained using the Cowling approximation to assess its accuracy. 
Many previous studies have used the Cowling approximation   \cite{mcdermott1983nonradial,ranea2018oscillation,sotani2011signatures,flores2014discriminating}, which lacks dissipation due to gravitational waves. The Cowling approximation introduces about a 20-30\% error in the $f$-mode frequency  \cite{yoshida1997accuracy,sotani2001density,chirenti2015fundamental,zhao2022universal}, which is significantly less accurate than that of the $\Omega_f-\bar I-\Lambda$ universal relations. A ~10\% error in the compositional $g$-mode frequency  \cite{zhao2022quasi} and ~19\% error in the discontinuity $g$-mode frequency  \cite{sotani2001density, zhao2022universal} from the Cowling approximation has also been found. However, the accuracy of the Cowling approximation for the and $p$-mode is not well quantified. In addition such comparisons have only been made for schematic EoSs such as piece-wise polytropes. Hence, another of our objective is to use physics-based EoSs that reproduce laboratory as well as observational data.   \\

Two classes of  EoSs are chosen for our study. The first one is based on non-relativistic zero-range Skyrme interactions using which calculations of the EoSs are performed in the Hartree-Fock approach; these models are labelled SHF. Care is taken to render these EoSs causal, at least for densities within the cores of the NSs. The second class of EoSs are relativistic mean-field models (RMF) that are naturally causal. A total of 35 EoSs are employed all of which reproduce nuclear systematics at near saturation density of $\rho_0\simeq 0.16\pm 0.01~{\rm fm}^{-3}$ and are able to to support a NS of $\geq 2M_\odot$. We also compare these EoSs with recent chiral effective field theoretical (EFT) calculations \cite{drischler2021limiting} for densities up to $2\rho_0$, the limiting density up to which such calculations are valid. Many of the SHF and RMF EoSs fall within the $\pm 2\sigma$ region of the  chiral EFT results.  \\

An important further objective of our work is to explore possible universal relations connecting the $f$- and $p$- mode frequencies with both nuclear and observational properties, such as the NSs mass, tidal deformability, etc.,  as well as EOS characteristics such as the slope of symmetry energy at $\rho_0$ and pressure at $\rho_0$ and $2\rho_0$.  In all of these cases, we find interesting correlations and proffer the underlying cause for such behaviors.   \\

This paper is organized as follows. Section II contains 
a description of non-radial oscillations in GR and the Cowling approximation to full GR.  The EoSs used in this work are described in Sec. III. Our results and discussion are contained in Sec. IV. 
A critical analysis of correlations including the dependence on the sample sizes of EoSs and findings in earlier works is provided in Sec. V.
Finally, Sec. VI presents a summary along with our conclusions. \\


\section{Non-radial oscillations in general relativity}
\label{sec:GRR}

NS oscillations that couple to gravitational radiation were first studied by Thorne {\it et~al.}  \cite{thorne1967non}. Such oscillations involve linear scalar variations of pressure and density. Only even-parity perturbations of the Regge-Wheeler metric are relevant. The appropriate line element is thus 
\begin{equation}
\begin{split}
ds^2 = - e^{\nu(r)} [1+r^l H_0(r)e^{i\omega t} Y_{lm}(\phi,\theta)]c^2 dt^2\\
+e^{\lambda(r)} [1-r^l H_0(r)e^{i\omega t} Y_{lm}(\phi,\theta)]dr^2 \\
+ [1-r^l K(r)e^{i\omega t}Y_{lm}(\phi,\theta)]r^2 d\Omega^2\\
-2i\omega r^{l+1}H_1(r)e^{i\omega t}Y_{lm}(\phi,\theta) dt~dr\,,
\end{split}
\end{equation}
where 
\begin{equation}
e^{\lambda(r)}=\frac{1}{1-2b(r)}
\end{equation}
and 
\begin{eqnarray}
 r\frac{d\nu(r)}{dr} &=& 2 e^{\lambda(r)} \textrm{Q}(r)\\
 \textrm{Q}(r) &=&b(r)+\frac{4\pi G r^2 p(r)}{c^4}
\end{eqnarray}
with the boundary condition $\nu(R)=-\lambda(R)$. Above, $b(r)= 2 G m(r) /(c^{2} r)$, $m(r)$ and $p(r)$ are the enclosed mass and pressure at radius $r$, respectively. The angular part is characterized by the angular quantum number $l$ and azimuthal quantum number $m$ in the spherical harmonics $Y_{lm}$; $m$ is degenerate for the nonrotating NSs we calculate here. Perturbations of the metric are described by the functions $H_0$, $H_1$, and $K$. The real part of the complex eigenvalue $\omega$ is the oscillation frequency, whereas the imaginary component is the inverse of the damping time (positive) or instability time (negative).

The Lagrangian displacement vector
\begin{eqnarray}
\xi^r &=& r^{l-1}e^{-\frac{\lambda}{2}}W Y^l_m e^{i\omega t} \label{eq:xi_radial}\\
\xi^\theta &=& -r^{l-2} V \partial_\theta Y_m^l e^{i\omega t}\\
\xi^\phi &=& -\frac{r^{l-2}}{ \sin^{2}\theta} V\partial_\phi Y_m^l e^{i\omega t} \,,
\end{eqnarray}
describes fluid perturbations inside the star with amplitudes $W$ and $V$ both of which have the dimension $[R]^{2-l}$, where $R$ is the radius of NS. In addition, an auxiliary function $X$, related to Lagrangian pressure variations,  is defined as 
\begin{equation}
\Delta p = -r^{l}e^{-\frac{\nu}{2}}X Y_m^l e^{i\omega t}\,.
\end{equation}

In some range of frequency, the ODEs governing NS oscillation can exhibit a singularity. In order to avoid such a singularity,  
Lindblom {\it et~al.}  \cite{Lindblom:1983,Detweiler:1985}   
choose the four degrees of freedom to be $H_1$, $K$, $W$, and $X$.  Evaluating the two remaining functions $H_0$ and $V$ in terms of them yields
\begin{eqnarray}
H_0&=&\left \{8\pi r^2 e^{-\nu/2}X-\left[(n_l+1)\textrm{Q}-\omega^2r^2 e^{-(\nu+\lambda)}\right]H_1  \right. \\
&+& \left. \left [ n_l-\omega^2 r^2 e^{-\nu}- \textrm{Q}(e^\lambda\textrm{Q}-1)  \right ] K \right \} (2b+n_l+\textrm{Q})^{-1}\label{eq:H0_def_in},\nonumber \\
V&=&\left[\frac{X}{\varepsilon+p}-\frac{\textrm{Q}}{r^2}e^{(\nu+\lambda)/2}W-e^{\nu/2}\frac{H_0}{2}\right]\frac{e^{\nu/2}}{\omega^2}\label{eq:V_def_in} \,,
\end{eqnarray}
where $n_l=(l-1)(l+2)/2$ and $\varepsilon$ is the local energy density. 
By expanding Einstein's equation to first-order, the homogeneous linear differential equations for $H_1$, $K$, $W$ and $X$ are  \cite{Detweiler:1985},
\begin{eqnarray}
r\frac{dH_1}{dr}&=&-[l+1+2b e^\lambda+4\pi r^2 e^\lambda(p-\varepsilon)] H_1\nonumber\\
&&+ e^\lambda[H_0+K-16\pi(\varepsilon+p)V]\,, \label{eq:ODE_DL1} \\
r\frac{dK}{dr}&=& H_0+(n_l+1)H_1 \nonumber \\
&&+[e^\lambda \textrm{Q}-l-1]K-8\pi(\varepsilon+p)e^{\lambda/2}W \,, \label{eq:ODE_DL2}\\
r\frac{dW}{dr}&=&-(l+1)[W+le^\frac{\lambda}{2}V] 
\nonumber \\
&&+r^2 e^{\lambda/2}\left[\frac{e^{-\nu/2}X}{ (\varepsilon+p) c_{\rm ad}^2}+\frac{H_0}{2}+K\right., \label{eq:ODE_DL3}\\
r\frac{dX}{dr}&=& \left.-lX+\frac{(\varepsilon+p)e^{\nu/2}}{2} \right.\nonumber \\
&&\left\{ \left. (1-e^\lambda \textrm{Q})H_0+(r^2\omega^2e^{-\nu}+n_l+1)H_1 \right. \right.\nonumber  \\
&&+(3e^\lambda\textrm{Q}-1)K
-\frac{4(n_l+1)e^\lambda\textrm{Q}}{r^2}V -2\left[\omega^2 e^{\lambda/2-\nu} \right. \nonumber\\
&&+4\pi(\varepsilon+p)e^{\lambda/2} \left.\left. -r^2\frac{d}{dr} \left(\frac{e^{{\lambda/2}} \textrm{Q}}{r^3} \right)\right]W \right\} \,, \label{eq:ODE_DL4}
\end{eqnarray}
where $c_{\rm ad}^2$ is the adiabatic sound speed of NS matter under oscillations. In this work, we approximate this speed of sound with the equilibrium sound speed $c_{\rm eq}^2=dp/d\varepsilon$. Generally, $c_{\rm ad}^2$ is usually slightly larger than the equilibrium sound speed $c_{\rm eq}^2$ due to lag in weak equilibrium \cite{wei2020lifting,jaikumar2021g}, and temperature equilibrium \cite{reisenegger1992new,kuan2022constraining}.\\

Perturbations at the center of the star $r=0$ are subject to
the boundary conditions 
\begin{eqnarray}
W(0)&=&1 \label{eq:BC_W}\\
X(0)&=&(\varepsilon_0+p_0)e^{\nu_0/2} \nonumber \\
&&\hspace*{-.2cm} \left \{ \left[ \frac{4\pi}{3}(\varepsilon_0+3p_0)-\frac{\omega^2}{l} e^{-\nu_0}\right]W(0)+\frac{K(0)}{2}\right \} \nonumber \\
\label{eq:BC_X0} \\
H_1(0)&=&\frac{lK(0)+8\pi(\varepsilon_0+p_0)W(0)}{n_l+1} \label{eq:BC_H}\\
X(R)&=&0 \,.
\end{eqnarray}
The last boundary condition above is 
obtained by first solving the two trial solutions with $K(0)=\pm(\varepsilon_0+p_0)$ and then constructing a linear combination to obtain the correct solution that satisfies the boundary condition $X(r=R)=0$.
The latter boundary condition corresponds to no pressure variations at the surface. By construction,  $H_0(0)=K(0)$. 


{Exterior solutions}

{Outside the NS, fluid perturbations $W$, $V$ and $X$ vanish and the metric perturbations follow Zerilli's differential equation
\begin{eqnarray}
{d^2Z}/{dr^{*2}}&=&(V_Z(r)-\omega^2)Z \,. \label{eq:zerilli}
\end{eqnarray}
The function $Z$ is defined through  \cite{fackerell1971solutions}
\begin{eqnarray}
\begin{pmatrix}
K(r)\\H_1(r)
\end{pmatrix}
&=&
\begin{pmatrix}
g(r) & 1\\
h(r) & k(r)
\end{pmatrix}
\begin{pmatrix}
{Z(r^*)/r}\\ {dZ(r^*)/dr^*}
\end{pmatrix},\nonumber \\
g(r)&=&\frac{n(n+1)+3n {b} +6 {b} ^2}{(n+3 {b} )}, \label{eq:zerilli_matching}\\
h(r)&=&\frac{(n-3n {b} -3 {b} ^2)}{(1-2 {b} )(n+3 {b} )},\nonumber \\
k(r)&\equiv&\frac{dr^*}{dr}=\frac{1}{1-2 {b} },\nonumber
\end{eqnarray}
and an effective potential 
\begin{eqnarray}
\hspace*{-.5cm}V_Z(r)=(1-2 {b} ) \frac{2n^2(n+1)+6n^2 {b} +18n {b} ^2+18 {b} ^3}{r^2(n+3 {b} )^2} \,,
\end{eqnarray}
where $b={GM}/{(c^2r)}$ since $m(r>R)=M$. We first solve Eqs. (\ref{eq:ODE_DL1}-\ref{eq:ODE_DL4}) from the stellar center to the surface. Then, we evaluate $Z(R)$ by the matching condition Eq. (\ref{eq:zerilli_matching}) and use it as the initial boundary condition to solve Eq. (\ref{eq:zerilli}) from the stellar surface to $r\gtrsim 100$ km. In the far-field limit, the solution $Z$ can be decomposed into incoming ($Z_+$) and outgoing ($Z_-$) spherical gravitational radiation: 
\begin{eqnarray}
\hspace*{-1cm}\begin{pmatrix}
Z(\omega)\\{dZ}/{dr^*}
\end{pmatrix}
&=&
\begin{pmatrix}
Z_-(\omega) \quad Z_+(\omega)\\ {dZ_-}/{dr^*} \quad {dZ_+}/{dr^*}
\end{pmatrix}
\begin{pmatrix}
A_-(\omega)\\A_+(\omega)
\end{pmatrix},\nonumber\\
Z_-&=&e^{-i\omega r^*}\left[\alpha_0+\frac{\alpha_1}{r}+\frac{\alpha_2}{r^2}+\mathcal{O}(r^{-3})\right]\nonumber,\\
\frac{dZ_-}{dr^*}&=&-i\omega e^{-i\omega r^*} [\alpha_0+\frac{\alpha_1}{r} \label{eq:zerilli_decompose}\\
&&+\frac{\alpha_2+{i\alpha_1}(1-2b)/\omega}{r^2}+\mathcal{O}(r^{-3})],\nonumber\\
\alpha_1&=&\frac{-i(n+1)\alpha_0}{\omega},\nonumber\\
\alpha_2&=&\frac{[-n(n+1)+iM\omega({3}/{2}+{3}/{n})]\alpha_0}{2\omega^2}\,,\nonumber
\end{eqnarray}
where $A_+(\omega)$ and $A_-(\omega)$ are the corresponding amplitudes, and $Z_+$ is the complex conjugate of $Z_-$.} 

{Numerical notes}

{To evaluate $A_+(\omega)$, we used the complex-valued ODE solver \emph{zvode} \cite{brown1989vode}. The solver \emph{zvode} uses implicit Adams method for non-stiff problems and a method based on backward differentiation formulas for stiff problems.  
The integration steps are automatically adjusted to yield a specified relative error tolerance. In order to take care of the imaginary part of $\omega$, which is over 1000 times smaller than the real part, we need to keep the relative error of our ODE solver to $10^{-6}$ for $H_1$, $K$, $W$, $X$ and $Z$. The quantity $A_-(\omega)$ is evaluated by Eq. (\ref{eq:zerilli_decompose}) at $r=25\omega^{-1}$ and $r=50\omega^{-1}$ to guarantee convergence within $10^{-6}$.} \\

{Up to this point, we have discussed the ODEs and boundary conditions for the initial value problem of all perturbation functions with a given oscillation frequency $\omega$. The eigenvalue problem is defined by the pure outgoing gravitational wave boundary condition at far-field, $A_+(\omega)=0$. 
In order to solve the eigenvalue problem, we use a complex root finding algorithm which takes about 8 Newton–Raphson iterations to converge.}  \\

{By solving the eigenvalue problem 
we can calculate even parity quasinormal modes of any order ($n=0,1,$ etc.), and angular number ($\ell-2, 3,$ etc.). 
Radial perturbation functions can have different number of turnovers (nodes) $n$ for different quasinormal modes. The $n=0$ modes that have no nodes are referred to as the $f$-modes, whereas those with $n$=1,2, etc., have the corresponding number of nodes and are termed as the $p$-modes.
While we calculate $f$-and $p$-modes of arbitrary order $n$, in this work we compute the $\ell=2$ quasinormal $f$-mode ($n=0$ of $W$ and $V$) and the lowest order $p$-mode ($n=1$ and $\ell=2$ of $W$ and $V$) 
as frequencies of higher order $p$-modes are too large compared with the detection limits. Henceforth, we will refer to the $p$-modes with $n=1$ as $p_1$ modes.  

The non-radial modes calculated by our codes are in agreement to four significant digits with the results given in the tables of the pioneering works of \cite{lindblom1983quadrupole,miniutti2003non,finn1987g}
for $f$- and $p$-modes as well as  the discontinuity $g$-mode \cite{zhao2022universal} and the compositional $g$-mode \cite{zhao2022quasi}.}
Profiles of the $f$- and $p$-modes will be presented in the section on results.

\subsection*{Relativistic Cowling approximation}
\label{sec:Cowling}

In Newtonian theory of stellar pulsations, 
the perturbation of the gravity field is neglected for fluid modes. This simplification is  known as the Cowling approximation \cite{cowling1941non}. 
In the relativistic theory, the perturbation of the GR metric is often neglected as well, which leads to the {\it relativistic} Cowling approximation~\cite{McD83}. The relativistic Cowling equations are obtained by setting $H_0=H_1=K=0$ in  Eqs. (\ref{eq:V_def_in}, \ref{eq:ODE_DL3}, \ref{eq:ODE_DL4}) and furthermore, dropping the term $-4\pi(\varepsilon+p)^2e^{(\nu+\lambda)/2}W$ in Eq. (\ref{eq:ODE_DL4}). These simplifications  lead to \cite{zhao2022quasi},
\begin{eqnarray}
\label{eq:ODE_DL3_cowling_UW}
\frac{dW}{d\ln r} & =& -(l+1)\left[W-l e^{\nu +\lambda/2} U\right] \nonumber \\
&& -\frac{e^{\lambda/2}(\omega r)^2}{c_{ad}^2}\left[U- \frac{e^{\lambda/2}\textrm{Q}}{(\omega r)^2} W\right ]\,,\\
\frac{dU}{d\ln r} && = e^{\lambda/2-\nu}\left[W -le^{\nu-\lambda/2}U\right] \,,\label{eq:ODE_DL4_cowling_UW}
\end{eqnarray}
where $W=e^{\lambda/2} r^{1-l} \xi^r$ and $U=-e^{-\nu}V=r^{-l} \omega^{-2} \delta p/(\varepsilon+p)$, $\xi^r$ are radial Lagrangian displacements defined in Eq. (\ref{eq:xi_radial}) and $\delta P$ is the Eulerian perturbation of pressure, which is related to the the Lagrangian perturbation by $\Delta P=\delta P-(\varepsilon+p) \frac{d\Phi}{dr} \xi^r$. The boundary conditions can be written explicitly as,
\begin{eqnarray}
\left.\frac{W}{U}\right|_{r=0}&=&l e^{\nu|_{r=0}} \\
\left.\frac{W}{U}\right|_{p=0}&=&\frac{\omega^2R^3}{GM}\sqrt{1-\frac{2GM}{c^2R}} \,.
\end{eqnarray}
These equations determine the eigenmode frequency of the oscillation in the relativistic Cowling approximation.


\section{Equations of state} 
\label{sec:NUC}

We have considered a diverse set of EoSs for charge neutral and $\beta$-equilibrated neutron star 
matter that are obtained using relativistic and non-relativistic mean-field models. Two different variants of the relativistic mean-field (RMF) models: (1) models with nonlinear self and/or mixed interaction terms with constant coupling strengths, and
(2) models with only linear interaction terms, but density-dependent coupling strengths have been employed.  The RMF models with density-independent coupling considered 
are BSR2, BSR6 \cite{PhysRevC.76.045801,PhysRevC.90.044305}, GM1 \cite{PhysRevLett.67.2414}, NL3 \cite{PhysRevC.55.540}, NL3$\omega\rho03$ \cite{Carriere_2003}, IOPB-I \cite{Kumar_ns,vishal}, G3 \cite{Kumar_param,vishal}, TM1 \cite{TM1}, and FSUG \cite{FsuG}. The RMF models with density-dependent couplings are DD2 \cite{PhysRevC.81.015803},DDH$\delta$ \cite{Gaitanos_2004}, and DDME2 \cite{DDME2}. The
non-relativistic mean-field models considered 
employ the Skyrme Hartree-Fock (SHF)
approach  and they are SKa, SKb~\cite{KOHLER1976301}, SkI2, SkI3,
SkI4, SkI5~\cite{REINHARD1995467}, SkI6 ~\cite{PhysRevC.53.740}, SLY2, 
SLY9~\cite{thesis}, SLY230a~\cite{CHABANAT1997710}
SLY4~\cite{CHABANAT1998231}, SkMP~\cite{PhysRevC.40.2834}, SKOp~\cite{REINHARD1999305},
KDE0V1~\cite{PhysRevC.72.014310}, SK255,
SK272~\cite{PhysRevC.67.034314}, Rs~\cite{PhysRevC.33.335}, BSk20, 
BSk21~\cite{PhysRevC.82.035804}, BSk22, BSk23,
BSk24, and BSk25~\cite{PhysRevC.88.024308}.

The EoS for the outer crust region is  taken to be the one obtained by Baym-Pethick-Sutherland \cite{BPS}. The EoSs for the inner crust in the case of RMF models are determined within a Thomas-Fermi approach up to the crust-core transition density. In the case of SHF models, the EoSs for the inner crust are constructed with the compressible liquid-drop model \cite{Gul2015,Fortin}. These unified EoSs models closely reproduce the properties of finite nuclei, nuclear matter and neutron stars \citep{Kumar_param,Kumar_ns, Fortin}. Moreover, these EoSs are causal \cite{Tuhin} and support a $2.0\,M_{\odot}$ star \citep{anto}. All of the employed unified EoSs are for $npe\mu$ matter, and apply from the neutron star crust to its core. \\

\begin{figure}[htbp!]
    \includegraphics[width=\linewidth]{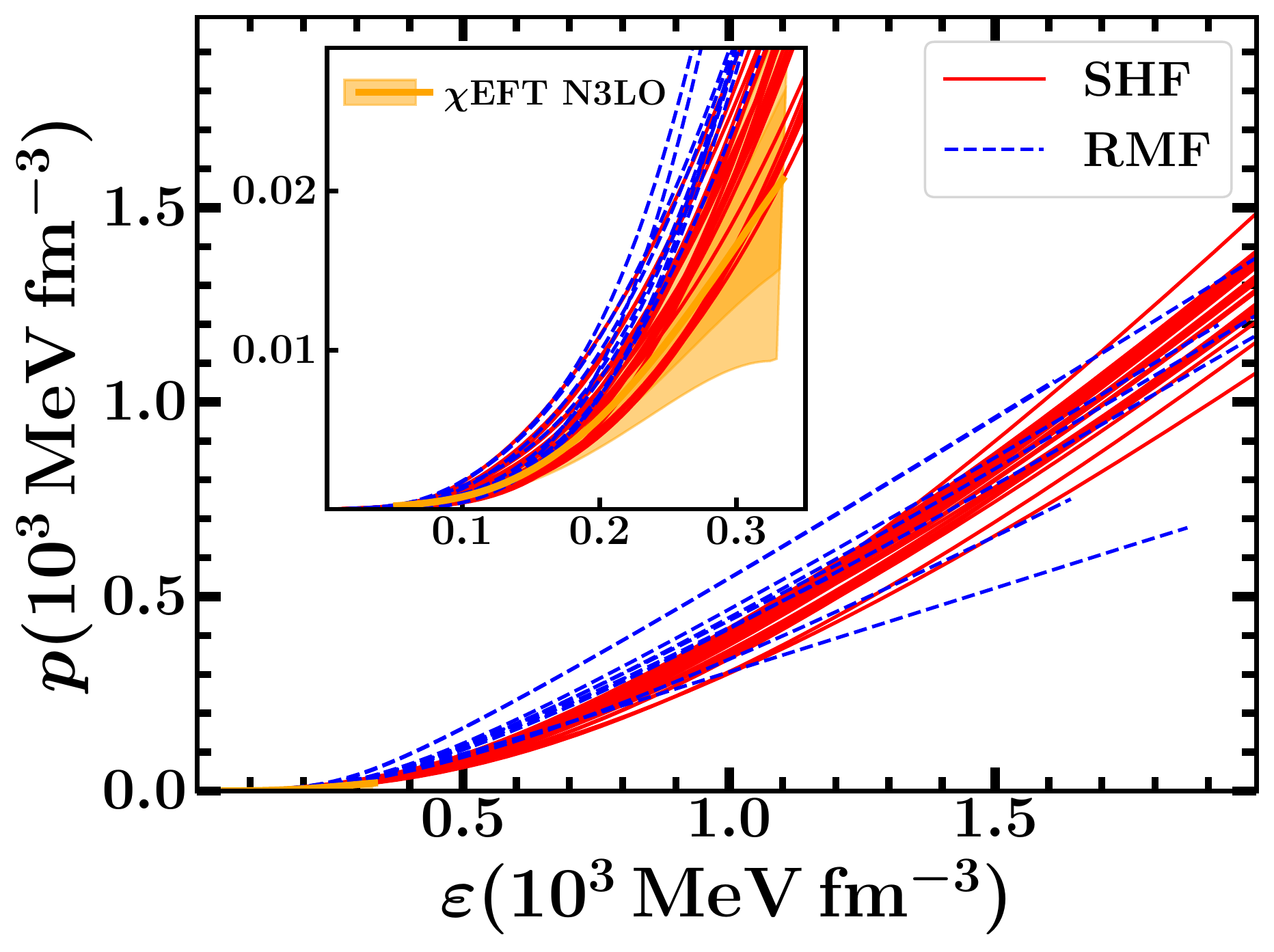}
    \caption{
Pressure vs energy density for a representative set of EoSs corrosponding to  the SHF  ( red solid ) and RMF (blue dashed )  models considered in the present work.
The orange region in the inset shows  results for the  Chiral EFT $\pm 1 \sigma$ and $\pm 2 \sigma$ bounds up to twice saturation density \cite{drischler2021limiting}.}
    \label{fig:pvse}
\end{figure}

The SHF and RMF EoSs used are displayed in terms of their pressure-energy density relations in Fig. \ref{fig:pvse}. The inset in this figure contrasts the SHF and RMF EOSs with the chiral EFT results of Ref. \cite{drischler2021limiting}  up to $2\rho_0$. Note that many SHF EOSs lie within the $2\sigma$ error estimates of the chiral EFT results, several RMF EoSs fall outside these limits. Such EoSs are nevertheless included in our analyses for the sake of comparison and completeness.  \\

The behaviour of EoS is often understood in terms of  various nuclear matter parameters for a given EoS. The lower order nuclear matter parameters, govern the behaviour of EoS  at lower densities,  are strongly correlated with several bulk properties of finite nuclei. The higher order nuclear matter
parameters are expected to be correlated with the bulk properties of NS.These parameters are  evaluated as follows.
For densities up to $2\rho_0$ - $3\rho_0$   and for neutron-proton asymmetry  $\delta =\frac{ \rho_n-\rho_p}{\rho},(\rho=\rho_n+\rho_p))$, where
$\rho_n$ and $\rho_p$ are the neutron and proton densities, respectively,
the energy per nucleon as a function of density $\rho$ 
and asymmetry $\delta$ can be expressed as,
\begin{eqnarray}
 e(\rho,\delta)&=&e_{\rm snm} (\rho) +e_{\rm sym} (\rho)\delta^2 + \mathcal {O}(\delta^4) \, ,\\
e_{\rm snm}(\rho)&=&e_0+\frac{K_0}{2} x^2+\frac{Q_0}{6}x^3+\mathcal{O}(x^4) \label{esnm}\\
e_{\rm sym}(\rho) &=&J_0+L_0 x+\frac{K_{\rm sym,0}}{2} x^2+\mathcal{O}(x^3),\label{esym} \,
\end{eqnarray}
where $x=\left(\frac{\rho-\rho_0}{3\rho_0}\right)$. 
The  binding energy $e_0$,   incompressibility $K_0$, and the skewness coefficient $Q_0$ can be determined from the energy per nucleon for the symmetric nuclear matter $e_{\rm snm}(\rho)$.  Likewise, the symmetry energy coefficient $J_0$, its slope $L_0$ and the curvature $K_{\rm sym,0}$ of the symmetry energy can be determined from $e_{\rm sym}(\rho)$. All the parameters appearing in the right hand side of
Eqs. (\ref{esnm}-\ref{esym}) are evaluated at the saturation density $\rho_0$. Once a functional form for $e(\rho,\delta)$ is known, the EoS for the $\beta$-equilibrated matter for near-nuclear densities can be  obtained by adding the ideal gas contributions from electrons and muons so that the conditions of chemical equilibrium and charge neutrality are satisfied. \\

The parametrization in Eqs. (27)-(29) is valid only for densities such that $x << 1$. 
These equations are employed here only to define the nuclear matter parameters. In constructing EoSs relevant for NS structure, the full equations from the various SHF and RMF models considered here have been used.

\section{Results and Discussion}
\label{results}

The values of the EoS parameters at saturation density show a diverse behaviour across the models \cite{Fortin}. In the present study, we shall investigate the correlations between the NS observables and saturation properties of matter expressed via parameters such as $K_0$, $Q_0$, $L_0$ and  $K_{\rm sym,0}$.  We shall also consider the correlations of NS properties with the presure of $\beta$-equilibrated matter at densities
in the range $1.0$ - $2.5\rho_0$. \\

\begin{figure}[htbp!]
    \includegraphics[width=\linewidth]{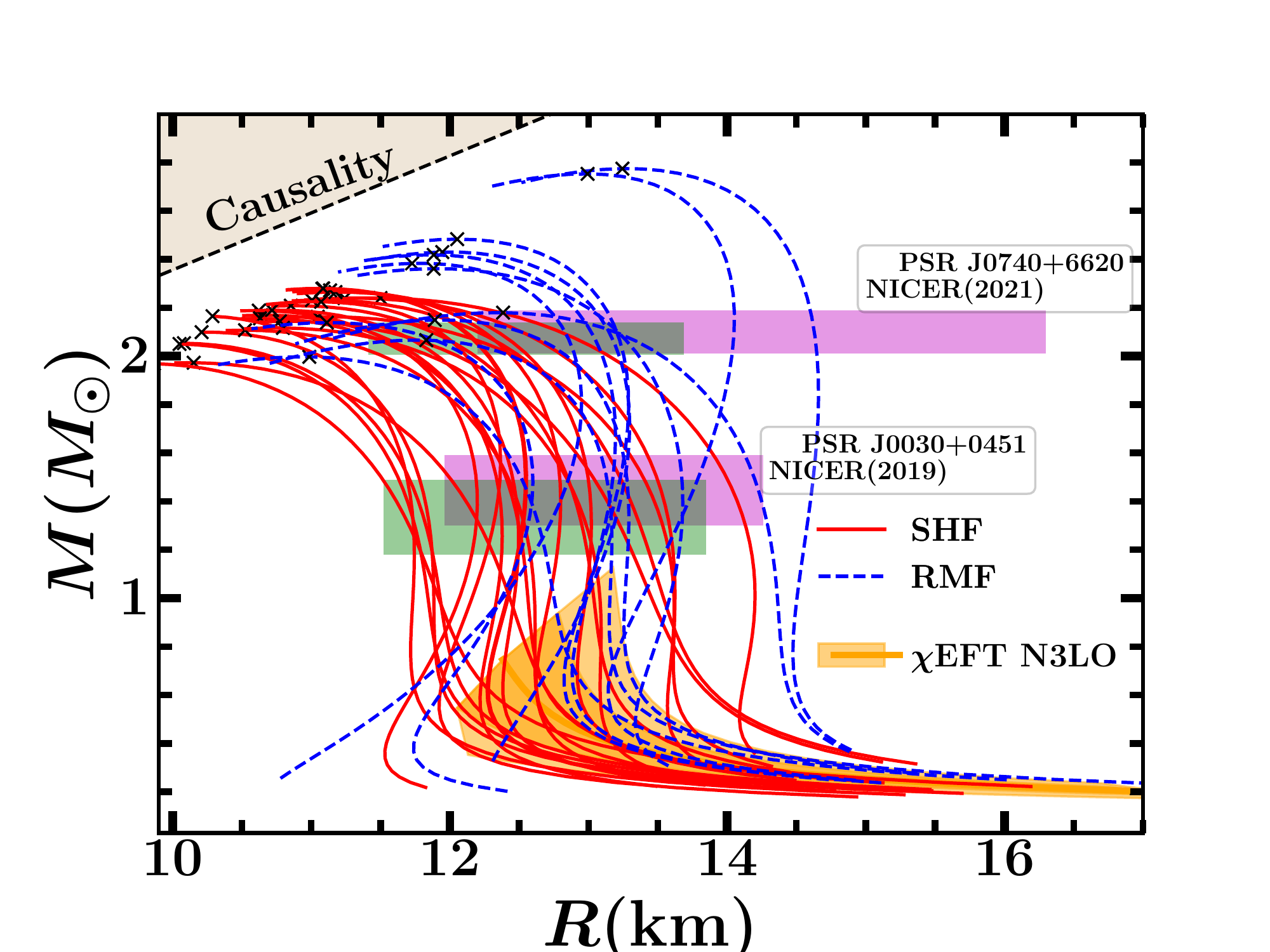}
    \caption{
The NS  mass-radius relationships obtained  for the EoSs for  SHF and RMF models  as shown in Fig.~\ref{fig:pvse}. The RMF models give higher masses than SHF models. 
{The lower 
boxes at slightly different central masses show the constraints on mass and radius from NICER (2019) data of PSR J0030+0451\cite{Miller_2019,Riley_2019}. The upper horizontal bands at the same central mass (bands shifted for clarity) show the NICER data with X-ray
Multi-Mirror (NICER XMM-Newton 2021) observations of PSR J0740+6620\cite{Fonseca_2021,Miller_2021}.} The Orange region shows results for the Chiral EFT 
$\pm 1 \sigma$ and $\pm 2 \sigma$ bounds up to twice saturation density \cite{drischler2021limiting}. 
}
    \label{fig:mr}
\end{figure}

\subsection{Masses and radii of neutron stars}

Figure \ref{fig:mr} 
shows mass-radius curves obtained by solving the 
TOV equations in general relativity ~\cite{PhysRev.55.364,PhysRev.55.374} with the various EoSs from SHF and RMF models used as input data. Results for the SHF models are shown by red lines, whereas, those from the RMF models are in blue dashed lines. The mass of a NS increases with increasing central density up to the maximum mass.
{ The `x's marked on the curves 
indicate the maximum masses of the various EoSs shown in Fig. \ref{fig:pvse}.} The orange region shows results of the chiral EFT calculations up to the maximum mass such EoSs can support. Note that without any extrapolation of the chiral EFT EoS beyond $2\rho_0$, the maximum mass falls well below even the canonical value of $1.4M_\odot$.
 \\

In Fig. \ref{fig:mr}, we have also shown the observational data and confirmed that all the SHF and RMF EoSs considered in this work could predict stars  $\geq 2M_{\odot}$. The lower boxes show the constraint on the mass and radius from the Neutron star Interior Composition Explorer (NICER) data of PSR J0030+0451. From the data analysis of this NS, Miller et al. reported that the mass-radius of 
the pulsar therein as 
$M = 1.44^{+0.15}_{-0.14} M_{\odot}$ and $R = 13.02^{+1.24 }_{-1.06} $ km~\cite{Miller_2019}, whereas Riley et al. gave it as $M=1.34^{+0.15}_{-0.16} M_{\odot}$ and $R = 12.71^{+1.14 }_{-1.19} $ km ~\cite{ Riley_2019}. 
{Cromartie et al. \cite{cromartie2020relativistic} and Antoniadis et al. \cite{anto} used radio observations from PSR J0740+6620 to determine the NS mass as $M=2.14^{+0.1}_{-0.09}$ $M_{\odot}$ and $M=2.01 \pm 0.04$ $M_\odot$, respectively. The model-averaged estimate of the mass of this pulsar 
by Fonseca et al. \cite{Fonseca_2021} gives the lower limit for the maximum mass of the NS as $2.08^{+0.07}_{-0.07}M_{\odot}$.
With NICER and X-ray Multi-Mirror (XMM-Newton) observations of 
the same pulsar, Miller et al. and Riley et al. infer the radius as $R = 13.7^{+2.6 }_{-1.5} $ km \cite{Miller_2021}  and  $R = 12.39^{+1.30 }_{-0.98} $ km  \cite{Riley_2021}, respectively.} \\

\subsection{$f$-mode and tidal deformabilities of a neutron star}

\begin{figure}[htbp!]
    \includegraphics[width=\linewidth]{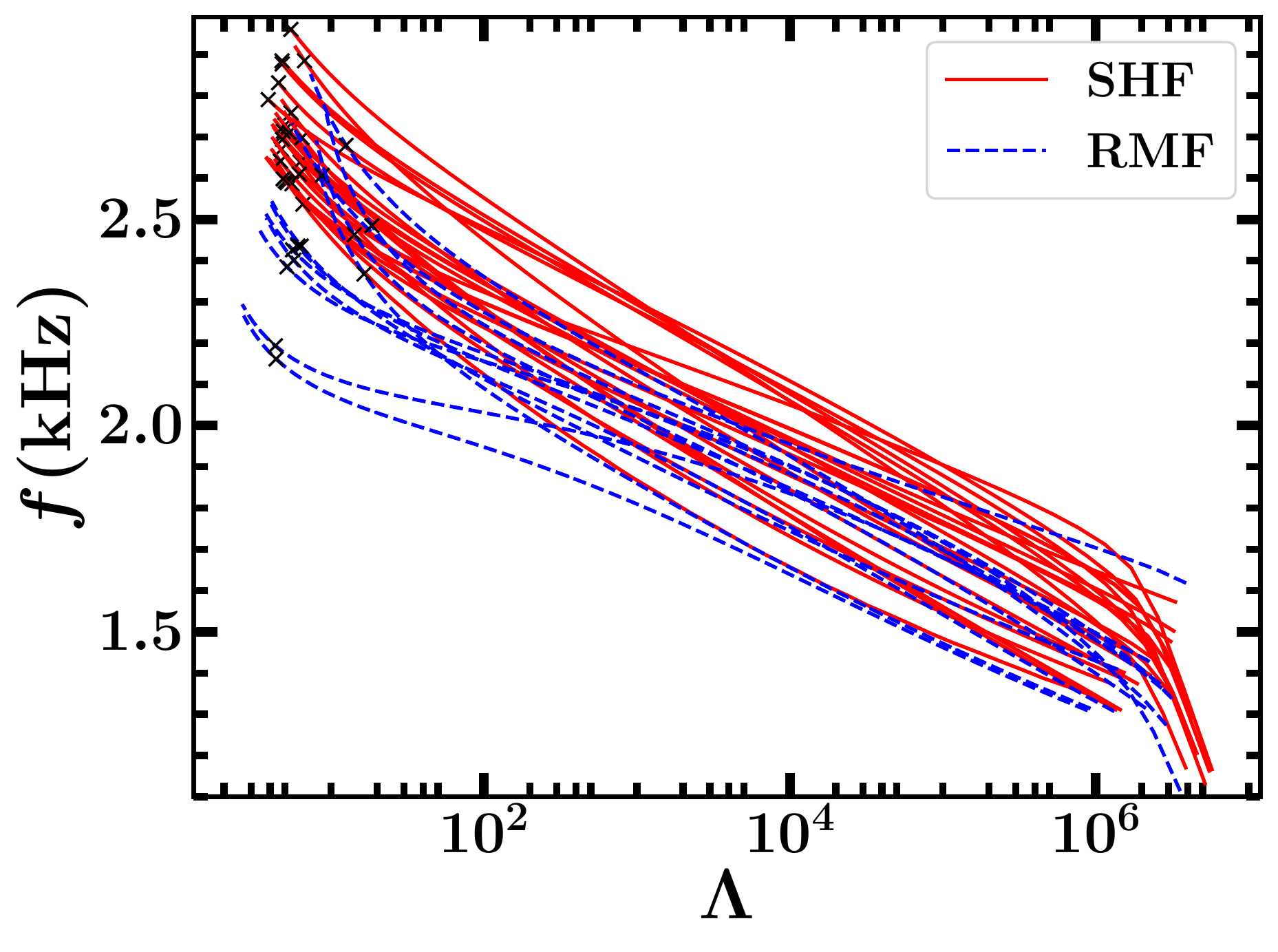}
    
    \caption{The Cowling $f$-mode frequency vs the tidal deformability parameter for the NSs obtained by varying the central density. 
    {The crosses on the curves indicate maximum mass configurations with the unstable branches extending toward the left.}}
    \label{fig:tidal}
\end{figure}
The historical event of the binary neutron star merger (BNS), GW170817, started a new era in multi-messenger astronomy.
This event gave important information about the tidal deformability of a NS. The later stage of the inspiring phase of the NS-NS merger creates a strong gravitational field, which deforms a NS’s multipolar structure.
This is quantified by the dimensionless tidal deformability parameter, 
\be
\Lambda=\frac{2}{3} k_2 C^{-5} \,,
\ee
where, $k_2$, is  the tidal Love number ~\cite{Hinderer_2009,PhysRevD.81.123016}. As $k_2 \propto R/M$, $\Lambda$ varies as the sixth power of $R$ \cite{Tuhin,zhao2018tidal}.
The tidal deformability of canonical neutron star $\Lambda_{1.4}=190 ^{+390}_{-120}$ is extracted from the combined tidal deformability of binary neutron stars \cite{Abbott:2020uma}. It is interesting to know how the tidal deformability helps constrain the $f$-mode oscillations, thereby inferring the properties and underlying EoS through an inverse approach. Figure \ref{fig:tidal} shows the dimensionless tidal deformability $\Lambda$ as a function of $f$-mode frequency obtained by varying the central density of the star. The later quantity is obtained within the Cowling approximation. Our results demonstrate that the tidal deformability has a strong dependence on $f$-mode frequency. Both the quantities $f$-mode and $\Lambda$ are anti-correlated to each other. \\

\begin{figure}[htbp!]
    \includegraphics[width=\linewidth]{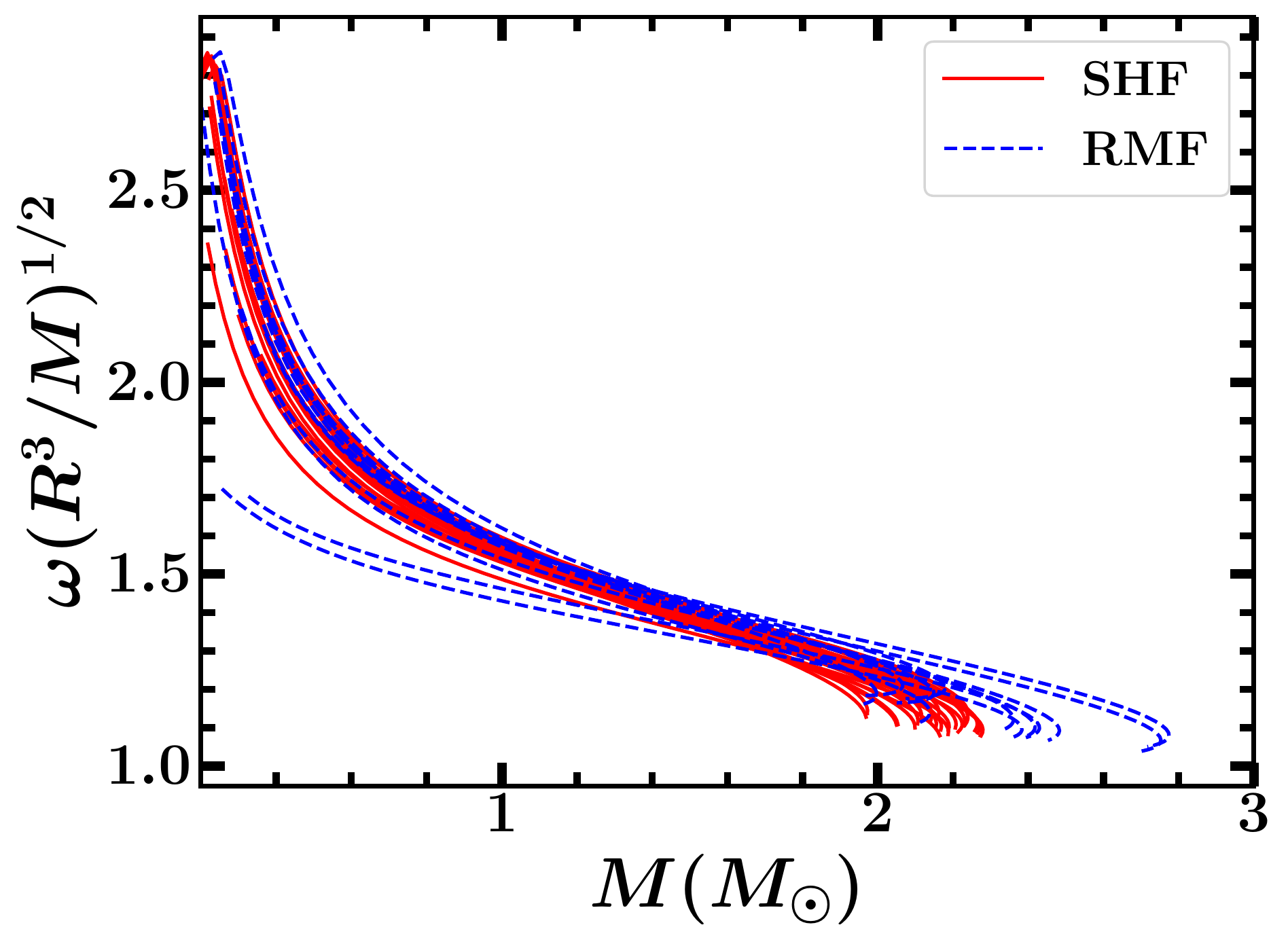}
   
    \caption{Normalised Cowling $f$-mode frequency vs NS  mass.
}
    \label{fig:norm}
\end{figure}

In Newtonian physics, the $f$-mode frequency is related to the stellar average density $M/R^3$~\cite{10.1046/j.1365-8711.1998.01840.x,CHIRENTI_2012}. In Fig.~\ref{fig:norm}, normalized
eigenvalues with average density $\omega (R^3/M)^{1/2}$, as a function of NS mass for each SHF and RMF EoSs are plotted. 
{We see that results for most of the EoSs 
roughly preserve the scaling relation  $1.1\lesssim \omega (R^3/M)^{1/2} \lesssim 1.6$ for $M > 1 M_\odot$, which is comparable to results for the Tolman VII (1.155) and Buchdal (1.237) EOSs, and larger than the incompressible EOS (0.894) 
\cite{zhao2022universal}.} Furthermore, it is interesting to notice that the normalized eigenvalues with the average density of $f$-modes are independent of the adopted EoSs. Therefore, it is not easy to distinguish the EoSs with the help of observation of stellar properties \cite{sotani11}. \\

Table \ref{tb:params1} summarizes results of the essential NS properties: canonical and maximum masses (in $ M_{\odot}$),  their corresponding radii R (in km), the dimensionless tidal deformability $\Lambda$, compactness $C$, $f$-mode frequency (in  kHz) and its damping time for QNMs. \\

\begin{table*}
\caption{Values of neutron star properties: maximum mass
($M_{\odot}$), radius at the  maximum mass $ R_{\rm {max}}$ (km) and radius at the canonical mass  $R_{1.4}$ (km), compactness
parameter $C_{1.4}$ and 
 the dimensionless tidal deformability for canonical mass, $\Lambda_{1.4}$. The  values of  the $f$-mode and $p_1$-mode frequencies (kHz) are obtained
 by using the Cowling approximation and  a full general relativistic treatment  for the EoSs considered in this work. The values of  damping time $\tau_{1.4}$ (s)  are also listed.}
 \label{params1}

\label{tb:params1}
\scalebox{1.0}{
\begin{tabular}{|l|ccccc|ccc|ccccc|c}
\hline
&&&&&&\multicolumn{3}{c|}{Cowling appr.}
&\multicolumn{5}{c|}{General Relativity}\\
\hline
\multicolumn{1}{|l|}{EoS}
&\multicolumn{1}{c}{$M_{\rm{max}}$}
&\multicolumn{1}{c}{$R_{\rm {max}}$}
&\multicolumn{1}{c}{$R_{1.4}$ }
&\multicolumn{1}{c}{$C_{1.4}$}
&\multicolumn{1}{c|}{$ \Lambda_{1.4}$}
&\multicolumn{1}{c}{$f_{\rm{max} }$}
&\multicolumn{1}{c}{ $f_{1.4}$ }
&\multicolumn{1}{c|}{ $\nu_{1.4}^{p_1}$ }
&\multicolumn{1}{c}{$f_{max}$}
&\multicolumn{1}{c}{$f_{1.4}$ }
&\multicolumn{1}{c}{$\tau_{1.4}^{f}$}
&\multicolumn{1}{c}{$\nu_{1.4}^{p_1}$ }
&\multicolumn{1}{c|}{$\tau_{1.4}^{p_1}$}\\
\hline
    SKa &      2.208 &     10.853 &     12.906 &      0.160 &          568.694 &                2.670 &                2.114 &              5.867 &           2.382 &           1.684 &             0.251 &                 5.391 &               4.666 \\  SKb &      2.188 &     10.621 &     12.198 &      0.169 &          474.683 &                2.702 &                2.212 &              6.551 &           2.415 &           1.752 &             0.232 &                 6.006 &               5.626 \\  SkI2 &      2.163 &     11.047 &     13.474 &      0.153 &          775.670 &                2.639 &                1.976 &              5.449 &           2.352 &           1.572 &             0.289 &                 4.964 &               4.401 \\ SkI3 &      2.240 &     11.241 &     13.545 &      0.153 &          789.865 &                2.586 &                1.979 &              5.613 &           2.302 &           1.565 &             0.291 &                 5.140 &               5.354 \\      SkI4 &      2.169 &     10.639 &     12.363 &      0.167 &          472.443 &                2.713 &                2.212 &              6.503 &           2.415 &           1.754 &             0.232 &                 5.973 &               5.514 \\      SkI5 &      2.240 &     11.499 &     14.074 &      0.147 &         1000.071 &                2.537 &                1.868 &              5.184 &           2.282 &           1.478 &             0.328 &                 4.718 &               4.909 \\      SkI6 &      2.190 &     10.717 &     12.481 &      0.166 &          490.162 &                2.694 &                2.192 &              6.443 &           2.397 &           1.737 &             0.236 &                 5.923 &               5.657 \\      SLY2 &      2.053 &     10.082 &     11.773 &      0.176 &          310.385 &                2.877 &                2.388 &              6.783 &           2.578 &           1.916 &             0.195 &                 6.243 &               4.245 \\   SLY230a &      2.099 &     10.209 &     11.822 &      0.175 &          332.692 &                2.831 &                2.371 &              6.960 &           2.513 &           1.892 &             0.200 &                 6.418 &               5.187 \\      SLY4 &      2.050 &     10.050 &     11.694 &      0.177 &          299.732 &                2.884 &                2.406 &              6.861 &           2.591 &           1.931 &             0.192 &                 6.314 &               4.273 \\      SLY9 &      2.156 &     10.630 &     12.457 &      0.166 &          453.543 &                2.728 &                2.223 &              6.351 &           2.436 &           1.770 &             0.227 &                 5.852 &               4.984 \\      SkMP &      2.107 &     10.522 &     12.487 &      0.166 &          482.124 &                2.758 &                2.183 &              6.051 &           2.467 &           1.747 &             0.234 &                 5.536 &               4.068 \\      SKOp &      1.973 &     10.152 &     12.116 &      0.171 &          363.930 &                2.884 &                2.294 &              6.163 &           2.587 &           1.855 &             0.208 &                 5.633 &               3.160 \\    KDE0v1 &      1.969 &      9.866 &     11.615 &      0.178 &          266.465 &                2.961 &                2.444 &              6.712 &           2.664 &           1.978 &             0.184 &                 6.169 &               3.406 \\     SK255 &      2.144 &     10.774 &     13.135 &      0.157 &          589.750 &                2.710 &                2.082 &              5.519 &           2.413 &           1.670 &             0.256 &                 5.059 &               3.787 \\     SK272 &      2.231 &     11.006 &     13.304 &      0.155 &          647.483 &                2.641 &                2.052 &              5.550 &           2.351 &           1.637 &             0.266 &                 5.103 &               4.480 \\        Rs &      2.116 &     10.796 &     12.921 &      0.160 &          594.582 &                2.698 &                2.084 &              5.697 &           2.424 &           1.667 &             0.257 &                 5.195 &               3.942 \\     BSk20 &      2.165 &     10.288 &     11.719 &      0.176 &          322.916 &                2.790 &                2.389 &              7.081 &           2.506 &           1.904 &             0.198 &                 6.536 &               5.560 \\     BSk21 &      2.278 &     11.080 &     12.539 &      0.165 &          520.829 &                2.599 &                2.189 &              6.787 &           2.314 &           1.719 &             0.241 &                 6.282 &               8.148 \\     BSk22 &      2.265 &     11.175 &     13.015 &      0.159 &          633.050 &                2.589 &                2.090 &              6.191 &           2.307 &           1.646 &             0.263 &                 5.711 &               6.638 \\     BSk23 &      2.271 &     11.136 &     12.799 &      0.162 &          576.473 &                2.593 &                2.135 &              6.459 &           2.307 &           1.678 &             0.253 &                 5.968 &               7.295 \\     BSk24 &      2.279 &     11.086 &     12.547 &      0.165 &          522.583 &                2.598 &                2.188 &              6.784 &           2.311 &           1.717 &             0.241 &                 6.280 &               8.173 \\     BSk25 &      2.224 &     11.072 &     12.344 &      0.167 &          485.592 &                2.609 &                2.228 &              7.135 &           2.314 &           1.744 &             0.234 &                 6.613 &               9.451 \\    
    \hline
    BSR2 &      2.383 &     11.728 &     12.796 &      0.162 &          743.777 &                2.435 &                2.025 &              6.117 &           2.142 &           1.583 &             0.285 &                 5.676 &               8.671 \\      BSR6 &      2.430 &     11.946 &     13.229 &      0.156 &          825.331 &                2.401 &                1.966 &              5.425 &           2.117 &           1.551 &             0.297 &                 5.027 &               6.546 \\       GM1 &      2.361 &     11.884 &     13.680 &      0.151 &          909.926 &                2.435 &                1.931 &              5.618 &           2.159 &           1.516 &             0.311 &                 5.158 &               6.727 \\       NL3 &      2.774 &     13.245 &     14.601 &      0.142 &         1284.080 &                2.162 &                1.793 &              5.095 &           1.909 &           1.400 &             0.368 &                 4.699 &               8.311 \\     NL3$\omega\rho$03 &      2.753 &     12.994 &     13.722 &      0.151 &          948.459 &                2.193 &                1.953 &              6.167 &           1.935 &           1.504 &             0.316 &                 5.833 &              16.933 \\       TM1 &      2.179 &     12.384 &     14.249 &      0.145 &         1059.437 &                2.368 &                1.855 &              5.139 &           2.086 &           1.465 &             0.334 &                 4.705 &               5.573 \\       DD2 &      2.418 &     11.884 &     13.151 &      0.157 &          695.868 &                2.426 &                2.062 &              6.183 &           2.148 &           1.612 &             0.274 &                 5.779 &               9.386 \\      DDH$\delta$ &      2.138 &     11.113 &     12.591 &      0.164 &          589.032 &                2.606 &                2.135 &              6.714 &           2.299 &           1.671 &             0.255 &                 6.165 &               7.855 \\     DDME2 &      2.483 &     12.053 &     13.190 &      0.157 &          715.231 &                2.385 &                2.058 &              6.293 &           2.106 &           1.601 &             0.278 &                 5.923 &              11.674 \\      IOPB-I &      2.149 &     11.891 &     13.288 &      0.156 &          694.552 &                2.463 &                2.048 &              5.849 &           2.139 &           1.612 &             0.274 &                 5.455 &               7.185 \\        G3 &      1.997 &     10.986 &     12.586 &      0.164 &          465.918 &                2.679 &                2.208 &              6.178 &           2.369 &           1.760 &             0.230 &                 5.724 &               4.817 \\ FSUG &      2.066 &     11.830 &     13.155 &      0.157 &          634.019 &                2.485 &                2.097 &    6.147 &           2.175 &           1.645 &             0.263 &                 5.787 &               8.172 \\ 
\hline
\end{tabular}}
\end{table*}

\subsection{Pearson's correlation coefficients}

Linear correlations between any given pair of quantities is 
measured by Pearson's correlation coefficient  
defined as~\cite{Benesty2009}  %
\begin{equation}
r(a,b)=\frac{\sigma_{ab}}{\sqrt{\sigma_{aa}\sigma_{bb}}}\, \\
\label{eq:cc}
\end{equation}
with the covariance, $\sigma_{ab}$, given by
\begin{equation}
\sigma_{ab}=\frac{1}{N_m}\sum_i a_i b_i -\left(\frac{1}{N_m}\sum_i a_i\right
)\left(\frac{1}{N_m}\sum_i b_i\right ) \,,
\end{equation}
where the index $i$ runs over the number of models $N_m$\cite{Brandt97}.
In what follows, $a_i$ and $b_i$ correspond to the  NS properties for a fixed mass 
obtained for the different models.   
A correlation coefficient close to unity in absolute value indicates a strong linear relation between the pair of quantities that are considered. \\

\begin{figure*}[htbp!]
\begin{minipage}{0.5\textwidth}
    \includegraphics[width=\linewidth]{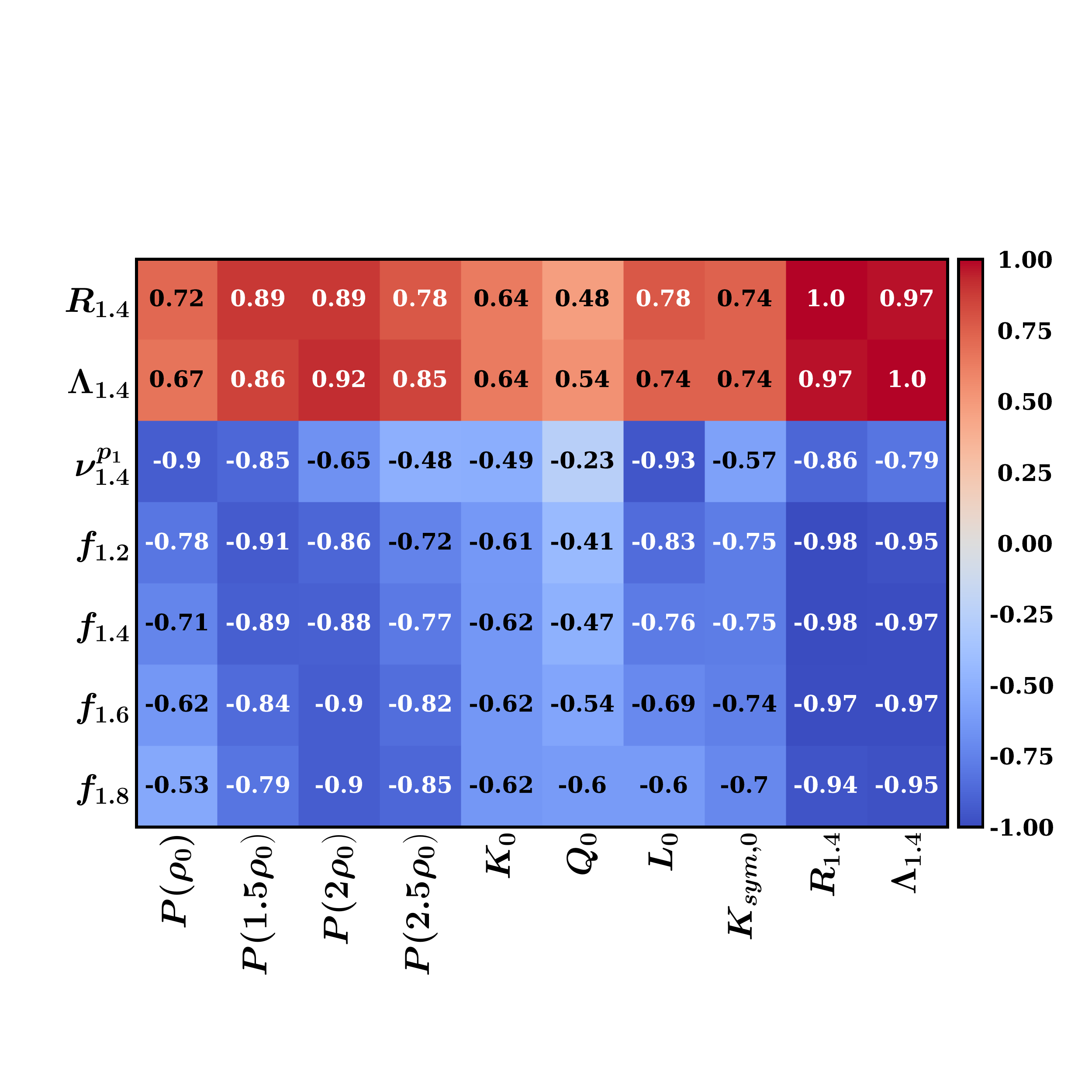}
    \subfloat{(a)Cowling approximation}
    \end{minipage}%
    \begin{minipage}{0.5\textwidth}
    \includegraphics[width=\linewidth]{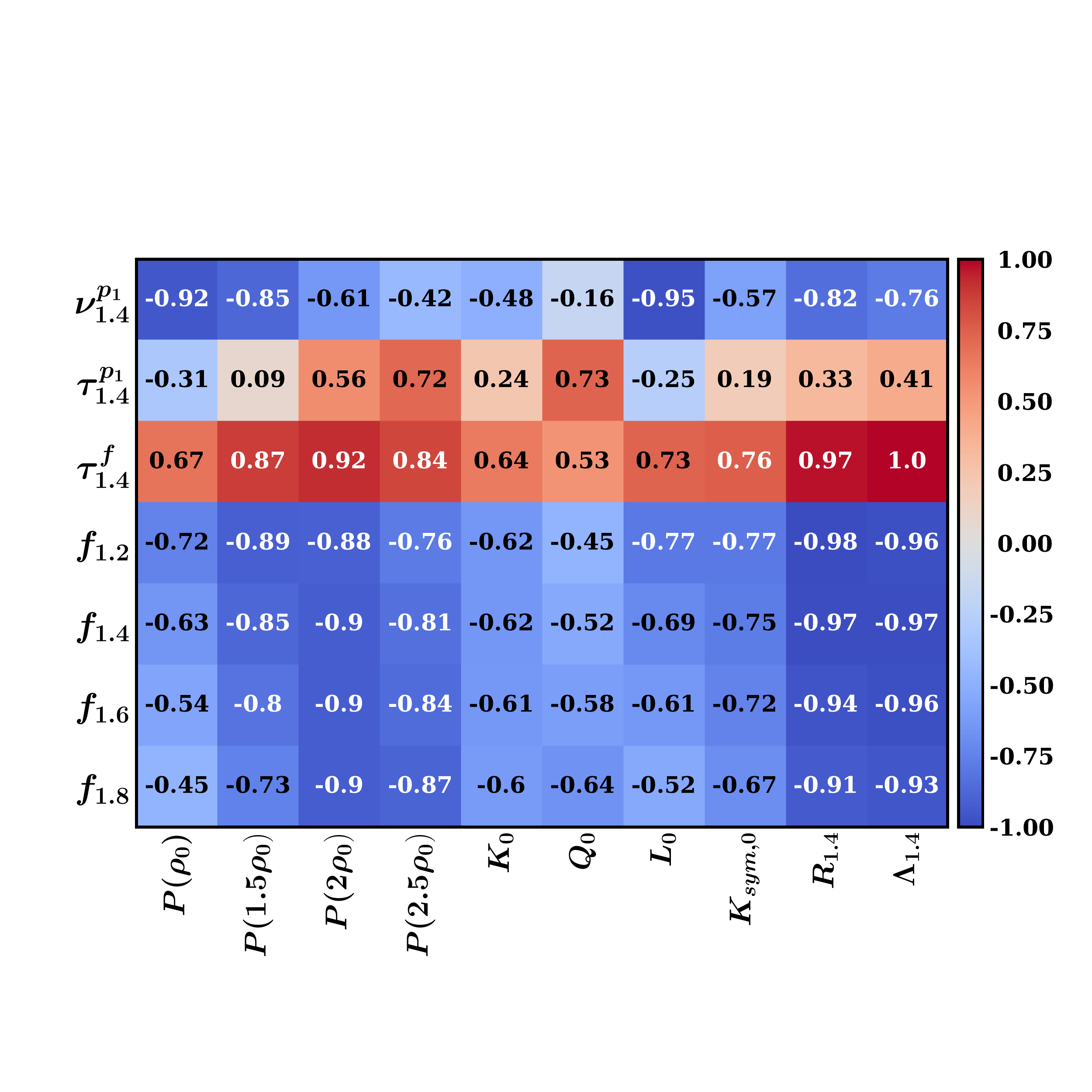}
    \subfloat{(b) Full GR}
\end{minipage}
\caption{ Heat map depicting  the correlation of nuclear matter parameters to neutron star properties.  The $f$- and $p_1$-mode frequencies are obtained within (a)  the Cowling approximation and (b)  full GR.   }
    \label{fig:oscillate}
\end{figure*}

We present the correlation matrices as heat maps in Figs.  \ref{fig:oscillate}(a) and \ref{fig:oscillate}(b) for NS properties, nuclear matter properties, and pressure for beta-equilibrated matter. The NS properties considered are the $f$- and $p_1$-mode frequencies, radii, and tidal deformability at fixed masses. For 
comparison, the $f$- and $p_1$-mode frequencies obtained with  the Cowling approximation and full GR are also shown. The inferences for Cowling approximation as shown in Fig. \ref{fig:oscillate}(a) are as follows: \\

\begin{enumerate}
\item
{ The $p-$ mode frequency for a NS with $1.4M_\odot$ is strongly correlated
with $L_0$.  The radius, tidal deformality and $f-$ mode frequency for a
NS with $1.4M_\odot$ are moderately correlated with $L_0$ and $K_{\rm sym,0}$ \citep{Tuhin,Tuhin2020}. 
consideration.
}

\item  
The beta-equilibrated matter
pressure corresponding to two times the saturation density has a substantial correlation with $f_{1.4}$, whereas the correlation becomes marginally stronger at higher masses; and

\item The $f$-mode frequency also strongly correlates with the corresponding tidal deformability and radius \cite{Pradhan2022a}. However, the $p_1$-mode  frequency is only moderately correlated with tidal deformability, but displays a strong correlation with the slope $L_{0}$ of symmetry energy curve.

\end{enumerate}

A similar analysis is made for $f$- and $p_1$-mode values obtained by the full GR approach as shown in Fig. \ref{fig:oscillate}(b). The damping time that appears only in full GR is correlated with various quantities associated with the EoS in the following manner:
\begin{enumerate}
 \item The damping time $\tau_{1.4}^f$ is strongly correlated with beta-equilibrated matter
 pressure at $2\rho_0$, $R_{1.4}$, and $\Lambda_{1.4}$; and 
\item The damping time $\tau_{1.4}^{p_1}$ is moderately correlated with beta-equilibrated matter pressure at $2.5\rho_0$, and $Q_{0}$. 
\end{enumerate}

\subsection{Correlations involving pressure}
\begin{figure*}[htbp!]
    \includegraphics[width=\linewidth]{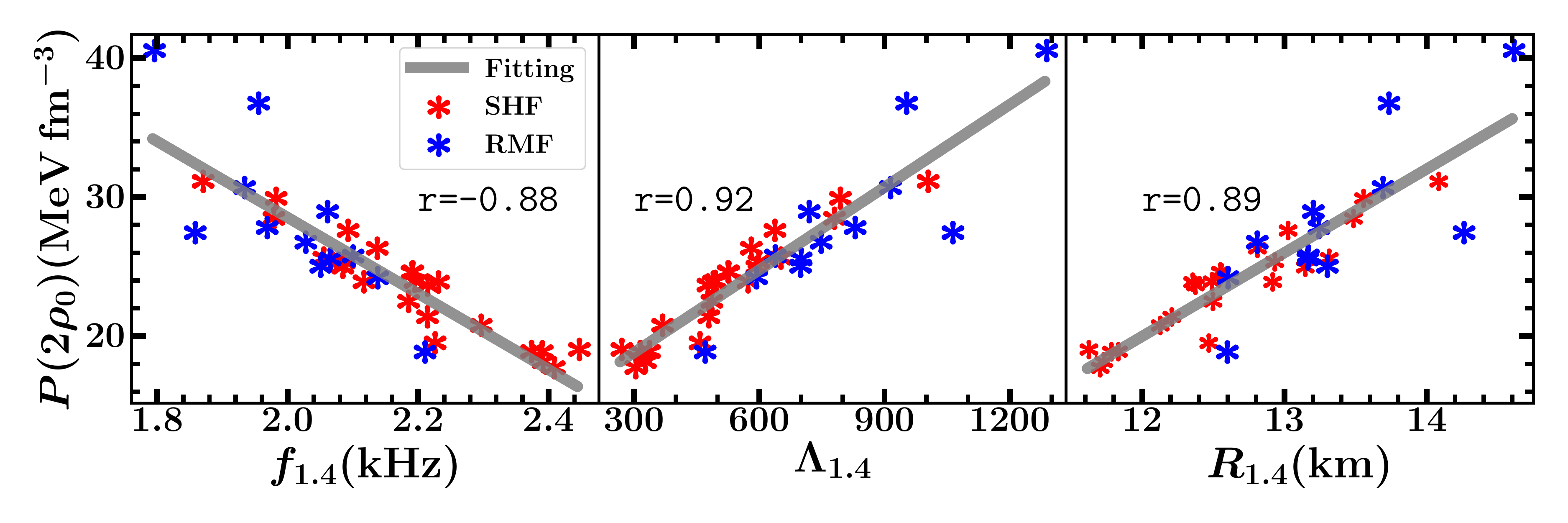}
   
    \caption{Pressure at twice the saturation density  plotted against $f$- mode frequency in Cowling approximation, tidal deformability and radius calculated for 1.4$M_{\odot}$ neutron  star using SHF and RMF models.   The  thick solid grey line gives the best fit. Values of Pearson's coefficient $r$ are as indicated. }
    \label{fcowfit}
\end{figure*}

The analysis of GW170817 has provided pivotal information about the tidal deformability and pressure as a function of density \cite{GW170817_radii}.
The behaviour of the pressure at twice nuclear saturation density is measured at $21.84^{+16.85}_{-10.61}$ MeV fm$^{-3}$ at the $90\%$ confidence limit \cite{GW170817_radii}. In Fig. \ref{fcowfit} we plot the $f_{1.4}$, $\Lambda_{1.4}$, and $R_{1.4}$  for NS with  $1.4M_\odot$ mass versus pressure at $2\rho_0$ for the sets of SHF (red stars)  and RMF (blue stars)  models considered. The thick solid grey line in the figure is obtained by linear regression and the correlation coefficients are displayed for each case considered. The
expressions obtained  by fitting the variation of  pressure $P(2\rho_0)$ with NS properties using linear regression are: 

\be
\frac{P(2\rho_{0})}{\text{MeV fm}^{-3}}= 83.24-27.35\,\, \frac{f_{1.4}}{\text{kHz}} \\ 
\ee

\be
\frac{P(2\rho_{0})}{\text{MeV fm}^{-3}}= 12.87+0.02\,\, \Lambda_{1.4}\\
\ee

\be
\frac{P(2\rho_{0})}{\text{MeV fm}^{-3}}= -52.27+6.02 \,. \frac{R_{1.4}}{\text{km}}\\
\ee

\begin{figure}[htbp!]
    \includegraphics[width=\linewidth]{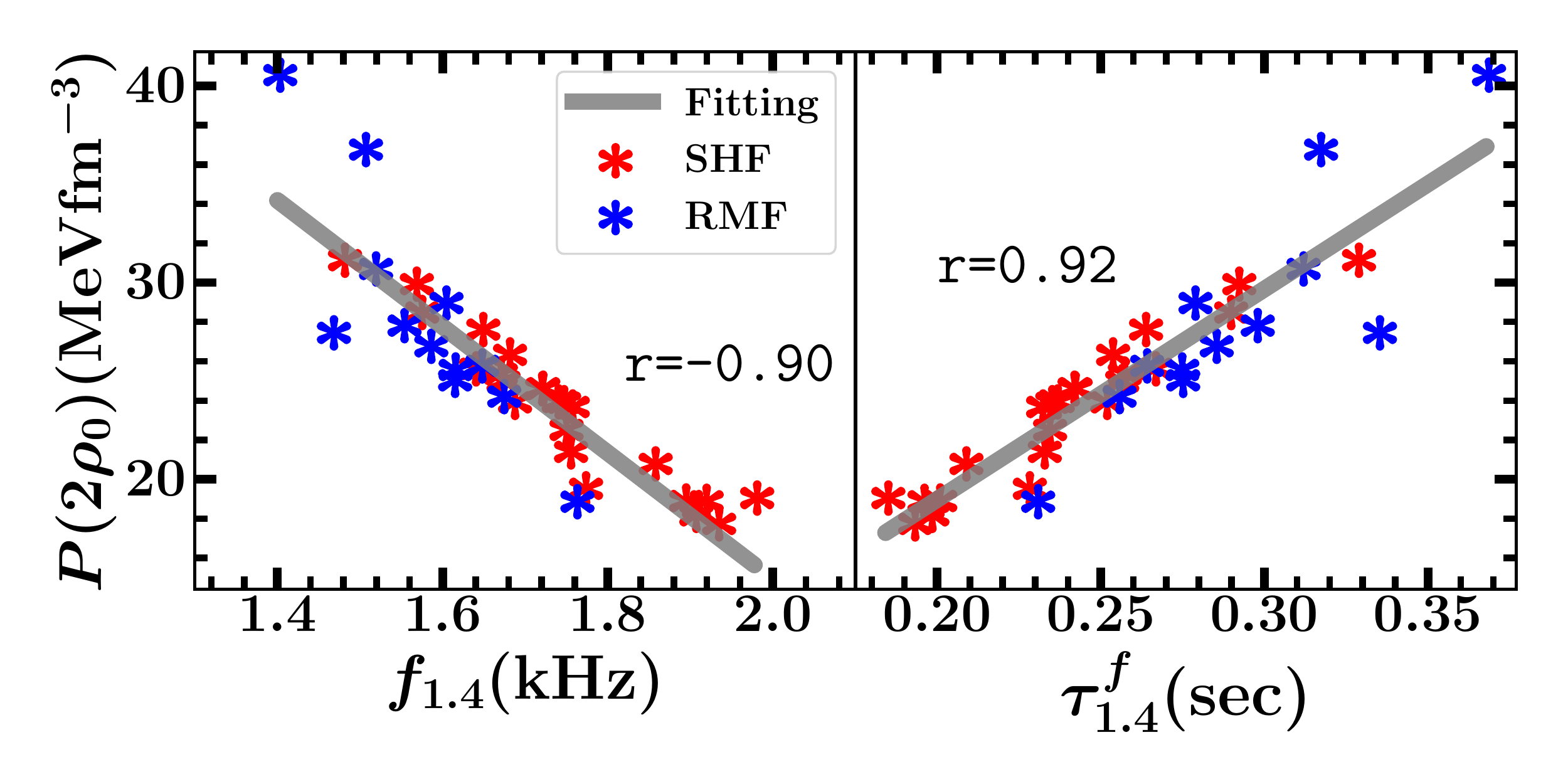}
   
    \caption{ Same as Fig \ref{fcowfit}, but , for the $f$- mode frequency   and damping time within full GR.   Values of Pearson's coefficient $r$ are as indicated.}
    \label{fgrfit}
\end{figure}

From the heat map  Fig. \ref{fig:oscillate}(a), we noticed that the correlations between the $f$-mode and pressure  $P(\rho_0)$ of beta-equilbrated matter decrease with the increasing
neutron star mass 
and are not significantly strong to make a meaningful statement.
However, the correlations of $f$-mode frequencies with $P(2\rho_0) $  for 
{beta-equilibrated matter}  are quite  strong   (r =  -0.86 to -0.90). The correlations are marginally higher for larger NS mass.  \\

Figure \ref{fgrfit} shows the frequency and damping time of $1.4M_\odot$ star, which is obtained with full GR calculations done for all EoSs considered in this work. We note that the correlation between $P(2\rho_0)$ and $f_{1.4}$ is almost identical as in the case of the Cowling approximation. 
Thus, we look into the correlations of the damping time (imaginary part of the eigenvalue) with the pressure $P(2\rho_0)$ of beta-equilibrated matter.
The behaviour is qualitatively similar to that for $\Lambda_{1.4}$ as shown in Fig. \ref{fcowfit} exhibiting a
strong correlations between $P(2\rho_0)$ and $\tau_{1.4}$. 
We have obtained following  expressions  from linear-regressions: 
\be
\frac{P(2\rho_{0})}{\text{MeV fm}^{-3}}= 79.10-32.09 \,\, \frac{f_{1.4}^{GR}}{\text{kHz}}  \\ 
\ee
\be 
\frac{P(2\rho_{0})}{\text{MeV fm}^{-3}}= -2.41+106.93 \,\, \frac{\tau_{1.4}^{f}}{\text{sec}}  \,. \\
\ee

\begin{figure}[htbp!]
    \includegraphics[width=\linewidth]{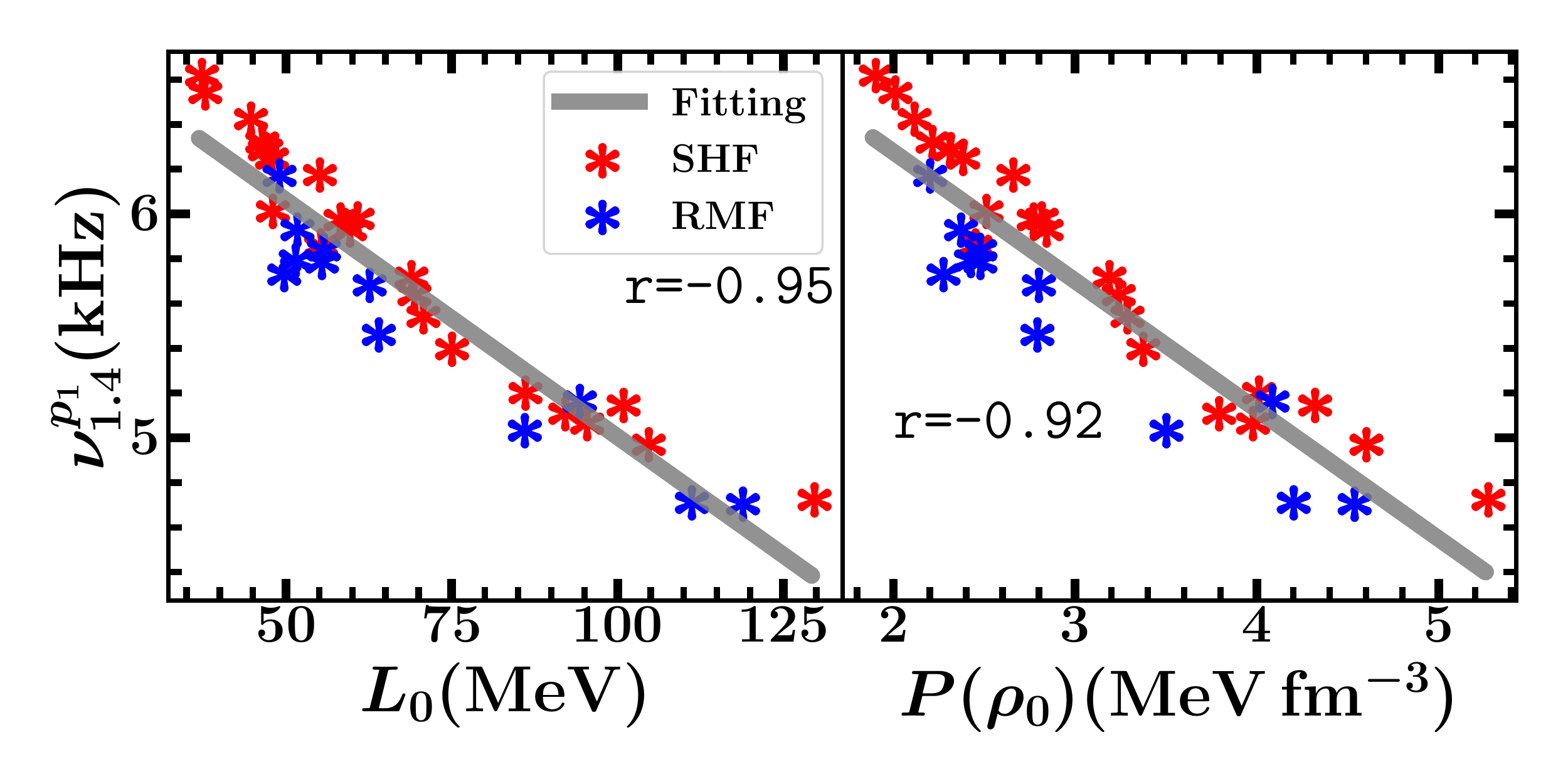}
    \caption{The $p_1$-mode frequency $\nu_{1.4}^{p1}$  vs symmetry energy  slope parameter  $L_{0}$  and  pressure at saturation density obtained  for NS mass $1.4M_{\odot}$  within  full GR.
    Values of Pearson's coefficient $r$ are as indicated.}
    \label{pmodefullgrfit}
\end{figure}

Finally, in Fig. \ref{pmodefullgrfit}, we plot the frequency of $p_1$-mode versus pressure at saturation density and slope of the symmetry energy $L_0$ for our representative sets of SHF  and RMF models. It is interesting to note that the $\nu_{1.4}^{p_1}-L_{0}$  correlation is stronger than the $f_{1.4}-L_{0}$ correlation.
Similar trends are
observed for the cases of   $\nu_{1.4}^{p_1}-P(\rho_{0})$ and  $f_{1.4}-P(\rho_{0})$ correlations  (see  Figs. \ref{fig:oscillate}(a) and \ref{fig:oscillate}(b)). These correlations indicate that the $p_1$-mode frequency is more sensitive to the isovector EoS parameters $L_{0}$ and pressure at saturation density. The correlations involving $p_1$-mode frequency take the form
\be
\frac{\nu_{1.4}^{p_1}}{\text{kHz}}= 7.12-0.02 \,\, \frac{L_{0}}{\text{MeV}} \\ 
\ee

\be
\frac{\nu_{1.4}^{p_1}}{\text{kHz}}= 7.43-0.58 \,\, \frac{P(\rho_0)}{\text{MeV fm}^{-3}} \,. \\
\ee

In summary, we notice from the various  correlations that the information containing the  $f$- and $p_1$-mode frequencies are complementary. These modes probe different density regions of the NS. The $f$-mode frequency is more sensitive to the region at $2\rho_0$, and the $p_1$-mode frequency is more sensitive to the region of the NS in the vicinity of the outer core. The reason for this is that the $p_1$-mode oscillation peaks stronger at the stellar surface than the $f$-mode, see Fig. \ref{fig:amplitudes}.

\section{Critical analysis of correlations}

Correlations of $R_{1.4}$ and $\Lambda_{1.4}$ with $L_0$ and $K_{\rm sym,0}$ have also been reported in 
several earlier investigations 
\citep{Pradhan2022a,Carson2019,Tsang2020,Reed2021,Reed2021A, Biswas:2021yge}.  The EoSs employed in these studies range 
from non-parametric~\cite{Reed2021A}, parametric~\cite{Tsang2020}, and physics-based models~\cite{Tuhin2020,Reed2021}. While in some cases, constraints imposed by bulk nuclear properties have been used~\cite{Tuhin2020,Tsang2020,Reed2021}, other cases have 
also imposed those provided by the bulk properties of 
nuclei through the constraints on nuclear matter parameters assuming them to be independent of each other. The sample sizes used in these works also differ. In what follows, we first explore the dependence on the  sample sizes for the EoSs used in our work, and thereafter recount the main findings in the above mentioned studies. \\ 


\subsection{Dependence on the sample sizes}

\begin{figure}[htbp!]
    \includegraphics[width=\linewidth]{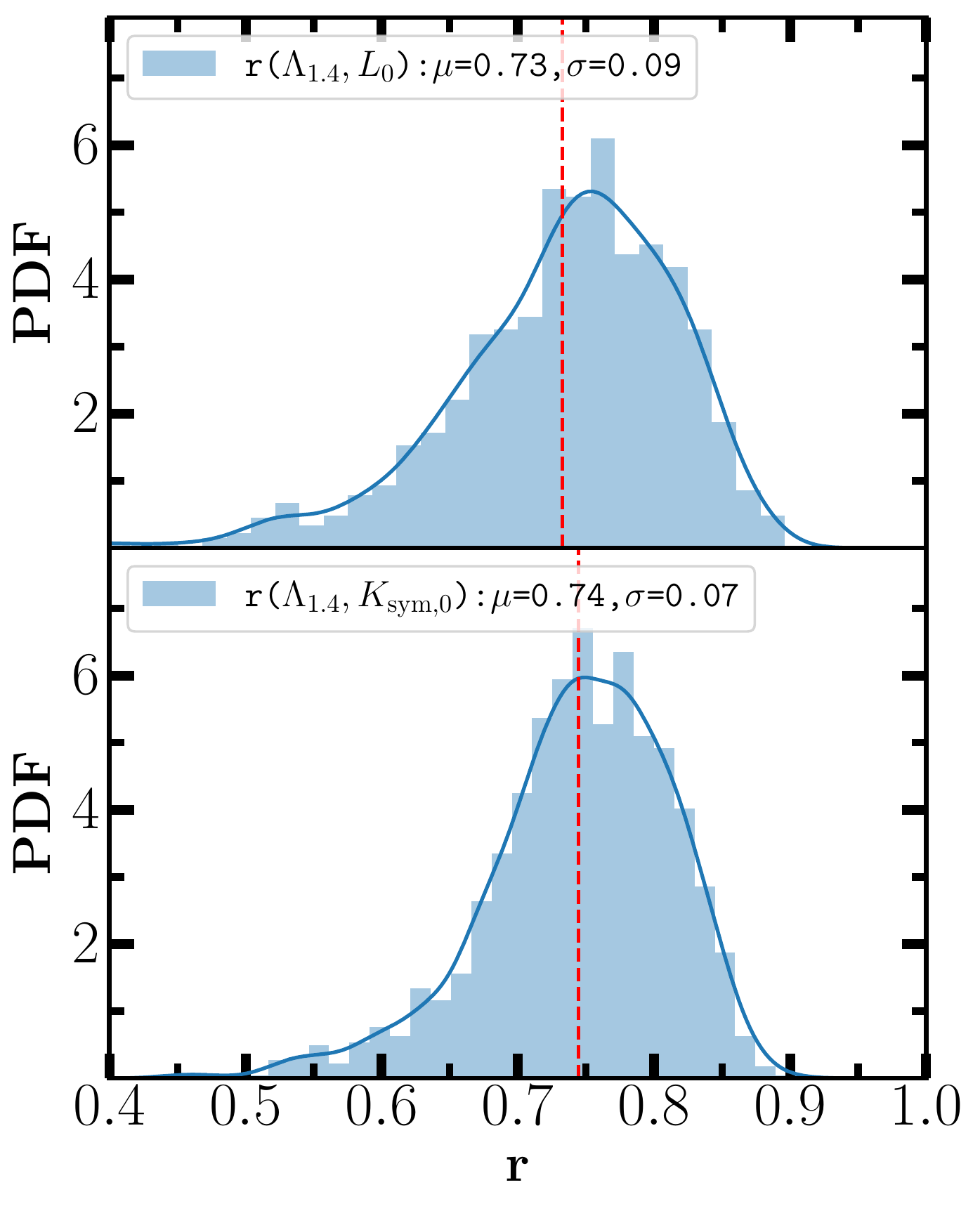}
   
    \caption{The distributions of correlation coefficients for $\Lambda_{1.4}$ with $L_0$ and $K_{\rm sym,0}$  obtained using 1500 samples. Each sample consists of a set of 20 models selected randomly from the pool of 35 SHF and RMF models considered in the previous section. The dashed red vertical line indicates the mean value.}
    \label{pdf}
\end{figure} 

The correlation systematics, up to a large  extent, is also driven by the selection criteria of models. Even not too large a number of models but  having diverse behaviour, if randomly selected, may serve the purpose\cite{Margueron2018}. The various models obtained by systematically varying one or few of the base model introduces a strong bias and usually result in large correlations under consideration.

\begin{table}[!htb] \setlength{\tabcolsep}{12pt}
\centering
\caption{The mean and standard deviations for the distributions of
correlation coefficients for $\Lambda_{1.4}$ and $R_{1.4}$  with $L_0$ and $K_{\rm sym,0}$ obtained with ensembles of  10, 20 and 30 models. Results for the correlations of $L_0$ and $K_{\rm sym,0}$ with $P(\frac{2}{3}\rho_0)$ and $P(2\rho_0)$ are also presented.}
\centering
\begin{tabular}{m{0.5 cm}m{0.5cm}m{1.1 cm}m{1.1 cm}m{1.1 cm}}
                  & \multicolumn{4}{l}{}  \\
\hline
\hline & & & \text{Number of EoSs}\\

                  & &   \text{10}&\text{20}  &\text{30} \\
\hline
\multirow{2}{*}{} & $L_{0}$ & 0.7$\pm$0.18 &  0.73$\pm$0.09 &  0.74$\pm$0.04\\           
                  $\Lambda_{1.4}$ & $K_{sym,0}$ &  0.73$\pm$0.15 &  0.74$\pm$0.07 & 0.74$\pm$0.03\\ 
                  \hline
\multirow{2}{*}{}  & $L_{0}$ &  0.76$\pm$0.15 &  0.78$\pm$0.07 &  0.78$\pm$0.03\\
$R_{1.4}$ & $K_{\rm sym,0}$ &  0.71$\pm$0.17 &  0.73$\pm$0.08  & 0.74$\pm$0.04\\
\hline
\multirow{2}{*}{} & $P(\frac{2}{3}\rho_{0})$ &  0.94$\pm$0.04 &  0.94$\pm$0.02 & 0.94$\pm$0.01\\
 $L_{0}$ &  $P(2\rho_{0})$ &  0.61$\pm$0.22 &  0.62$\pm$0.12 &  0.62$\pm$0.06\\
\hline
\multirow{2}{*}{} & $P(\frac{2}{3}\rho_{0})$ &  0.34$\pm$0.36 &  0.38$\pm$0.19 & 0.39$\pm$0.09\\
$K_{\rm sym,0}$  & $P(2\rho_{0})$ &  0.74$\pm$0.15 &  0.73$\pm$0.08 &  0.73$\pm$0.04\\
\hline
\hline
\end{tabular}
\label{probability}
\end{table}

We investigate how the selection of EoSs affects the correlations of $\Lambda_{1.4}$ and $R_{1.4}$ with $L_{0}$ and $K_{\rm sym,0}$. For this purpose,  sets of 10, 20 and 30 EoS  randomly selected from our
pool of 35 EoSs corresponding to the SHF and RMF models are constructed. The distribution of correlation coefficients are studied using 1500 such samples.
In Fig. \ref{pdf}, results obtained for sets of 20 EoSs are displayed.
The mean and standard deviations for the distributions of correlation coefficients corresponding to different  number of EoS models 
are presented in Table \ref{probability}.
The standard deviations are found to decrease with increase in number of EoSs in a  sample, but the mean values show only marginal changes. \\

The distributions of correlation coefficients for the correlations of $L_{0}$ and $K_{\rm sym,0}$ with the pressures $P(\frac{2}{3}\rho_{0})$ and $P(2\rho_{ 0})$ are also subjected to a similar analysis. The trend for $K_{\rm sym,0}- P$
correlations is opposite to that obtained for the $L_{0} - P$ correlations. The $K_{\rm sym,0}$ is better
correlated with $P(2\rho_{0})$ than with $P(\frac{2}{3}\rho_{0})$ since as the density increases, the contribution to the pressure from the $K_{\rm sym,0}$ increases faster compared to the ones for $L_{0}$.

\subsection{Correlation findings in earlier works}


The correlations of  various NS properties with nuclear matter parameters were studied using  large set  of RMF models in Ref. \cite{Pradhan2022a}.
The  coupling constants that appear  in the   effective Lagrangian were  determined  from various nuclear matter parameters. 
Values of these nuclear matter parameters  were drawn  from  uniform
distributions with their ranges guided by available data on  finite
nuclei assuming them to be uncorrelated with each other.  The correlations of $R_{1.4}$ and $\Lambda_{1.4}$ with $L_0$   were found to be very weak. A similar  investigation performed in Ref. \cite{Carson2019}  using
metamodels \cite{Margueron2018} also yielded weak correlations  of the combined tidal deformability
$\tilde\Lambda$ with $L_0$. However, the  correlations of $\tilde\Lambda$ with $K_{\rm sym,0}$  was stronger than those with $L_0$. The values of $\tilde\Lambda $ are  expected to be
strongly correlated with $\Lambda_{1.4}$ \cite{Malik2018,De18}. The $\Lambda_{1.4}$
in Ref. \cite{Tsang2020} was found to be moderately sensitive to the $L_0$
and $K_{\rm sym,0}$ within a metamodel. The
nuclear matter parameters in Ref. \cite{Tsang2020}  were considered to have Gaussian distributions
with  the mean and standard deviations obtained  by combining the  results
from the SHF and RMF models fitted to some selected properties of finite nuclei. The parameters were further constrained
using  current  estimates for $\Lambda_{1.4}=190^{+390}_{-120}$ that 
led to very weak correlations in the values of $L_0$ and $K_{\rm sym,0}$. Reference \cite{Biswas:2021yge} presented an interesting study on the
correlations of $R_{1.4}$  with $L_0$ and $K_{\rm sym,0}$ obtained
by combining the impact of PREX-II and Radio/NICER/XMM-Newton’s
mass–radius measurement of PSR J0740+6620 on the dense matter EoS. The EoSs employed  in Ref. \cite{Biswas:2021yge} were obtained using  Taylor
expansion up to $1.25\rho_0$  and piece wise polytrope (PP) were used at
higher densities.
It was shown that the $R_{1.4}$ - $L_0$ correlations somewhat  improves  with the  increase in the $L_0$ - $K_{\rm sym,0}$ correlations
which is enforced  by demanding  higher accuracy in the neutron-skin 
measurement of PREX-II for the $^{208}$Pb nucleus. The strong correlations of $\Lambda_{1.4}$ with
$L_0$ reported in Refs. \cite{Tuhin2020,AgrawalCRC}  for the
SHF models stem from the  model parameters informed by the bulk properties of
finite nuclei. Similar conclusions were drawn for the RMF models fitted
to  data on finite nuclei \cite{Reed2021}. \\

\begin{table}[!htb] \setlength{\tabcolsep}{12pt}
\caption{
Overview of the correlations of $R_{1.4}$ and $\Lambda_{1.4}$ with
the symmetry energy parameters $L_0$ and $K_{\rm sym,0}$ obtained using
different models. The model parameters  are determined  either in
terms of the nuclear matter parameters or by fitting directly some
selected properties of finite nuclei. The  nuclear matter parameters
are  assumed to have uniform (U), Gaussian (G) or Hybrid  (H)
distributions. Furthermore,  these parameters may be uncorrelated(Unc)
or  correlated (Cor); see the text for details.
The  `W', `M' and `S' denote the  weak ($r \leq 0.5$), moderate ($0.5 < r \leq 
0.75$) and strong ($r \geq 0.75$), correlations. \\
}
\begin{tabular}{m{1.8cm}m{1.8cm}m{0.2cm}m{0.2cm}m{0.5cm}} 
\hline \hline
Model & Dist. NMPs& $L_0$ & $K_{\rm sym,0}$ & Ref.\\
\hline
RMF& U-Unc& W & --   & \cite{Pradhan2022a}\\
Metamodel& U-Unc& W & S &\cite{Carson2019}\\
Metamodel & G-Unc & M & M &\cite{Tsang2020}\\
Taylor+PP & H-Unc & W & W & \cite{Biswas:2021yge} \\
SHF & G-Unc & M & M & \cite{Tuhin2020} \\
SHF& G-Cor &  S & S & \cite{Tuhin2020}\\
SHF& Finite nuclei&  S & S& \cite{AgrawalCRC}\\
RMF & Finite nuclei &  S & S &\cite{Reed2021} \\
\hline \hline 

\end{tabular}
\label{tab:corr}
\end{table}


The outcomes described above are summarized in Table \ref{tab:corr} for the sake of convenience. It seems that the correlations of $R_{1.4}$ and $\Lambda_{1.4}$ with various symmetry energy parameters are sensitive to the joint  distribution of the nuclear matter parameters  that enter into the investigation. The  nuclear matter parameters drawn from uncorrelated uniform distributions tend to yield weaker correlation of $R_{1.4}$ and $\Lambda_{1.4}$ with $L_0$ and  $K_{\rm sym,0}$. On the other extreme, models fitted to finite nuclei yield stronger correlations. The bulk   properties of finite nuclei not only constrain the values of few low order nuclear matter parameters, but also  introduce the correlations among them \cite{Tuhin2020}. 
For instance, the  values of iso-scalar giant monopole resonance energies
impose stringent constraints on the slope of incompressibility coefficient
at crossing density ($\sim 0.7\rho_0$)  that can be expressed in terms of
$K_0$ and $Q_0$ \cite{Khan2012,Khan2013}.
The correlations between $L_0$ and $K_{\rm sym,0}$ have also been extensively studied using mean field models such as SHF and RMF  \cite{Tews2017,Mondal2017,Li2020} and these correlations were found to be strong. The existence of universal correlations among various symmetry energy parameters $ J_0, L_0 $ and $K_{\rm sym,0}$ which are independent of the details of the nuclear forces used in the calculation has been demonstrated \cite{Holt2018}.
Given the above conflicting findings in the literature, it appears that
the dependence of $R_{1.4}$ and $\Lambda_{1.4}$ on $L_0$ and other
higher order nuclear matter parameters are not yet 
conclusive. Detailed analysis must be done in this regard for arriving at some meaningful conclusions.
\\\
\color{black}
\section{Damping times of QNMs in General Relativity}
\begin{figure}[htbp!]
    \includegraphics[width=\linewidth]{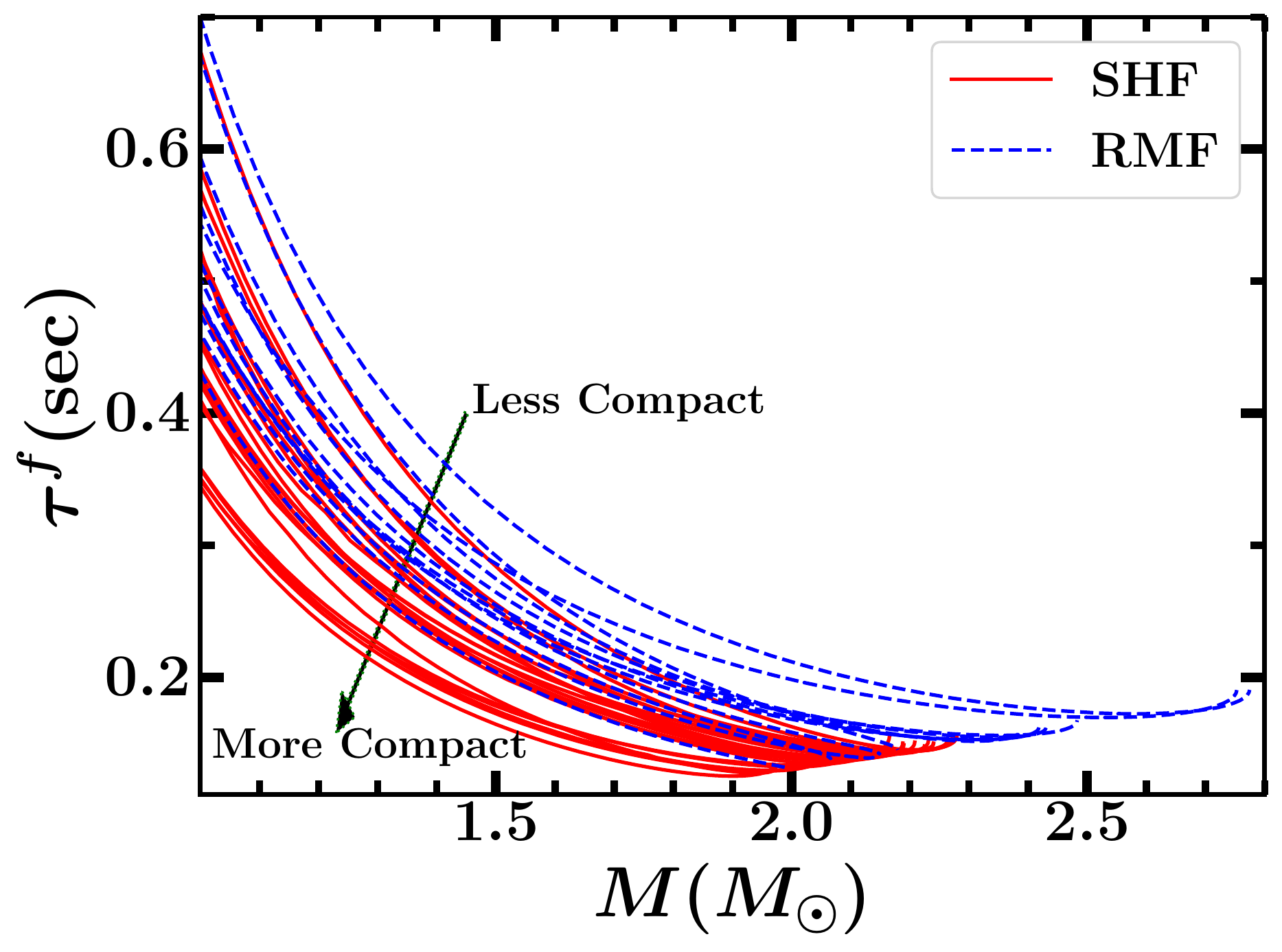}
   
    \caption{Variation of the damping time for $f$-mode oscillations vs neutron star mass.}
    \label{damp}
\end{figure} 

We have computed the damping time of the 
$f$- modes 
constructed from 
the EoSs of RMF and SHF models. The dependence of the damping time on the NS mass is shown in Fig. \ref{damp}. For a typical NS, the frequency of the $f$-mode is around 1-3 kHz, whereas their damping times $\tau_f$ can be a few 
{tenths of} a seconds \cite{Kokkotas_1999}. The damping time decreases as the neutron star mass increases. The variation of the curves shows the nature of the softer and stiffest EoSs, which can be understood from Table \ref{tb:params1}. For example, the soft EoS, such as G3, has the shortest damping time, 0.230 s, and the highest $f$-mode frequency, 1.760 kHz of $1.4M_{\odot}$. The nature of a short damping time would be to enable sizeable emission of gravitational radiation \cite{Dscale,Burgio11}. \\

\begin{figure}[htbp!]
    \includegraphics[width=\linewidth]{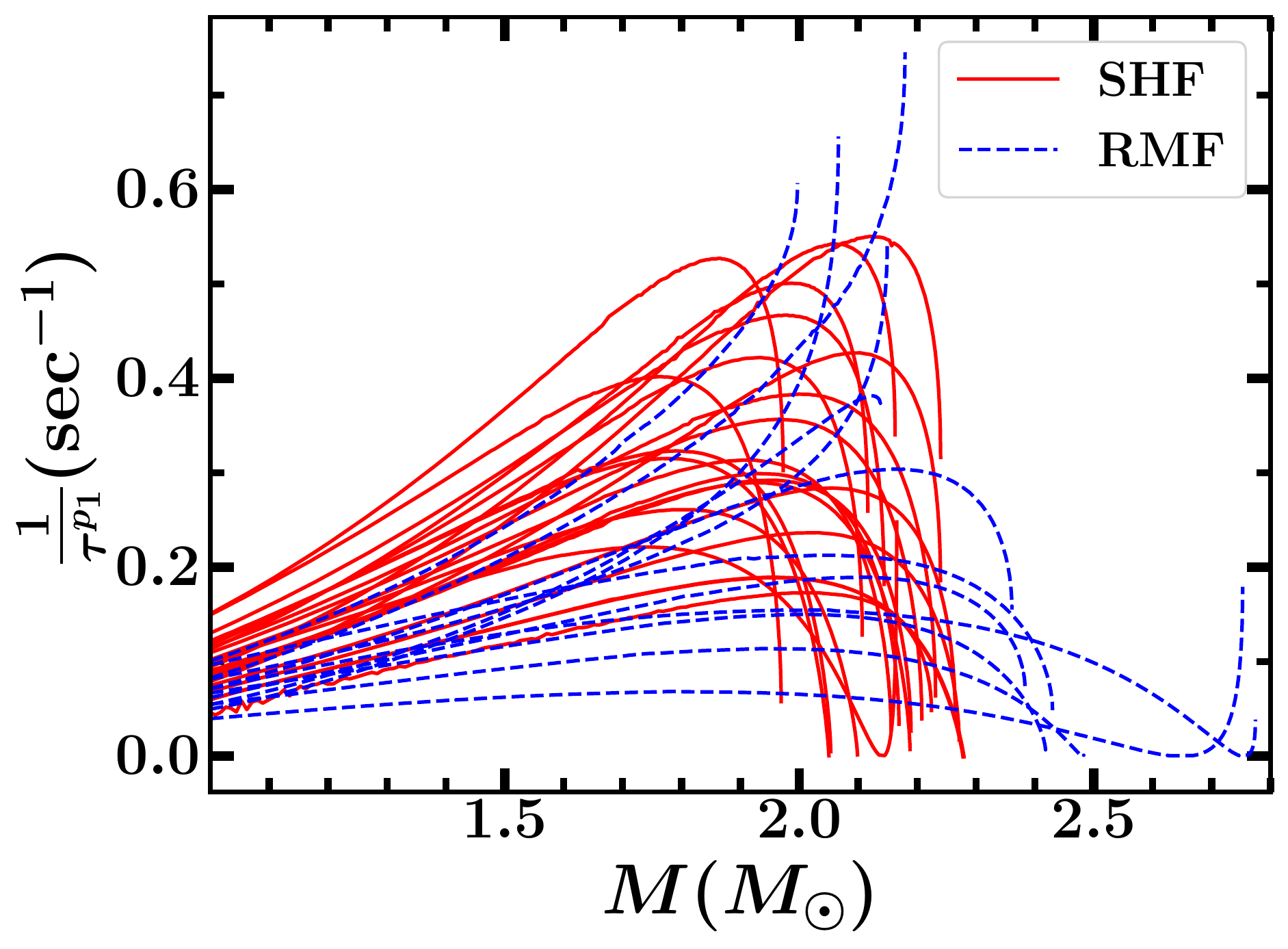}
   
    \caption{
    {Variation of the imaginary part $\omega$ of $p_{1}$-mode oscillations (inverse damping time) vs neutron star mass.}}
    \label{dampp}
\end{figure} 
%
{Fig. \ref{dampp} shows the the imaginary part of the eigen-frequency (inverse damping time) of the $p_{1}$-mode. We found the damping times of the $p_{1}$-mode $\tau^{p_1}$ is significantly longer than that of the $f$-mode $\tau^{f}$. The damping times are sensitive to the structure of neutron stars and are smallest for stars slightly lower than the maximum mass configurations. Results for $1.4M_\odot$ stars can also be seen in the rightmost column of Table \ref{tb:params1}.}

\subsection*{Accuracy of the Cowling approximation}

\begin{figure*}[htbp!]
    \centering
    \begin{minipage}{0.5\textwidth}
        \includegraphics[width=\linewidth]{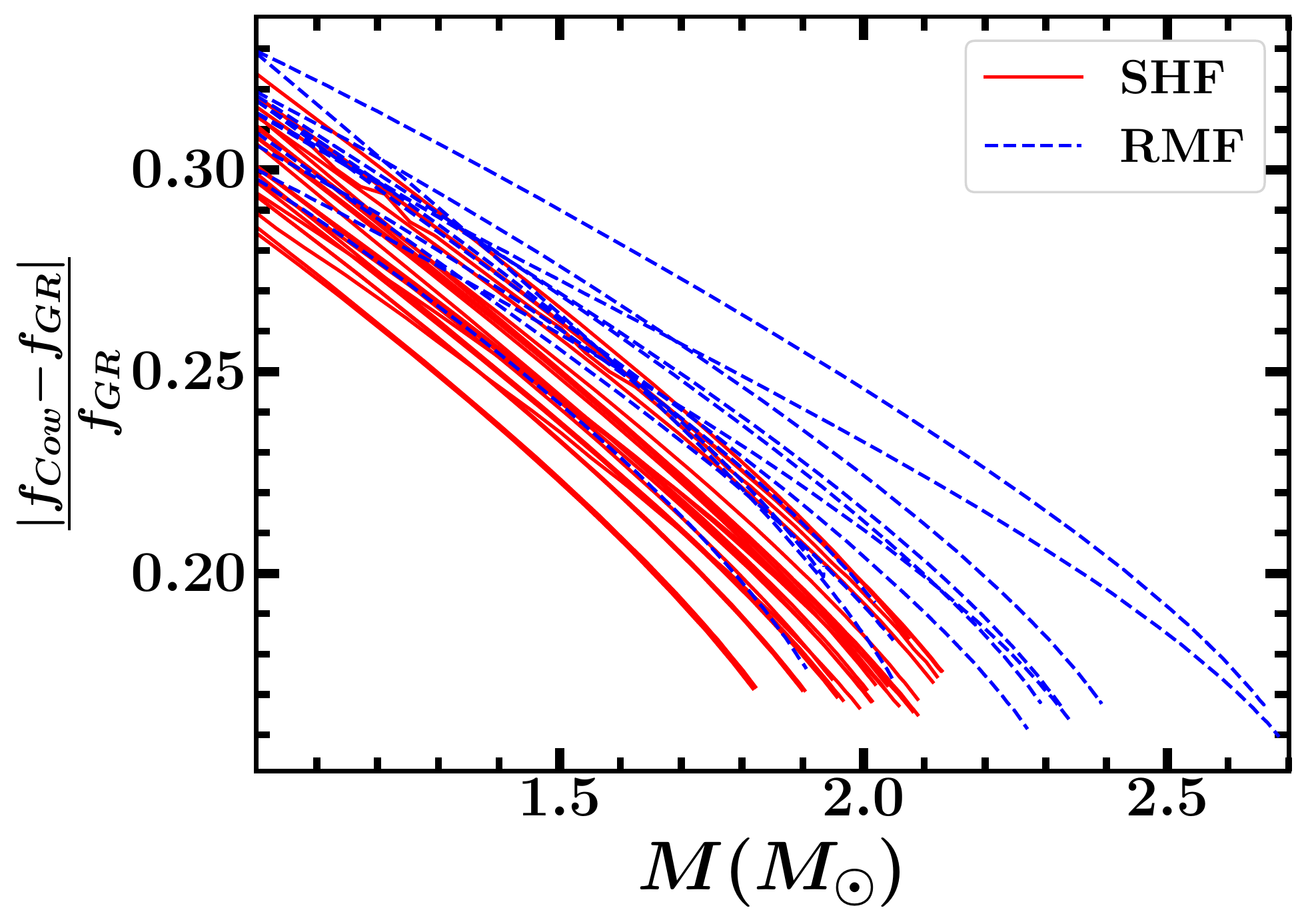}
        \subfloat{(a) $f$-mode}
    \end{minipage}%
    \begin{minipage}{0.5\textwidth}
        \includegraphics[width=\linewidth]{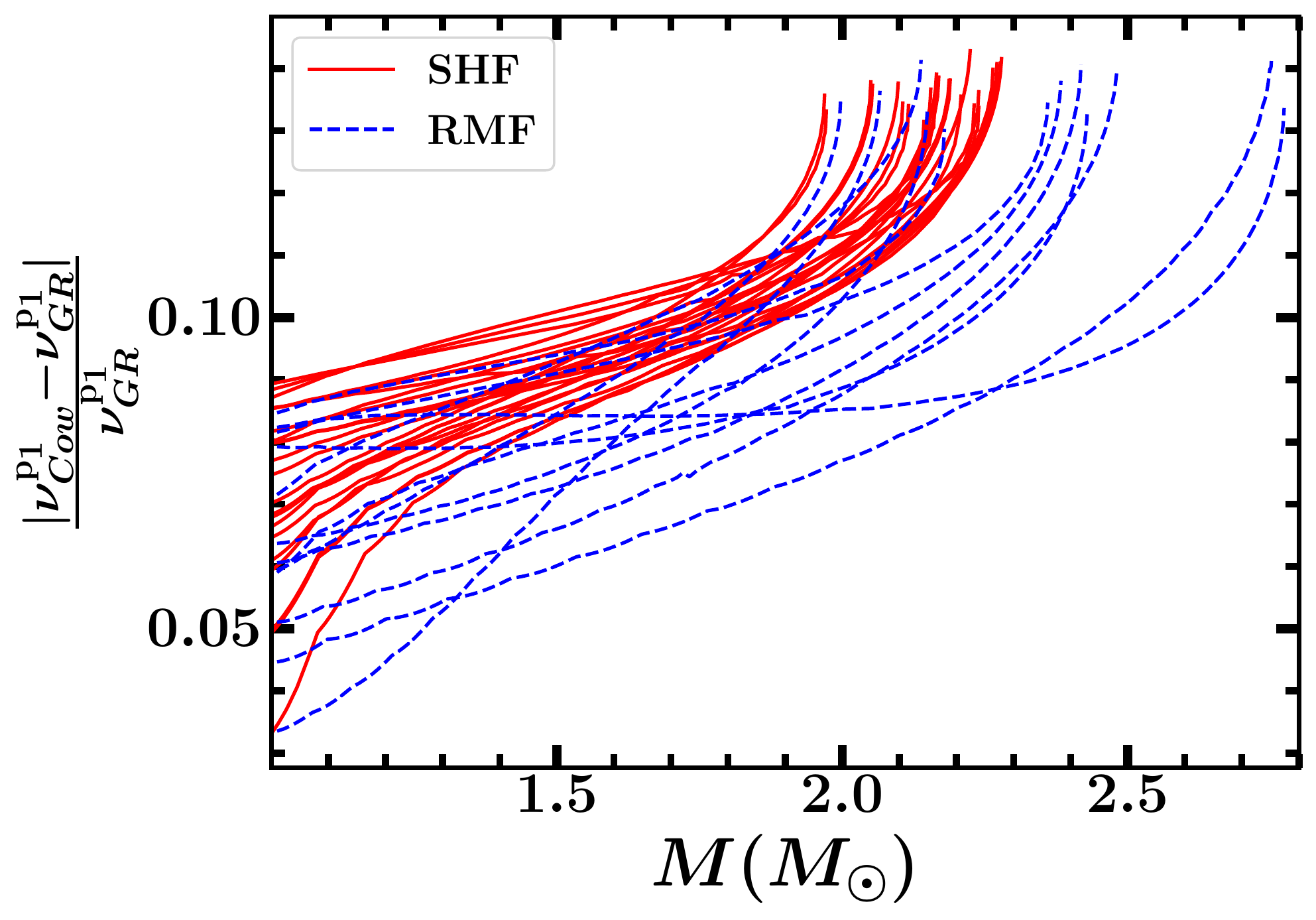}
        \subfloat{(b) $p_1$-mode}
     \end{minipage}
    \caption{ Relative deviations of the $f$- and $p_1$-mode frequencies obtained using  the Cowling approximation with those for full GR. 
    }
    \label{fig:error}
\end{figure*}

\begin{figure*}[htbp!]
    \centering
    \begin{minipage}{0.5\textwidth}
        \includegraphics[width=\linewidth]{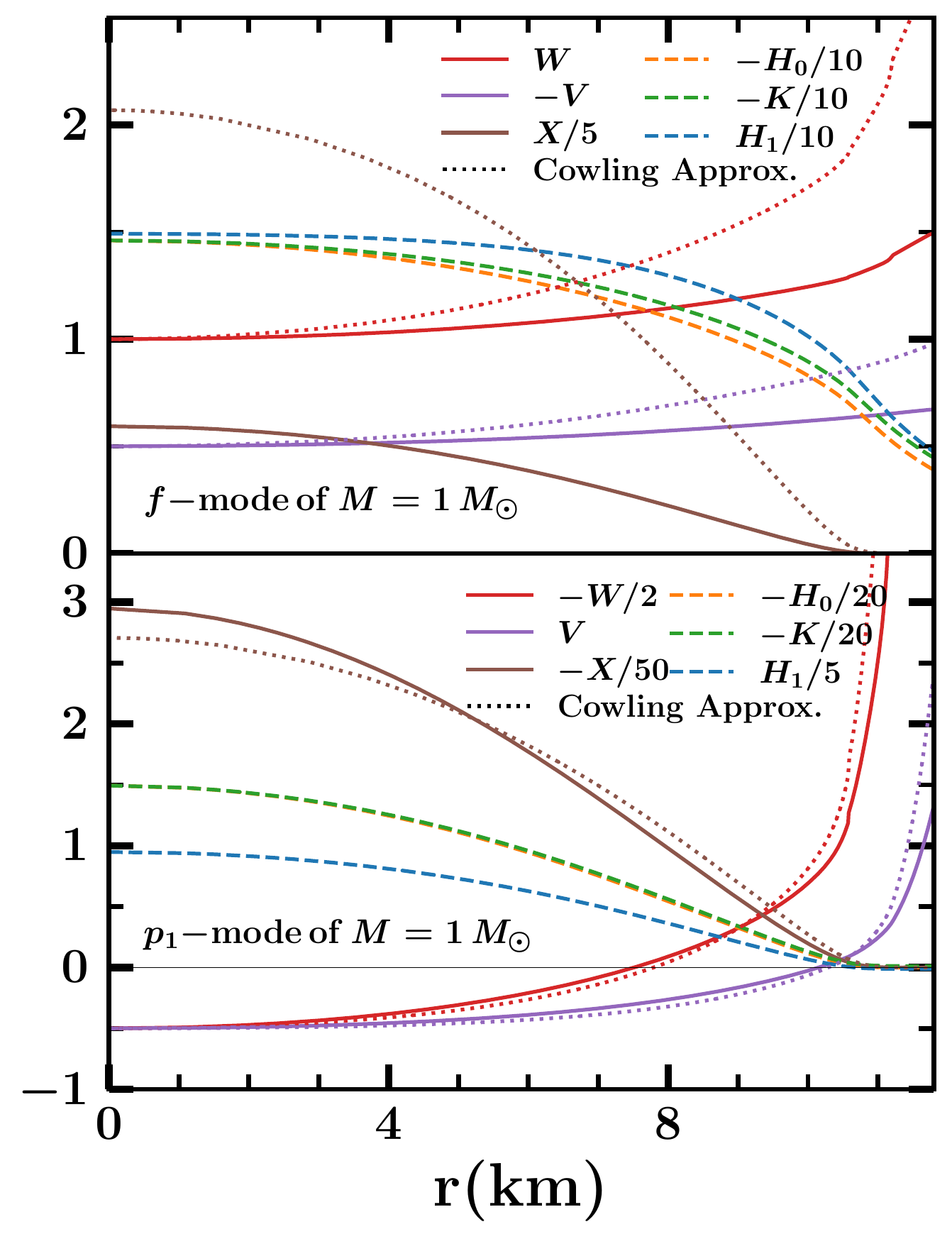}
    \end{minipage}%
    \begin{minipage}{0.5\textwidth}
        \includegraphics[width=\linewidth]{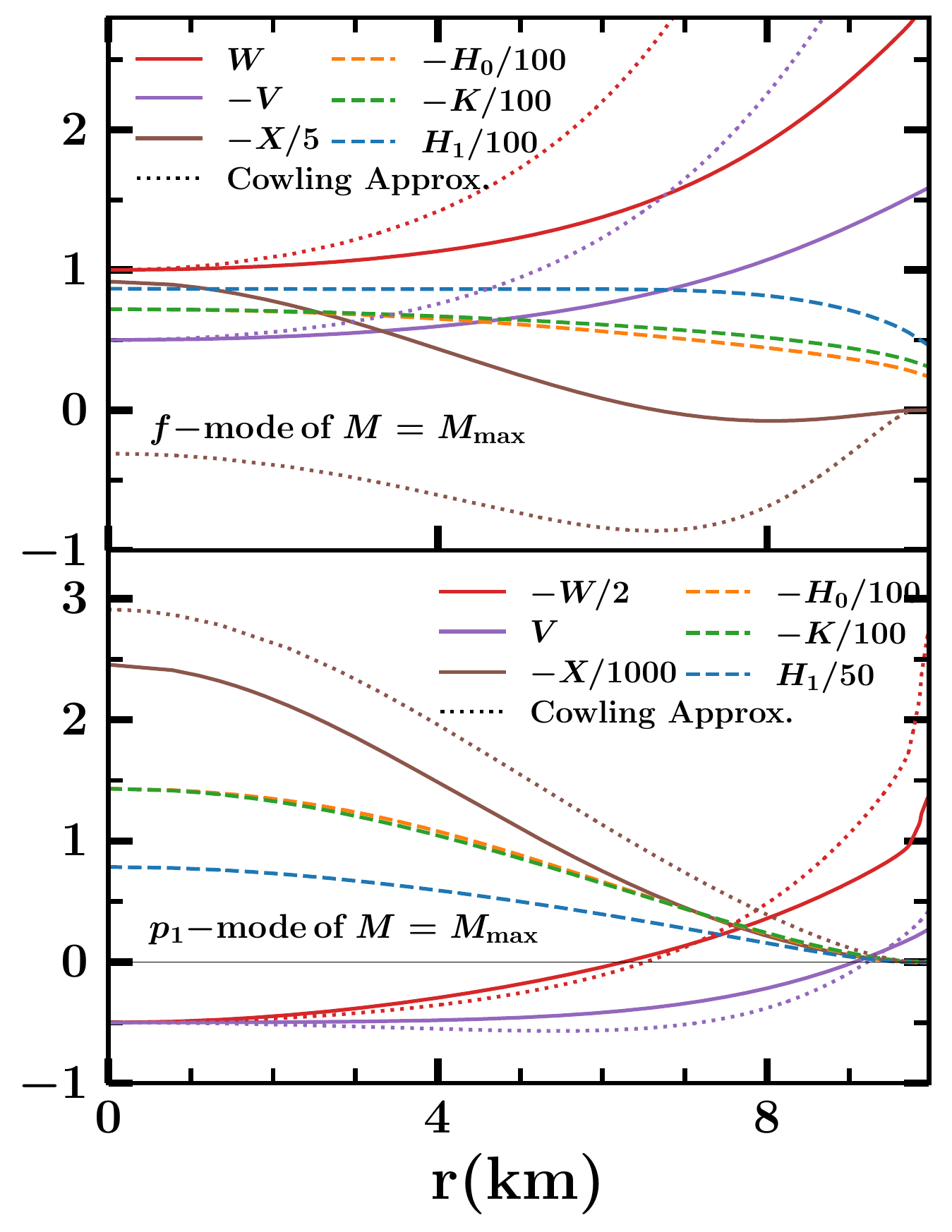}
    \end{minipage}
    \caption{Fluid and metric perturbation (real part) amplitude profiles of $f$-mode (top panels) and $p_1$-mode (bottom panels) oscillations for a NS with 1$M_\odot$ (left panels) and $M_{max}$ (right panels). $W$ and $V$ are dimensionless, $H_0$, $H_1$ and $K$ are in units of $\varepsilon_s=152.26$ MeV fm$^{-3}$, and $X$ is in units of $\varepsilon_s^2$. While the SLY4 EoS is used as an example in this figure, all EoSs have qualitatively similar behavior. } 
    \label{fig:amplitudes}
\end{figure*}

The frequencies calculated by the Cowling approximation and full GR  show significant
differences although their qualitative trends are similar. 
For the $\ell=2$ modes considered, We calculate the relative error in the $f$-mode frequencies using
\be
\frac{\left| f_{Cow}-f_{GR} \right|}{f_{GR}}
\ee

Similar calculation has done in case of $p_1$-mode as depicted in Fig. \ref{fig:error}(b). The Cowling approximation has an error of around $15-30\%$ for $f$-mode frequencies as shown in
Fig. \ref{fig:error}(a).
The error exhibits a linear trend with increasing mass with a more rapid variation occurring as the maximum is reached. These features were also noted in ~\cite{PhysRevD.91.044034,1997ApJ...490..779Y}.
\\

The $f$-mode oscillations have fluid perturbations ($W$ and $V$) that peak near the stellar surface,  whereas the 
metric perturbations ($H_0$, $H_1$ and $K$) peak close to
the stellar center; see top two panels in Fig. \ref{fig:amplitudes}. The Cowling approximation neglects metric perturbations which introduce a smaller error in NSs with higher mass, as massive NSs have a stronger fluid perturbation peak at the stellar surface and couples weakly to the metric perturbation at the stellar center.  Fluid perturbations of the $p_1$-mode have a radial node 
{within the NS}; 
see the bottom two panels in Fig. \ref{fig:amplitudes}. Also, the fluid perturbation is much weaker at the stellar center which results in a much higher peak at the stellar surface. As a result, the $p_1$-mode oscillation frequency calculated by 
the Cowling approximation has an error of around 10\%, less than the error in the $f$-mode frequency. However, unlike 
for the $f$-mode, fluid perturbations peak 
at the stellar surface the more massive NSs exhibiting a higher peak than stars of lesser mass. Therefore, the error in the Cowling approximation for $p_1$-modes increases with increasing mass. Since the 
{$p$-modes} with more radial nodes generally have weak fluid perturbations in the stellar center, the accuracy of the Cowling approximation increases with the order of the mode~\cite{lindblom1990accuracy}. Therefore, the error of the Cowling approximation in 
{$p$-modes of any order is less than 15\%}, which corresponds to $p_1$-mode for stable NS with maximum mass.\\ 

{Since the $f-mode$ frequency can be estimated 
by the $1\%$-accurate universal relations between the $f$-mode frequency vs tidal deformability and vs momentum of inertia \cite{lau2010inferring,zhao2022universal}, the 30\% error of the Cowling approximation is unacceptable. Due to the lack of such a tight universal relation for $p$-modes, a 15$\%$ error from the Cowling approximation is regarded as 
tolerable 
for 
explorative purposes. For example, the Cowling approximation can greatly simplify the equations especially in non-linear full GR simulations \cite{rosofsky2019probing,shashank2021f}.}

\section{Summary and Conclusions}
\label{sec:concs}

We have used a representative set of EoSs from non-relativistic and relativistic mean-field models to explore the possibility of strong correlations of $f$-mode, $p_1$-mode frequencies and damping time with the properties of nuclear matter and neutron stars. Most of the EoSs are consistent with that for the chiral EFT model upto $2\rho_0$. Our chosen EoSs also satisfy 
experimental data on the bulk properties of finite nuclei and are also compatible with the various properties of NS observables such as maximum mass, radius, and tidal deformability. For our correlation systematics, the values of nuclear matter incompressibility $K_0$, skewness parameter  $Q_0$  for symmetric nuclear matter, and the slope of the symmetry energy coefficients $L_0$ as well as pressure  for the $\beta$-equilibrated matter in the range of $\rho_0$-$2.5\rho_0$ are employed. 
We also present a critical analysis of the correlations of nuclear matter parameters, 
such as aspects of the 
density-dependent symmetry energy, with the tidal deformability 

The differences in the $f$- and $p_1$-mode frequencies calculated using full GR and  
the Cowling approximation are also studied. \\

Our principal findings are as follow.  The $f$-mode frequencies corresponding to NS masses in the range $1.2-1.8 M_\odot$ are strongly correlated with the radius and tidal deformability of a $1.4 M_\odot$ neutron  star. The pressure at $2\rho_0$ for 
$\beta$-equilibrated matter is also strongly correlated with 
the $f$-mode frequencies. This correlation grows marginally with increase in the NS mass. The $L_0$ is found to be moderately correlated with $f$-mode frequencies whereas
it decreases with increasing NS mass. The trends of the correlations for the $f$-mode frequencies obtained with the Cowling approximation and full GR are quite similar. \\

The frequency of the $p_1$-mode for a NS with mass
$1.4 M_\odot$ is strongly correlated with $L_0$ and pressure at $\rho_0$. However, the correlation significantly decreases with pressure at higher densities. This can be understood as being due to the $p$-mode oscillation peaking stronger at the stellar surface than  the $f$-mode, see Fig. \ref{fig:amplitudes}. The correlation of the $p_1$-mode frequency with the canonical radius and tidal deformability are weaker than those for the $f$-mode frequency. These results imply  that the $f$- and $p_1$-mode frequencies are sensitive to the behavior of EoS at different densities. \\

The correlation of the damping time for $f$-mode exhibits
the same behavior as those for the $f$-mode frequencies. However, the damping time for the $p_1$-mode displays opposite trends. For instance, the damping time for the $p_1$-mode is only weakly correlated with $L_0$ and pressure for $\beta$-equlibrated matter at saturation density $\rho_0$. \\

Our examination of the accuracy and its mass dependence of the Cowling approximation reveals that the error 
in the $f$-mode frequency decreases from about $30\%$ for a $1M_\odot$ NS to about $15\%$ for a NS with the highest mass, which is consistent with previous calculations \cite{Pradhan_2022,zhao2022universal}. Recently, calculation of the Newtonian Kelvin frequency has been shown to be a simpler and more accurate approximation than the Cowling approximation \cite{zhao2022universal}. However, the Cowling approximation gives reasonable results for the
$p_1$-mode with errors $\lesssim 15\%$. Unlike non-linear simulation of the $f$-mode oscillation which requires full GR \cite{rosofsky2019probing,shashank2021f}, numerical simulation of the $p_1$-mode could be simplified with the Cowling approximation, which is sufficiently accurate for most purposes. \\

Our work here has concentrated on NS matter in which the strongly interacting components are nucleons only. In work to follow, we will be considering the possible presence of quarks treated  via the Gibbs construction and the case in which a nucleon-to-quark crossover transition occurs.

\begin{acknowledgments}
T.Z. and M. P. are supported by the Department of Energy, Grant No. DE-FG02-93ER40756. 
B.K. and B.K.A. acknowledge partial support from the Department of Science
and Technology, Government of India with grant no. CRG/2021/000101. \\

\end{acknowledgments}

\newpage

\bibliography{GRRvsCOW}

\begin{thebibliography}{121}%
\makeatletter
\providecommand \@ifxundefined [1]{%
 \@ifx{#1\undefined}
}%
\providecommand \@ifnum [1]{%
 \ifnum #1\expandafter \@firstoftwo
 \else \expandafter \@secondoftwo
 \fi
}%
\providecommand \@ifx [1]{%
 \ifx #1\expandafter \@firstoftwo
 \else \expandafter \@secondoftwo
 \fi
}%
\providecommand \natexlab [1]{#1}%
\providecommand \enquote  [1]{``#1''}%
\providecommand \bibnamefont  [1]{#1}%
\providecommand \bibfnamefont [1]{#1}%
\providecommand \citenamefont [1]{#1}%
\providecommand \href@noop [0]{\@secondoftwo}%
\providecommand \href [0]{\begingroup \@sanitize@url \@href}%
\providecommand \@href[1]{\@@startlink{#1}\@@href}%
\providecommand \@@href[1]{\endgroup#1\@@endlink}%
\providecommand \@sanitize@url [0]{\catcode `\\12\catcode `\$12\catcode
  `\&12\catcode `\#12\catcode `\^12\catcode `\_12\catcode `\%12\relax}%
\providecommand \@@startlink[1]{}%
\providecommand \@@endlink[0]{}%
\providecommand \url  [0]{\begingroup\@sanitize@url \@url }%
\providecommand \@url [1]{\endgroup\@href {#1}{\urlprefix }}%
\providecommand \urlprefix  [0]{URL }%
\providecommand \Eprint [0]{\href }%
\providecommand \doibase [0]{http://dx.doi.org/}%
\providecommand \selectlanguage [0]{\@gobble}%
\providecommand \bibinfo  [0]{\@secondoftwo}%
\providecommand \bibfield  [0]{\@secondoftwo}%
\providecommand \translation [1]{[#1]}%
\providecommand \BibitemOpen [0]{}%
\providecommand \bibitemStop [0]{}%
\providecommand \bibitemNoStop [0]{.\EOS\space}%
\providecommand \EOS [0]{\spacefactor3000\relax}%
\providecommand \BibitemShut  [1]{\csname bibitem#1\endcsname}%
\let\auto@bib@innerbib\@empty
\bibitem [{\citenamefont {Kokkotas}\ and\ \citenamefont
  {Schmidt}(1999)}]{Kokkotas_1999}%
  \BibitemOpen
  \bibfield  {author} {\bibinfo {author} {\bibfnamefont {K.~D.}\ \bibnamefont
  {Kokkotas}}\ and\ \bibinfo {author} {\bibfnamefont {B.~G.}\ \bibnamefont
  {Schmidt}},\ }\href {\doibase 10.12942/lrr-1999-2} {\bibfield  {journal}
  {\bibinfo  {journal} {Living Reviews in Relativity}\ }\textbf {\bibinfo
  {volume} {2}} (\bibinfo {year} {1999}),\ 10.12942/lrr-1999-2}\BibitemShut
  {NoStop}%
\bibitem [{\citenamefont {Zhao}\ \emph {et~al.}(2022)\citenamefont {Zhao},
  \citenamefont {Constantinou}, \citenamefont {Jaikumar},\ and\ \citenamefont
  {Prakash}}]{zhao2022quasi}%
  \BibitemOpen
  \bibfield  {author} {\bibinfo {author} {\bibfnamefont {T.}~\bibnamefont
  {Zhao}}, \bibinfo {author} {\bibfnamefont {C.}~\bibnamefont {Constantinou}},
  \bibinfo {author} {\bibfnamefont {P.}~\bibnamefont {Jaikumar}}, \ and\
  \bibinfo {author} {\bibfnamefont {M.}~\bibnamefont {Prakash}},\ }\href
  {\doibase 10.1103/PhysRevD.105.103025} {\bibfield  {journal} {\bibinfo
  {journal} {Phys. Rev. D}\ }\textbf {\bibinfo {volume} {105}},\ \bibinfo
  {pages} {103025} (\bibinfo {year} {2022})}\BibitemShut {NoStop}%
\bibitem [{\citenamefont {Zhao}\ and\ \citenamefont
  {Lattimer}(2022)}]{zhao2022universal}%
  \BibitemOpen
  \bibfield  {author} {\bibinfo {author} {\bibfnamefont {T.}~\bibnamefont
  {Zhao}}\ and\ \bibinfo {author} {\bibfnamefont {J.~M.}\ \bibnamefont
  {Lattimer}},\ }\href@noop {} {\bibfield  {journal} {\bibinfo  {journal}
  {arXiv preprint arXiv:2204.03037}\ } (\bibinfo {year} {2022})}\BibitemShut
  {NoStop}%
\bibitem [{\citenamefont {Ferrari}\ \emph {et~al.}(2003)\citenamefont
  {Ferrari}, \citenamefont {Miniutti},\ and\ \citenamefont
  {Pons}}]{ferrari2003gravitational}%
  \BibitemOpen
  \bibfield  {author} {\bibinfo {author} {\bibfnamefont {V.}~\bibnamefont
  {Ferrari}}, \bibinfo {author} {\bibfnamefont {G.}~\bibnamefont {Miniutti}}, \
  and\ \bibinfo {author} {\bibfnamefont {J.~A.}\ \bibnamefont {Pons}},\
  }\href@noop {} {\bibfield  {journal} {\bibinfo  {journal} {Monthly Notices of
  the Royal Astronomical Society}\ }\textbf {\bibinfo {volume} {342}},\
  \bibinfo {pages} {629} (\bibinfo {year} {2003})}\BibitemShut {NoStop}%
\bibitem [{\citenamefont {Shibagaki}\ \emph {et~al.}(2020)\citenamefont
  {Shibagaki}, \citenamefont {Kuroda}, \citenamefont {Kotake},\ and\
  \citenamefont {Takiwaki}}]{shibagaki2020new}%
  \BibitemOpen
  \bibfield  {author} {\bibinfo {author} {\bibfnamefont {S.}~\bibnamefont
  {Shibagaki}}, \bibinfo {author} {\bibfnamefont {T.}~\bibnamefont {Kuroda}},
  \bibinfo {author} {\bibfnamefont {K.}~\bibnamefont {Kotake}}, \ and\ \bibinfo
  {author} {\bibfnamefont {T.}~\bibnamefont {Takiwaki}},\ }\href@noop {}
  {\bibfield  {journal} {\bibinfo  {journal} {Monthly Notices of the Royal
  Astronomical Society: Letters}\ }\textbf {\bibinfo {volume} {493}},\ \bibinfo
  {pages} {L138} (\bibinfo {year} {2020})}\BibitemShut {NoStop}%
\bibitem [{\citenamefont {Radice}\ \emph {et~al.}(2019)\citenamefont {Radice},
  \citenamefont {Morozova}, \citenamefont {Burrows}, \citenamefont
  {Vartanyan},\ and\ \citenamefont {Nagakura}}]{radice2019characterizing}%
  \BibitemOpen
  \bibfield  {author} {\bibinfo {author} {\bibfnamefont {D.}~\bibnamefont
  {Radice}}, \bibinfo {author} {\bibfnamefont {V.}~\bibnamefont {Morozova}},
  \bibinfo {author} {\bibfnamefont {A.}~\bibnamefont {Burrows}}, \bibinfo
  {author} {\bibfnamefont {D.}~\bibnamefont {Vartanyan}}, \ and\ \bibinfo
  {author} {\bibfnamefont {H.}~\bibnamefont {Nagakura}},\ }\href@noop {}
  {\bibfield  {journal} {\bibinfo  {journal} {The Astrophysical Journal
  Letters}\ }\textbf {\bibinfo {volume} {876}},\ \bibinfo {pages} {L9}
  (\bibinfo {year} {2019})}\BibitemShut {NoStop}%
\bibitem [{\citenamefont {Li}\ \emph {et~al.}(2011)\citenamefont {Li},
  \citenamefont {Chornock}, \citenamefont {Leaman}, \citenamefont {Filippenko},
  \citenamefont {Poznanski}, \citenamefont {Wang}, \citenamefont
  {Ganeshalingam},\ and\ \citenamefont {Mannucci}}]{li2011nearby}%
  \BibitemOpen
  \bibfield  {author} {\bibinfo {author} {\bibfnamefont {W.}~\bibnamefont
  {Li}}, \bibinfo {author} {\bibfnamefont {R.}~\bibnamefont {Chornock}},
  \bibinfo {author} {\bibfnamefont {J.}~\bibnamefont {Leaman}}, \bibinfo
  {author} {\bibfnamefont {A.~V.}\ \bibnamefont {Filippenko}}, \bibinfo
  {author} {\bibfnamefont {D.}~\bibnamefont {Poznanski}}, \bibinfo {author}
  {\bibfnamefont {X.}~\bibnamefont {Wang}}, \bibinfo {author} {\bibfnamefont
  {M.}~\bibnamefont {Ganeshalingam}}, \ and\ \bibinfo {author} {\bibfnamefont
  {F.}~\bibnamefont {Mannucci}},\ }\href@noop {} {\bibfield  {journal}
  {\bibinfo  {journal} {Monthly Notices of the Royal Astronomical Society}\
  }\textbf {\bibinfo {volume} {412}},\ \bibinfo {pages} {1473} (\bibinfo {year}
  {2011})}\BibitemShut {NoStop}%
\bibitem [{\citenamefont {Stergioulas}\ \emph {et~al.}(2011)\citenamefont
  {Stergioulas}, \citenamefont {Bauswein}, \citenamefont {Zagkouris},\ and\
  \citenamefont {Janka}}]{stergioulas2011gravitational}%
  \BibitemOpen
  \bibfield  {author} {\bibinfo {author} {\bibfnamefont {N.}~\bibnamefont
  {Stergioulas}}, \bibinfo {author} {\bibfnamefont {A.}~\bibnamefont
  {Bauswein}}, \bibinfo {author} {\bibfnamefont {K.}~\bibnamefont {Zagkouris}},
  \ and\ \bibinfo {author} {\bibfnamefont {H.-T.}\ \bibnamefont {Janka}},\
  }\href@noop {} {\bibfield  {journal} {\bibinfo  {journal} {Monthly Notices of
  the Royal Astronomical Society}\ }\textbf {\bibinfo {volume} {418}},\
  \bibinfo {pages} {427} (\bibinfo {year} {2011})}\BibitemShut {NoStop}%
\bibitem [{\citenamefont {Lioutas}\ \emph {et~al.}(2021)\citenamefont
  {Lioutas}, \citenamefont {Bauswein},\ and\ \citenamefont
  {Stergioulas}}]{lioutas2021frequency}%
  \BibitemOpen
  \bibfield  {author} {\bibinfo {author} {\bibfnamefont {G.}~\bibnamefont
  {Lioutas}}, \bibinfo {author} {\bibfnamefont {A.}~\bibnamefont {Bauswein}}, \
  and\ \bibinfo {author} {\bibfnamefont {N.}~\bibnamefont {Stergioulas}},\
  }\href@noop {} {\bibfield  {journal} {\bibinfo  {journal} {arXiv preprint
  arXiv:2102.12455}\ } (\bibinfo {year} {2021})}\BibitemShut {NoStop}%
\bibitem [{\citenamefont {Ng}\ \emph {et~al.}(2020)\citenamefont {Ng},
  \citenamefont {Cheong}, \citenamefont {Lin},\ and\ \citenamefont
  {Li}}]{ng2020gravitational}%
  \BibitemOpen
  \bibfield  {author} {\bibinfo {author} {\bibfnamefont {H.~H.-Y.}\
  \bibnamefont {Ng}}, \bibinfo {author} {\bibfnamefont {P.~C.-K.}\ \bibnamefont
  {Cheong}}, \bibinfo {author} {\bibfnamefont {L.-M.}\ \bibnamefont {Lin}}, \
  and\ \bibinfo {author} {\bibfnamefont {T.~G.~F.}\ \bibnamefont {Li}},\
  }\href@noop {} {\bibfield  {journal} {\bibinfo  {journal} {arXiv preprint
  arXiv:2012.08263}\ } (\bibinfo {year} {2020})}\BibitemShut {NoStop}%
\bibitem [{\citenamefont {Reitze}\ \emph {et~al.}(2019)\citenamefont {Reitze},
  \citenamefont {Adhikari}, \citenamefont {Ballmer}, \citenamefont {Barish},
  \citenamefont {Barsotti}, \citenamefont {Billingsley}, \citenamefont {Brown},
  \citenamefont {Chen}, \citenamefont {Coyne}, \citenamefont {Eisenstein} \emph
  {et~al.}}]{reitze2019cosmic}%
  \BibitemOpen
  \bibfield  {author} {\bibinfo {author} {\bibfnamefont {D.}~\bibnamefont
  {Reitze}}, \bibinfo {author} {\bibfnamefont {R.~X.}\ \bibnamefont
  {Adhikari}}, \bibinfo {author} {\bibfnamefont {S.}~\bibnamefont {Ballmer}},
  \bibinfo {author} {\bibfnamefont {B.}~\bibnamefont {Barish}}, \bibinfo
  {author} {\bibfnamefont {L.}~\bibnamefont {Barsotti}}, \bibinfo {author}
  {\bibfnamefont {G.}~\bibnamefont {Billingsley}}, \bibinfo {author}
  {\bibfnamefont {D.~A.}\ \bibnamefont {Brown}}, \bibinfo {author}
  {\bibfnamefont {Y.}~\bibnamefont {Chen}}, \bibinfo {author} {\bibfnamefont
  {D.}~\bibnamefont {Coyne}}, \bibinfo {author} {\bibfnamefont
  {R.}~\bibnamefont {Eisenstein}},  \emph {et~al.},\ }\href@noop {} {\bibfield
  {journal} {\bibinfo  {journal} {arXiv preprint arXiv:1907.04833}\ } (\bibinfo
  {year} {2019})}\BibitemShut {NoStop}%
\bibitem [{\citenamefont {Punturo}\ \emph {et~al.}(2010)\citenamefont
  {Punturo}, \citenamefont {Abernathy}, \citenamefont {Acernese}, \citenamefont
  {Allen}, \citenamefont {Andersson}, \citenamefont {Arun}, \citenamefont
  {Barone}, \citenamefont {Barr}, \citenamefont {Barsuglia}, \citenamefont
  {Beker} \emph {et~al.}}]{punturo2010einstein}%
  \BibitemOpen
  \bibfield  {author} {\bibinfo {author} {\bibfnamefont {M.}~\bibnamefont
  {Punturo}}, \bibinfo {author} {\bibfnamefont {M.}~\bibnamefont {Abernathy}},
  \bibinfo {author} {\bibfnamefont {F.}~\bibnamefont {Acernese}}, \bibinfo
  {author} {\bibfnamefont {B.}~\bibnamefont {Allen}}, \bibinfo {author}
  {\bibfnamefont {N.}~\bibnamefont {Andersson}}, \bibinfo {author}
  {\bibfnamefont {K.}~\bibnamefont {Arun}}, \bibinfo {author} {\bibfnamefont
  {F.}~\bibnamefont {Barone}}, \bibinfo {author} {\bibfnamefont
  {B.}~\bibnamefont {Barr}}, \bibinfo {author} {\bibfnamefont {M.}~\bibnamefont
  {Barsuglia}}, \bibinfo {author} {\bibfnamefont {M.}~\bibnamefont {Beker}},
  \emph {et~al.},\ }\href@noop {} {\bibfield  {journal} {\bibinfo  {journal}
  {Classical and Quantum Gravity}\ }\textbf {\bibinfo {volume} {27}},\ \bibinfo
  {pages} {194002} (\bibinfo {year} {2010})}\BibitemShut {NoStop}%
\bibitem [{\citenamefont {Abbott}\ \emph {et~al.}(2021)\citenamefont {Abbott},
  \citenamefont {Abbott}, \citenamefont {Abraham}, \citenamefont {Acernese},
  \citenamefont {Ackley}, \citenamefont {Adams}, \citenamefont {Adams},
  \citenamefont {Adhikari}, \citenamefont {Adya}, \citenamefont {Affeldt} \emph
  {et~al.}}]{abbott2021population}%
  \BibitemOpen
  \bibfield  {author} {\bibinfo {author} {\bibfnamefont {R.}~\bibnamefont
  {Abbott}}, \bibinfo {author} {\bibfnamefont {T.}~\bibnamefont {Abbott}},
  \bibinfo {author} {\bibfnamefont {S.}~\bibnamefont {Abraham}}, \bibinfo
  {author} {\bibfnamefont {F.}~\bibnamefont {Acernese}}, \bibinfo {author}
  {\bibfnamefont {K.}~\bibnamefont {Ackley}}, \bibinfo {author} {\bibfnamefont
  {A.}~\bibnamefont {Adams}}, \bibinfo {author} {\bibfnamefont
  {C.}~\bibnamefont {Adams}}, \bibinfo {author} {\bibfnamefont
  {R.}~\bibnamefont {Adhikari}}, \bibinfo {author} {\bibfnamefont
  {V.}~\bibnamefont {Adya}}, \bibinfo {author} {\bibfnamefont {C.}~\bibnamefont
  {Affeldt}},  \emph {et~al.},\ }\href@noop {} {\bibfield  {journal} {\bibinfo
  {journal} {The Astrophysical journal letters}\ }\textbf {\bibinfo {volume}
  {913}},\ \bibinfo {pages} {L7} (\bibinfo {year} {2021})}\BibitemShut
  {NoStop}%
\bibitem [{\citenamefont {Hinderer}\ \emph {et~al.}(2016)\citenamefont
  {Hinderer}, \citenamefont {Taracchini}, \citenamefont {Foucart},
  \citenamefont {Buonanno}, \citenamefont {Steinhoff}, \citenamefont {Duez},
  \citenamefont {Kidder}, \citenamefont {Pfeiffer}, \citenamefont {Scheel},
  \citenamefont {Szilagyi} \emph {et~al.}}]{hinderer2016effects}%
  \BibitemOpen
  \bibfield  {author} {\bibinfo {author} {\bibfnamefont {T.}~\bibnamefont
  {Hinderer}}, \bibinfo {author} {\bibfnamefont {A.}~\bibnamefont
  {Taracchini}}, \bibinfo {author} {\bibfnamefont {F.}~\bibnamefont {Foucart}},
  \bibinfo {author} {\bibfnamefont {A.}~\bibnamefont {Buonanno}}, \bibinfo
  {author} {\bibfnamefont {J.}~\bibnamefont {Steinhoff}}, \bibinfo {author}
  {\bibfnamefont {M.}~\bibnamefont {Duez}}, \bibinfo {author} {\bibfnamefont
  {L.~E.}\ \bibnamefont {Kidder}}, \bibinfo {author} {\bibfnamefont {H.~P.}\
  \bibnamefont {Pfeiffer}}, \bibinfo {author} {\bibfnamefont {M.~A.}\
  \bibnamefont {Scheel}}, \bibinfo {author} {\bibfnamefont {B.}~\bibnamefont
  {Szilagyi}},  \emph {et~al.},\ }\href@noop {} {\bibfield  {journal} {\bibinfo
   {journal} {Physical review letters}\ }\textbf {\bibinfo {volume} {116}},\
  \bibinfo {pages} {181101} (\bibinfo {year} {2016})}\BibitemShut {NoStop}%
\bibitem [{\citenamefont {Steinhoff}\ \emph {et~al.}(2016)\citenamefont
  {Steinhoff}, \citenamefont {Hinderer}, \citenamefont {Buonanno},\ and\
  \citenamefont {Taracchini}}]{steinhoff2016dynamical}%
  \BibitemOpen
  \bibfield  {author} {\bibinfo {author} {\bibfnamefont {J.}~\bibnamefont
  {Steinhoff}}, \bibinfo {author} {\bibfnamefont {T.}~\bibnamefont {Hinderer}},
  \bibinfo {author} {\bibfnamefont {A.}~\bibnamefont {Buonanno}}, \ and\
  \bibinfo {author} {\bibfnamefont {A.}~\bibnamefont {Taracchini}},\
  }\href@noop {} {\bibfield  {journal} {\bibinfo  {journal} {Physical Review
  D}\ }\textbf {\bibinfo {volume} {94}},\ \bibinfo {pages} {104028} (\bibinfo
  {year} {2016})}\BibitemShut {NoStop}%
\bibitem [{\citenamefont {Schmidt}\ and\ \citenamefont
  {Hinderer}(2019)}]{schmidt2019frequency}%
  \BibitemOpen
  \bibfield  {author} {\bibinfo {author} {\bibfnamefont {P.}~\bibnamefont
  {Schmidt}}\ and\ \bibinfo {author} {\bibfnamefont {T.}~\bibnamefont
  {Hinderer}},\ }\href@noop {} {\bibfield  {journal} {\bibinfo  {journal}
  {Physical Review D}\ }\textbf {\bibinfo {volume} {100}},\ \bibinfo {pages}
  {021501} (\bibinfo {year} {2019})}\BibitemShut {NoStop}%
\bibitem [{\citenamefont {Wynn}\ and\ \citenamefont
  {Dong}(1999)}]{wynn1999resonant}%
  \BibitemOpen
  \bibfield  {author} {\bibinfo {author} {\bibfnamefont {C.~H.}\ \bibnamefont
  {Wynn}}\ and\ \bibinfo {author} {\bibfnamefont {L.}~\bibnamefont {Dong}},\
  }\href@noop {} {\bibfield  {journal} {\bibinfo  {journal} {Monthly Notices of
  the Royal Astronomical Society}\ }\textbf {\bibinfo {volume} {308}},\
  \bibinfo {pages} {153} (\bibinfo {year} {1999})}\BibitemShut {NoStop}%
\bibitem [{\citenamefont {Steinhoff}\ \emph {et~al.}(2021)\citenamefont
  {Steinhoff}, \citenamefont {Hinderer}, \citenamefont {Dietrich},\ and\
  \citenamefont {Foucart}}]{steinhoff2021spin}%
  \BibitemOpen
  \bibfield  {author} {\bibinfo {author} {\bibfnamefont {J.}~\bibnamefont
  {Steinhoff}}, \bibinfo {author} {\bibfnamefont {T.}~\bibnamefont {Hinderer}},
  \bibinfo {author} {\bibfnamefont {T.}~\bibnamefont {Dietrich}}, \ and\
  \bibinfo {author} {\bibfnamefont {F.}~\bibnamefont {Foucart}},\ }\href@noop
  {} {\bibfield  {journal} {\bibinfo  {journal} {arXiv preprint
  arXiv:2103.06100}\ } (\bibinfo {year} {2021})}\BibitemShut {NoStop}%
\bibitem [{\citenamefont {Andersson}\ and\ \citenamefont
  {Kokkotas}(1998{\natexlab{a}})}]{andersson1998towards}%
  \BibitemOpen
  \bibfield  {author} {\bibinfo {author} {\bibfnamefont {N.}~\bibnamefont
  {Andersson}}\ and\ \bibinfo {author} {\bibfnamefont {K.~D.}\ \bibnamefont
  {Kokkotas}},\ }\href@noop {} {\bibfield  {journal} {\bibinfo  {journal}
  {Monthly Notices of the Royal Astronomical Society}\ }\textbf {\bibinfo
  {volume} {299}},\ \bibinfo {pages} {1059} (\bibinfo {year}
  {1998}{\natexlab{a}})}\BibitemShut {NoStop}%
\bibitem [{\citenamefont {Lau}\ \emph {et~al.}(2010)\citenamefont {Lau},
  \citenamefont {Leung},\ and\ \citenamefont {Lin}}]{lau2010inferring}%
  \BibitemOpen
  \bibfield  {author} {\bibinfo {author} {\bibfnamefont {H.}~\bibnamefont
  {Lau}}, \bibinfo {author} {\bibfnamefont {P.}~\bibnamefont {Leung}}, \ and\
  \bibinfo {author} {\bibfnamefont {L.}~\bibnamefont {Lin}},\ }\href@noop {}
  {\bibfield  {journal} {\bibinfo  {journal} {The Astrophysical Journal}\
  }\textbf {\bibinfo {volume} {714}},\ \bibinfo {pages} {1234} (\bibinfo {year}
  {2010})}\BibitemShut {NoStop}%
\bibitem [{\citenamefont {Chan}\ \emph {et~al.}(2014)\citenamefont {Chan},
  \citenamefont {Sham}, \citenamefont {Leung},\ and\ \citenamefont
  {Lin}}]{chan2014multipolar}%
  \BibitemOpen
  \bibfield  {author} {\bibinfo {author} {\bibfnamefont {T.}~\bibnamefont
  {Chan}}, \bibinfo {author} {\bibfnamefont {Y.-H.}\ \bibnamefont {Sham}},
  \bibinfo {author} {\bibfnamefont {P.}~\bibnamefont {Leung}}, \ and\ \bibinfo
  {author} {\bibfnamefont {L.-M.}\ \bibnamefont {Lin}},\ }\href@noop {}
  {\bibfield  {journal} {\bibinfo  {journal} {Physical Review D}\ }\textbf
  {\bibinfo {volume} {90}},\ \bibinfo {pages} {124023} (\bibinfo {year}
  {2014})}\BibitemShut {NoStop}%
\bibitem [{\citenamefont {Sotani}\ and\ \citenamefont
  {Kumar}(2021)}]{sotani2021universal}%
  \BibitemOpen
  \bibfield  {author} {\bibinfo {author} {\bibfnamefont {H.}~\bibnamefont
  {Sotani}}\ and\ \bibinfo {author} {\bibfnamefont {B.}~\bibnamefont {Kumar}},\
  }\href@noop {} {\bibfield  {journal} {\bibinfo  {journal} {Physical Review
  D}\ }\textbf {\bibinfo {volume} {104}},\ \bibinfo {pages} {123002} (\bibinfo
  {year} {2021})}\BibitemShut {NoStop}%
\bibitem [{\citenamefont {Lindblom}\ and\ \citenamefont
  {Detweiler}(1983)}]{lindblom1983quadrupole}%
  \BibitemOpen
  \bibfield  {author} {\bibinfo {author} {\bibfnamefont {L.}~\bibnamefont
  {Lindblom}}\ and\ \bibinfo {author} {\bibfnamefont {S.~L.}\ \bibnamefont
  {Detweiler}},\ }\href@noop {} {\bibfield  {journal} {\bibinfo  {journal} {The
  Astrophysical Journal Supplement Series}\ }\textbf {\bibinfo {volume} {53}},\
  \bibinfo {pages} {73} (\bibinfo {year} {1983})}\BibitemShut {NoStop}%
\bibitem [{\citenamefont {Detweiler}\ and\ \citenamefont
  {Lindblom}(1985)}]{detweiler1985nonradial}%
  \BibitemOpen
  \bibfield  {author} {\bibinfo {author} {\bibfnamefont {S.}~\bibnamefont
  {Detweiler}}\ and\ \bibinfo {author} {\bibfnamefont {L.}~\bibnamefont
  {Lindblom}},\ }\href@noop {} {\bibfield  {journal} {\bibinfo  {journal} {The
  Astrophysical Journal}\ }\textbf {\bibinfo {volume} {292}},\ \bibinfo {pages}
  {12} (\bibinfo {year} {1985})}\BibitemShut {NoStop}%
\bibitem [{\citenamefont {Kojima}(1993)}]{kojima1993normal}%
  \BibitemOpen
  \bibfield  {author} {\bibinfo {author} {\bibfnamefont {Y.}~\bibnamefont
  {Kojima}},\ }\href@noop {} {\bibfield  {journal} {\bibinfo  {journal} {The
  Astrophysical Journal}\ }\textbf {\bibinfo {volume} {414}},\ \bibinfo {pages}
  {247} (\bibinfo {year} {1993})}\BibitemShut {NoStop}%
\bibitem [{\citenamefont {Kr{\"u}ger}\ and\ \citenamefont
  {Kokkotas}(2020)}]{kruger2020dynamics}%
  \BibitemOpen
  \bibfield  {author} {\bibinfo {author} {\bibfnamefont {C.~J.}\ \bibnamefont
  {Kr{\"u}ger}}\ and\ \bibinfo {author} {\bibfnamefont {K.~D.}\ \bibnamefont
  {Kokkotas}},\ }\href@noop {} {\bibfield  {journal} {\bibinfo  {journal}
  {Physical Review D}\ }\textbf {\bibinfo {volume} {102}},\ \bibinfo {pages}
  {064026} (\bibinfo {year} {2020})}\BibitemShut {NoStop}%
\bibitem [{\citenamefont {McDermott}\ \emph {et~al.}(1983)\citenamefont
  {McDermott}, \citenamefont {Van~Horn},\ and\ \citenamefont
  {Scholl}}]{mcdermott1983nonradial}%
  \BibitemOpen
  \bibfield  {author} {\bibinfo {author} {\bibfnamefont {P.}~\bibnamefont
  {McDermott}}, \bibinfo {author} {\bibfnamefont {H.~M.}\ \bibnamefont
  {Van~Horn}}, \ and\ \bibinfo {author} {\bibfnamefont {J.}~\bibnamefont
  {Scholl}},\ }\href@noop {} {\bibfield  {journal} {\bibinfo  {journal} {The
  Astrophysical Journal}\ }\textbf {\bibinfo {volume} {268}},\ \bibinfo {pages}
  {837} (\bibinfo {year} {1983})}\BibitemShut {NoStop}%
\bibitem [{\citenamefont {Ranea-Sandoval}\ \emph {et~al.}(2018)\citenamefont
  {Ranea-Sandoval}, \citenamefont {Guilera}, \citenamefont {Mariani},\ and\
  \citenamefont {Orsaria}}]{ranea2018oscillation}%
  \BibitemOpen
  \bibfield  {author} {\bibinfo {author} {\bibfnamefont {I.~F.}\ \bibnamefont
  {Ranea-Sandoval}}, \bibinfo {author} {\bibfnamefont {O.~M.}\ \bibnamefont
  {Guilera}}, \bibinfo {author} {\bibfnamefont {M.}~\bibnamefont {Mariani}}, \
  and\ \bibinfo {author} {\bibfnamefont {M.~G.}\ \bibnamefont {Orsaria}},\
  }\href@noop {} {\bibfield  {journal} {\bibinfo  {journal} {Journal of
  Cosmology and Astroparticle Physics}\ }\textbf {\bibinfo {volume} {2018}},\
  \bibinfo {pages} {031} (\bibinfo {year} {2018})}\BibitemShut {NoStop}%
\bibitem [{\citenamefont {Sotani}\ \emph
  {et~al.}(2011{\natexlab{a}})\citenamefont {Sotani}, \citenamefont {Yasutake},
  \citenamefont {Maruyama},\ and\ \citenamefont
  {Tatsumi}}]{sotani2011signatures}%
  \BibitemOpen
  \bibfield  {author} {\bibinfo {author} {\bibfnamefont {H.}~\bibnamefont
  {Sotani}}, \bibinfo {author} {\bibfnamefont {N.}~\bibnamefont {Yasutake}},
  \bibinfo {author} {\bibfnamefont {T.}~\bibnamefont {Maruyama}}, \ and\
  \bibinfo {author} {\bibfnamefont {T.}~\bibnamefont {Tatsumi}},\ }\href@noop
  {} {\bibfield  {journal} {\bibinfo  {journal} {Physical Review D}\ }\textbf
  {\bibinfo {volume} {83}},\ \bibinfo {pages} {024014} (\bibinfo {year}
  {2011}{\natexlab{a}})}\BibitemShut {NoStop}%
\bibitem [{\citenamefont {Flores}\ and\ \citenamefont
  {Lugones}(2014)}]{flores2014discriminating}%
  \BibitemOpen
  \bibfield  {author} {\bibinfo {author} {\bibfnamefont {C.~V.}\ \bibnamefont
  {Flores}}\ and\ \bibinfo {author} {\bibfnamefont {G.}~\bibnamefont
  {Lugones}},\ }\href@noop {} {\bibfield  {journal} {\bibinfo  {journal}
  {Classical and Quantum Gravity}\ }\textbf {\bibinfo {volume} {31}},\ \bibinfo
  {pages} {155002} (\bibinfo {year} {2014})}\BibitemShut {NoStop}%
\bibitem [{\citenamefont {Yoshida}\ and\ \citenamefont
  {Kojima}(1997)}]{yoshida1997accuracy}%
  \BibitemOpen
  \bibfield  {author} {\bibinfo {author} {\bibfnamefont {S.}~\bibnamefont
  {Yoshida}}\ and\ \bibinfo {author} {\bibfnamefont {Y.}~\bibnamefont
  {Kojima}},\ }\href@noop {} {\bibfield  {journal} {\bibinfo  {journal}
  {Monthly Notices of the Royal Astronomical Society}\ }\textbf {\bibinfo
  {volume} {289}},\ \bibinfo {pages} {117} (\bibinfo {year}
  {1997})}\BibitemShut {NoStop}%
\bibitem [{\citenamefont {Sotani}\ \emph {et~al.}(2001)\citenamefont {Sotani},
  \citenamefont {Tominaga},\ and\ \citenamefont {Maeda}}]{sotani2001density}%
  \BibitemOpen
  \bibfield  {author} {\bibinfo {author} {\bibfnamefont {H.}~\bibnamefont
  {Sotani}}, \bibinfo {author} {\bibfnamefont {K.}~\bibnamefont {Tominaga}}, \
  and\ \bibinfo {author} {\bibfnamefont {K.-i.}\ \bibnamefont {Maeda}},\
  }\href@noop {} {\bibfield  {journal} {\bibinfo  {journal} {Physical Review
  D}\ }\textbf {\bibinfo {volume} {65}},\ \bibinfo {pages} {024010} (\bibinfo
  {year} {2001})}\BibitemShut {NoStop}%
\bibitem [{\citenamefont {Chirenti}\ \emph
  {et~al.}(2015{\natexlab{a}})\citenamefont {Chirenti}, \citenamefont
  {de~Souza},\ and\ \citenamefont {Kastaun}}]{chirenti2015fundamental}%
  \BibitemOpen
  \bibfield  {author} {\bibinfo {author} {\bibfnamefont {C.}~\bibnamefont
  {Chirenti}}, \bibinfo {author} {\bibfnamefont {G.~H.}\ \bibnamefont
  {de~Souza}}, \ and\ \bibinfo {author} {\bibfnamefont {W.}~\bibnamefont
  {Kastaun}},\ }\href@noop {} {\bibfield  {journal} {\bibinfo  {journal}
  {Physical Review D}\ }\textbf {\bibinfo {volume} {91}},\ \bibinfo {pages}
  {044034} (\bibinfo {year} {2015}{\natexlab{a}})}\BibitemShut {NoStop}%
\bibitem [{\citenamefont {Drischler}\ \emph {et~al.}(2021)\citenamefont
  {Drischler}, \citenamefont {Han}, \citenamefont {Lattimer}, \citenamefont
  {Prakash}, \citenamefont {Reddy},\ and\ \citenamefont
  {Zhao}}]{drischler2021limiting}%
  \BibitemOpen
  \bibfield  {author} {\bibinfo {author} {\bibfnamefont {C.}~\bibnamefont
  {Drischler}}, \bibinfo {author} {\bibfnamefont {S.}~\bibnamefont {Han}},
  \bibinfo {author} {\bibfnamefont {J.~M.}\ \bibnamefont {Lattimer}}, \bibinfo
  {author} {\bibfnamefont {M.}~\bibnamefont {Prakash}}, \bibinfo {author}
  {\bibfnamefont {S.}~\bibnamefont {Reddy}}, \ and\ \bibinfo {author}
  {\bibfnamefont {T.}~\bibnamefont {Zhao}},\ }\href@noop {} {\bibfield
  {journal} {\bibinfo  {journal} {Physical Review C}\ }\textbf {\bibinfo
  {volume} {103}},\ \bibinfo {pages} {045808} (\bibinfo {year}
  {2021})}\BibitemShut {NoStop}%
\bibitem [{\citenamefont {Thorne}\ and\ \citenamefont
  {Campolattaro}(1967)}]{thorne1967non}%
  \BibitemOpen
  \bibfield  {author} {\bibinfo {author} {\bibfnamefont {K.~S.}\ \bibnamefont
  {Thorne}}\ and\ \bibinfo {author} {\bibfnamefont {A.}~\bibnamefont
  {Campolattaro}},\ }\href@noop {} {\bibfield  {journal} {\bibinfo  {journal}
  {The Astrophysical Journal}\ }\textbf {\bibinfo {volume} {149}},\ \bibinfo
  {pages} {591} (\bibinfo {year} {1967})}\BibitemShut {NoStop}%
\bibitem [{\citenamefont {{Lindblom}}\ and\ \citenamefont
  {{Detweiler}}(1983{\natexlab{a}})}]{Lindblom:1983}%
  \BibitemOpen
  \bibfield  {author} {\bibinfo {author} {\bibfnamefont {L.}~\bibnamefont
  {{Lindblom}}}\ and\ \bibinfo {author} {\bibfnamefont {S.~L.}\ \bibnamefont
  {{Detweiler}}},\ }\href {\doibase 10.1086/190884} {\bibfield  {journal}
  {\bibinfo  {journal} {Astrophys. J. Suppl.}\ }\textbf {\bibinfo {volume}
  {53}},\ \bibinfo {pages} {73} (\bibinfo {year}
  {1983}{\natexlab{a}})}\BibitemShut {NoStop}%
\bibitem [{\citenamefont {{Detweiler}}\ and\ \citenamefont
  {{Lindblom}}(1985)}]{Detweiler:1985}%
  \BibitemOpen
  \bibfield  {author} {\bibinfo {author} {\bibfnamefont {S.}~\bibnamefont
  {{Detweiler}}}\ and\ \bibinfo {author} {\bibfnamefont {L.}~\bibnamefont
  {{Lindblom}}},\ }\href {\doibase 10.1086/163127} {\bibfield  {journal}
  {\bibinfo  {journal} {\apj}\ }\textbf {\bibinfo {volume} {292}},\ \bibinfo
  {pages} {12} (\bibinfo {year} {1985})}\BibitemShut {NoStop}%
\bibitem [{\citenamefont {Wei}\ \emph {et~al.}(2020)\citenamefont {Wei},
  \citenamefont {Salinas}, \citenamefont {Kl{\"a}hn}, \citenamefont
  {Jaikumar},\ and\ \citenamefont {Barry}}]{wei2020lifting}%
  \BibitemOpen
  \bibfield  {author} {\bibinfo {author} {\bibfnamefont {W.}~\bibnamefont
  {Wei}}, \bibinfo {author} {\bibfnamefont {M.}~\bibnamefont {Salinas}},
  \bibinfo {author} {\bibfnamefont {T.}~\bibnamefont {Kl{\"a}hn}}, \bibinfo
  {author} {\bibfnamefont {P.}~\bibnamefont {Jaikumar}}, \ and\ \bibinfo
  {author} {\bibfnamefont {M.}~\bibnamefont {Barry}},\ }\href@noop {}
  {\bibfield  {journal} {\bibinfo  {journal} {The Astrophysical Journal}\
  }\textbf {\bibinfo {volume} {904}},\ \bibinfo {pages} {187} (\bibinfo {year}
  {2020})}\BibitemShut {NoStop}%
\bibitem [{\citenamefont {Jaikumar}\ \emph {et~al.}(2021)\citenamefont
  {Jaikumar}, \citenamefont {Semposki}, \citenamefont {Prakash},\ and\
  \citenamefont {Constantinou}}]{jaikumar2021g}%
  \BibitemOpen
  \bibfield  {author} {\bibinfo {author} {\bibfnamefont {P.}~\bibnamefont
  {Jaikumar}}, \bibinfo {author} {\bibfnamefont {A.}~\bibnamefont {Semposki}},
  \bibinfo {author} {\bibfnamefont {M.}~\bibnamefont {Prakash}}, \ and\
  \bibinfo {author} {\bibfnamefont {C.}~\bibnamefont {Constantinou}},\
  }\href@noop {} {\bibfield  {journal} {\bibinfo  {journal} {Phys. Rev. D}\
  }\textbf {\bibinfo {volume} {103}},\ \bibinfo {pages} {123009} (\bibinfo
  {year} {2021})}\BibitemShut {NoStop}%
\bibitem [{\citenamefont {Reisenegger}\ and\ \citenamefont
  {Goldreich}(1992)}]{reisenegger1992new}%
  \BibitemOpen
  \bibfield  {author} {\bibinfo {author} {\bibfnamefont {A.}~\bibnamefont
  {Reisenegger}}\ and\ \bibinfo {author} {\bibfnamefont {P.}~\bibnamefont
  {Goldreich}},\ }\href@noop {} {\bibfield  {journal} {\bibinfo  {journal}
  {Astrophysical Journal}\ }\textbf {\bibinfo {volume} {395}},\ \bibinfo
  {pages} {240} (\bibinfo {year} {1992})}\BibitemShut {NoStop}%
\bibitem [{\citenamefont {Kuan}\ \emph {et~al.}(2022)\citenamefont {Kuan},
  \citenamefont {Kr{\"u}ger}, \citenamefont {Suvorov},\ and\ \citenamefont
  {Kokkotas}}]{kuan2022constraining}%
  \BibitemOpen
  \bibfield  {author} {\bibinfo {author} {\bibfnamefont {H.-J.}\ \bibnamefont
  {Kuan}}, \bibinfo {author} {\bibfnamefont {C.~J.}\ \bibnamefont
  {Kr{\"u}ger}}, \bibinfo {author} {\bibfnamefont {A.~G.}\ \bibnamefont
  {Suvorov}}, \ and\ \bibinfo {author} {\bibfnamefont {K.~D.}\ \bibnamefont
  {Kokkotas}},\ }\href@noop {} {\bibfield  {journal} {\bibinfo  {journal}
  {arXiv preprint arXiv:2204.08492}\ } (\bibinfo {year} {2022})}\BibitemShut
  {NoStop}%
\bibitem [{\citenamefont {Fackerell}(1971)}]{fackerell1971solutions}%
  \BibitemOpen
  \bibfield  {author} {\bibinfo {author} {\bibfnamefont {E.~D.}\ \bibnamefont
  {Fackerell}},\ }\href@noop {} {\bibfield  {journal} {\bibinfo  {journal} {The
  Astrophysical Journal}\ }\textbf {\bibinfo {volume} {166}},\ \bibinfo {pages}
  {197} (\bibinfo {year} {1971})}\BibitemShut {NoStop}%
\bibitem [{\citenamefont {Brown}\ \emph {et~al.}(1989)\citenamefont {Brown},
  \citenamefont {Byrne},\ and\ \citenamefont {Hindmarsh}}]{brown1989vode}%
  \BibitemOpen
  \bibfield  {author} {\bibinfo {author} {\bibfnamefont {P.~N.}\ \bibnamefont
  {Brown}}, \bibinfo {author} {\bibfnamefont {G.~D.}\ \bibnamefont {Byrne}}, \
  and\ \bibinfo {author} {\bibfnamefont {A.~C.}\ \bibnamefont {Hindmarsh}},\
  }\href@noop {} {\bibfield  {journal} {\bibinfo  {journal} {SIAM journal on
  scientific and statistical computing}\ }\textbf {\bibinfo {volume} {10}},\
  \bibinfo {pages} {1038} (\bibinfo {year} {1989})}\BibitemShut {NoStop}%
\bibitem [{\citenamefont {Miniutti}\ \emph {et~al.}(2003)\citenamefont
  {Miniutti}, \citenamefont {Pons}, \citenamefont {Berti}, \citenamefont
  {Gualtieri},\ and\ \citenamefont {Ferrari}}]{miniutti2003non}%
  \BibitemOpen
  \bibfield  {author} {\bibinfo {author} {\bibfnamefont {G.}~\bibnamefont
  {Miniutti}}, \bibinfo {author} {\bibfnamefont {J.}~\bibnamefont {Pons}},
  \bibinfo {author} {\bibfnamefont {E.}~\bibnamefont {Berti}}, \bibinfo
  {author} {\bibfnamefont {L.}~\bibnamefont {Gualtieri}}, \ and\ \bibinfo
  {author} {\bibfnamefont {V.}~\bibnamefont {Ferrari}},\ }\href@noop {}
  {\bibfield  {journal} {\bibinfo  {journal} {Monthly Notices of the Royal
  Astronomical Society}\ }\textbf {\bibinfo {volume} {338}},\ \bibinfo {pages}
  {389} (\bibinfo {year} {2003})}\BibitemShut {NoStop}%
\bibitem [{\citenamefont {Finn}(1987)}]{finn1987g}%
  \BibitemOpen
  \bibfield  {author} {\bibinfo {author} {\bibfnamefont {L.~S.}\ \bibnamefont
  {Finn}},\ }\href@noop {} {\bibfield  {journal} {\bibinfo  {journal} {Monthly
  Notices of the Royal Astronomical Society}\ }\textbf {\bibinfo {volume}
  {227}},\ \bibinfo {pages} {265} (\bibinfo {year} {1987})}\BibitemShut
  {NoStop}%
\bibitem [{\citenamefont {Cowling}(1941)}]{cowling1941non}%
  \BibitemOpen
  \bibfield  {author} {\bibinfo {author} {\bibfnamefont {T.~G.}\ \bibnamefont
  {Cowling}},\ }\href@noop {} {\bibfield  {journal} {\bibinfo  {journal}
  {Monthly Notices of the Royal Astronomical Society}\ }\textbf {\bibinfo
  {volume} {101}},\ \bibinfo {pages} {367} (\bibinfo {year}
  {1941})}\BibitemShut {NoStop}%
\bibitem [{\citenamefont {{McDermott}}\ \emph {et~al.}(1983)\citenamefont
  {{McDermott}}, \citenamefont {{van Horn}},\ and\ \citenamefont
  {{Scholl}}}]{McD83}%
  \BibitemOpen
  \bibfield  {author} {\bibinfo {author} {\bibfnamefont {P.~N.}\ \bibnamefont
  {{McDermott}}}, \bibinfo {author} {\bibfnamefont {H.~M.}\ \bibnamefont {{van
  Horn}}}, \ and\ \bibinfo {author} {\bibfnamefont {J.~F.}\ \bibnamefont
  {{Scholl}}},\ }\href {\doibase 10.1086/161006} {\bibfield  {journal}
  {\bibinfo  {journal} {\apj}\ }\textbf {\bibinfo {volume} {268}},\ \bibinfo
  {pages} {837} (\bibinfo {year} {1983})}\BibitemShut {NoStop}%
\bibitem [{\citenamefont {Dhiman}\ \emph {et~al.}(2007)\citenamefont {Dhiman},
  \citenamefont {Kumar},\ and\ \citenamefont {Agrawal}}]{PhysRevC.76.045801}%
  \BibitemOpen
  \bibfield  {author} {\bibinfo {author} {\bibfnamefont {S.~K.}\ \bibnamefont
  {Dhiman}}, \bibinfo {author} {\bibfnamefont {R.}~\bibnamefont {Kumar}}, \
  and\ \bibinfo {author} {\bibfnamefont {B.~K.}\ \bibnamefont {Agrawal}},\
  }\href {\doibase 10.1103/PhysRevC.76.045801} {\bibfield  {journal} {\bibinfo
  {journal} {Phys. Rev. C}\ }\textbf {\bibinfo {volume} {76}},\ \bibinfo
  {pages} {045801} (\bibinfo {year} {2007})}\BibitemShut {NoStop}%
\bibitem [{\citenamefont {Chen}\ and\ \citenamefont
  {Piekarewicz}(2014{\natexlab{a}})}]{PhysRevC.90.044305}%
  \BibitemOpen
  \bibfield  {author} {\bibinfo {author} {\bibfnamefont {W.-C.}\ \bibnamefont
  {Chen}}\ and\ \bibinfo {author} {\bibfnamefont {J.}~\bibnamefont
  {Piekarewicz}},\ }\href {\doibase 10.1103/PhysRevC.90.044305} {\bibfield
  {journal} {\bibinfo  {journal} {Phys. Rev. C}\ }\textbf {\bibinfo {volume}
  {90}},\ \bibinfo {pages} {044305} (\bibinfo {year}
  {2014}{\natexlab{a}})}\BibitemShut {NoStop}%
\bibitem [{\citenamefont {Glendenning}\ and\ \citenamefont
  {Moszkowski}(1991)}]{PhysRevLett.67.2414}%
  \BibitemOpen
  \bibfield  {author} {\bibinfo {author} {\bibfnamefont {N.~K.}\ \bibnamefont
  {Glendenning}}\ and\ \bibinfo {author} {\bibfnamefont {S.~A.}\ \bibnamefont
  {Moszkowski}},\ }\href {\doibase 10.1103/PhysRevLett.67.2414} {\bibfield
  {journal} {\bibinfo  {journal} {Phys. Rev. Lett.}\ }\textbf {\bibinfo
  {volume} {67}},\ \bibinfo {pages} {2414} (\bibinfo {year}
  {1991})}\BibitemShut {NoStop}%
\bibitem [{\citenamefont {Lalazissis}\ \emph {et~al.}(1997)\citenamefont
  {Lalazissis}, \citenamefont {K\"onig},\ and\ \citenamefont
  {Ring}}]{PhysRevC.55.540}%
  \BibitemOpen
  \bibfield  {author} {\bibinfo {author} {\bibfnamefont {G.~A.}\ \bibnamefont
  {Lalazissis}}, \bibinfo {author} {\bibfnamefont {J.}~\bibnamefont {K\"onig}},
  \ and\ \bibinfo {author} {\bibfnamefont {P.}~\bibnamefont {Ring}},\ }\href
  {\doibase 10.1103/PhysRevC.55.540} {\bibfield  {journal} {\bibinfo  {journal}
  {Phys. Rev. C}\ }\textbf {\bibinfo {volume} {55}},\ \bibinfo {pages} {540}
  (\bibinfo {year} {1997})}\BibitemShut {NoStop}%
\bibitem [{\citenamefont {Carriere}\ \emph {et~al.}(2003)\citenamefont
  {Carriere}, \citenamefont {Horowitz},\ and\ \citenamefont
  {Piekarewicz}}]{Carriere_2003}%
  \BibitemOpen
  \bibfield  {author} {\bibinfo {author} {\bibfnamefont {J.}~\bibnamefont
  {Carriere}}, \bibinfo {author} {\bibfnamefont {C.~J.}\ \bibnamefont
  {Horowitz}}, \ and\ \bibinfo {author} {\bibfnamefont {J.}~\bibnamefont
  {Piekarewicz}},\ }\href {\doibase 10.1086/376515} {\bibfield  {journal}
  {\bibinfo  {journal} {The Astrophysical Journal}\ }\textbf {\bibinfo {volume}
  {593}},\ \bibinfo {pages} {463} (\bibinfo {year} {2003})}\BibitemShut
  {NoStop}%
\bibitem [{\citenamefont {{Kumar}}\ \emph {et~al.}(2018)\citenamefont
  {{Kumar}}, \citenamefont {{Patra}},\ and\ \citenamefont
  {{Agrawal}}}]{Kumar_ns}%
  \BibitemOpen
  \bibfield  {author} {\bibinfo {author} {\bibfnamefont {B.}~\bibnamefont
  {{Kumar}}}, \bibinfo {author} {\bibfnamefont {S.~K.}\ \bibnamefont
  {{Patra}}}, \ and\ \bibinfo {author} {\bibfnamefont {B.~K.}\ \bibnamefont
  {{Agrawal}}},\ }\href {\doibase 10.1103/PhysRevC.97.045806} {\bibfield
  {journal} {\bibinfo  {journal} {\prc}\ }\textbf {\bibinfo {volume} {97}},\
  \bibinfo {eid} {045806} (\bibinfo {year} {2018})},\ \Eprint
  {http://arxiv.org/abs/1711.04940} {arXiv:1711.04940 [nucl-th]} \BibitemShut
  {NoStop}%
\bibitem [{\citenamefont {Parmar}\ \emph {et~al.}(2022)\citenamefont {Parmar},
  \citenamefont {Das}, \citenamefont {Kumar}, \citenamefont {Sharma},\ and\
  \citenamefont {Patra}}]{vishal}%
  \BibitemOpen
  \bibfield  {author} {\bibinfo {author} {\bibfnamefont {V.}~\bibnamefont
  {Parmar}}, \bibinfo {author} {\bibfnamefont {H.~C.}\ \bibnamefont {Das}},
  \bibinfo {author} {\bibfnamefont {A.}~\bibnamefont {Kumar}}, \bibinfo
  {author} {\bibfnamefont {M.~K.}\ \bibnamefont {Sharma}}, \ and\ \bibinfo
  {author} {\bibfnamefont {S.~K.}\ \bibnamefont {Patra}},\ }\href {\doibase
  10.1103/PhysRevD.105.043017} {\bibfield  {journal} {\bibinfo  {journal}
  {Phys. Rev. D}\ }\textbf {\bibinfo {volume} {105}},\ \bibinfo {pages}
  {043017} (\bibinfo {year} {2022})}\BibitemShut {NoStop}%
\bibitem [{\citenamefont {{Kumar}}\ \emph {et~al.}(2017)\citenamefont
  {{Kumar}}, \citenamefont {{Singh}}, \citenamefont {{Agrawal}},\ and\
  \citenamefont {{Patra}}}]{Kumar_param}%
  \BibitemOpen
  \bibfield  {author} {\bibinfo {author} {\bibfnamefont {B.}~\bibnamefont
  {{Kumar}}}, \bibinfo {author} {\bibfnamefont {S.~K.}\ \bibnamefont
  {{Singh}}}, \bibinfo {author} {\bibfnamefont {B.~K.}\ \bibnamefont
  {{Agrawal}}}, \ and\ \bibinfo {author} {\bibfnamefont {S.~K.}\ \bibnamefont
  {{Patra}}},\ }\href {\doibase 10.1016/j.nuclphysa.2017.07.001} {\bibfield
  {journal} {\bibinfo  {journal} {NuPhA}\ }\textbf {\bibinfo {volume} {966}},\
  \bibinfo {pages} {197} (\bibinfo {year} {2017})},\ \Eprint
  {http://arxiv.org/abs/1705.02621} {arXiv:1705.02621 [nucl-th]} \BibitemShut
  {NoStop}%
\bibitem [{\citenamefont {Sugahara}\ and\ \citenamefont {Toki}(1994)}]{TM1}%
  \BibitemOpen
  \bibfield  {author} {\bibinfo {author} {\bibfnamefont {Y.}~\bibnamefont
  {Sugahara}}\ and\ \bibinfo {author} {\bibfnamefont {H.}~\bibnamefont
  {Toki}},\ }\href {\doibase https://doi.org/10.1016/0375-9474(94)90923-7}
  {\bibfield  {journal} {\bibinfo  {journal} {Nuclear Physics A}\ }\textbf
  {\bibinfo {volume} {579}},\ \bibinfo {pages} {557} (\bibinfo {year}
  {1994})}\BibitemShut {NoStop}%
\bibitem [{\citenamefont {Chen}\ and\ \citenamefont
  {Piekarewicz}(2014{\natexlab{b}})}]{FsuG}%
  \BibitemOpen
  \bibfield  {author} {\bibinfo {author} {\bibfnamefont {W.-C.}\ \bibnamefont
  {Chen}}\ and\ \bibinfo {author} {\bibfnamefont {J.}~\bibnamefont
  {Piekarewicz}},\ }\href {\doibase 10.1103/PhysRevC.90.044305} {\bibfield
  {journal} {\bibinfo  {journal} {Phys. Rev. C}\ }\textbf {\bibinfo {volume}
  {90}},\ \bibinfo {pages} {044305} (\bibinfo {year}
  {2014}{\natexlab{b}})}\BibitemShut {NoStop}%
\bibitem [{\citenamefont {Typel}\ \emph {et~al.}(2010)\citenamefont {Typel},
  \citenamefont {R\"opke}, \citenamefont {Kl\"ahn}, \citenamefont {Blaschke},\
  and\ \citenamefont {Wolter}}]{PhysRevC.81.015803}%
  \BibitemOpen
  \bibfield  {author} {\bibinfo {author} {\bibfnamefont {S.}~\bibnamefont
  {Typel}}, \bibinfo {author} {\bibfnamefont {G.}~\bibnamefont {R\"opke}},
  \bibinfo {author} {\bibfnamefont {T.}~\bibnamefont {Kl\"ahn}}, \bibinfo
  {author} {\bibfnamefont {D.}~\bibnamefont {Blaschke}}, \ and\ \bibinfo
  {author} {\bibfnamefont {H.~H.}\ \bibnamefont {Wolter}},\ }\href {\doibase
  10.1103/PhysRevC.81.015803} {\bibfield  {journal} {\bibinfo  {journal} {Phys.
  Rev. C}\ }\textbf {\bibinfo {volume} {81}},\ \bibinfo {pages} {015803}
  (\bibinfo {year} {2010})}\BibitemShut {NoStop}%
\bibitem [{\citenamefont {Gaitanos}\ \emph {et~al.}(2004)\citenamefont
  {Gaitanos}, \citenamefont {Toro}, \citenamefont {Typel}, \citenamefont
  {Baran}, \citenamefont {Fuchs}, \citenamefont {Greco},\ and\ \citenamefont
  {Wolter}}]{Gaitanos_2004}%
  \BibitemOpen
  \bibfield  {author} {\bibinfo {author} {\bibfnamefont {T.}~\bibnamefont
  {Gaitanos}}, \bibinfo {author} {\bibfnamefont {M.~D.}\ \bibnamefont {Toro}},
  \bibinfo {author} {\bibfnamefont {S.}~\bibnamefont {Typel}}, \bibinfo
  {author} {\bibfnamefont {V.}~\bibnamefont {Baran}}, \bibinfo {author}
  {\bibfnamefont {C.}~\bibnamefont {Fuchs}}, \bibinfo {author} {\bibfnamefont
  {V.}~\bibnamefont {Greco}}, \ and\ \bibinfo {author} {\bibfnamefont
  {H.}~\bibnamefont {Wolter}},\ }\href {\doibase
  10.1016/j.nuclphysa.2003.12.001} {\bibfield  {journal} {\bibinfo  {journal}
  {Nuclear Physics A}\ }\textbf {\bibinfo {volume} {732}},\ \bibinfo {pages}
  {24} (\bibinfo {year} {2004})}\BibitemShut {NoStop}%
\bibitem [{\citenamefont {Lalazissis}\ \emph {et~al.}(2005)\citenamefont
  {Lalazissis}, \citenamefont {Nik\ifmmode \check{s}\else
  \v{s}\fi{}i\ifmmode~\acute{c}\else \'{c}\fi{}}, \citenamefont {Vretenar},\
  and\ \citenamefont {Ring}}]{DDME2}%
  \BibitemOpen
  \bibfield  {author} {\bibinfo {author} {\bibfnamefont {G.~A.}\ \bibnamefont
  {Lalazissis}}, \bibinfo {author} {\bibfnamefont {T.}~\bibnamefont
  {Nik\ifmmode \check{s}\else \v{s}\fi{}i\ifmmode~\acute{c}\else \'{c}\fi{}}},
  \bibinfo {author} {\bibfnamefont {D.}~\bibnamefont {Vretenar}}, \ and\
  \bibinfo {author} {\bibfnamefont {P.}~\bibnamefont {Ring}},\ }\href {\doibase
  10.1103/PhysRevC.71.024312} {\bibfield  {journal} {\bibinfo  {journal} {Phys.
  Rev. C}\ }\textbf {\bibinfo {volume} {71}},\ \bibinfo {pages} {024312}
  (\bibinfo {year} {2005})}\BibitemShut {NoStop}%
\bibitem [{\citenamefont {Köhler}(1976)}]{KOHLER1976301}%
  \BibitemOpen
  \bibfield  {author} {\bibinfo {author} {\bibfnamefont {H.}~\bibnamefont
  {Köhler}},\ }\href {\doibase https://doi.org/10.1016/0375-9474(76)90008-7}
  {\bibfield  {journal} {\bibinfo  {journal} {Nuclear Physics A}\ }\textbf
  {\bibinfo {volume} {258}},\ \bibinfo {pages} {301} (\bibinfo {year}
  {1976})}\BibitemShut {NoStop}%
\bibitem [{\citenamefont {Reinhard}\ and\ \citenamefont
  {Flocard}(1995)}]{REINHARD1995467}%
  \BibitemOpen
  \bibfield  {author} {\bibinfo {author} {\bibfnamefont {P.-G.}\ \bibnamefont
  {Reinhard}}\ and\ \bibinfo {author} {\bibfnamefont {H.}~\bibnamefont
  {Flocard}},\ }\href {\doibase https://doi.org/10.1016/0375-9474(94)00770-N}
  {\bibfield  {journal} {\bibinfo  {journal} {Nuclear Physics A}\ }\textbf
  {\bibinfo {volume} {584}},\ \bibinfo {pages} {467} (\bibinfo {year}
  {1995})}\BibitemShut {NoStop}%
\bibitem [{\citenamefont {Nazarewicz}\ \emph {et~al.}(1996)\citenamefont
  {Nazarewicz}, \citenamefont {Dobaczewski}, \citenamefont {Werner},
  \citenamefont {Maruhn}, \citenamefont {Reinhard}, \citenamefont {Rutz},
  \citenamefont {Chinn}, \citenamefont {Umar},\ and\ \citenamefont
  {Strayer}}]{PhysRevC.53.740}%
  \BibitemOpen
  \bibfield  {author} {\bibinfo {author} {\bibfnamefont {W.}~\bibnamefont
  {Nazarewicz}}, \bibinfo {author} {\bibfnamefont {J.}~\bibnamefont
  {Dobaczewski}}, \bibinfo {author} {\bibfnamefont {T.~R.}\ \bibnamefont
  {Werner}}, \bibinfo {author} {\bibfnamefont {J.~A.}\ \bibnamefont {Maruhn}},
  \bibinfo {author} {\bibfnamefont {P.-G.}\ \bibnamefont {Reinhard}}, \bibinfo
  {author} {\bibfnamefont {K.}~\bibnamefont {Rutz}}, \bibinfo {author}
  {\bibfnamefont {C.~R.}\ \bibnamefont {Chinn}}, \bibinfo {author}
  {\bibfnamefont {A.~S.}\ \bibnamefont {Umar}}, \ and\ \bibinfo {author}
  {\bibfnamefont {M.~R.}\ \bibnamefont {Strayer}},\ }\href {\doibase
  10.1103/PhysRevC.53.740} {\bibfield  {journal} {\bibinfo  {journal} {Phys.
  Rev. C}\ }\textbf {\bibinfo {volume} {53}},\ \bibinfo {pages} {740} (\bibinfo
  {year} {1996})}\BibitemShut {NoStop}%
\bibitem [{\citenamefont {Chabanat}()}]{thesis}%
  \BibitemOpen
  \bibfield  {author} {\bibinfo {author} {\bibfnamefont {E.}~\bibnamefont
  {Chabanat}},\ }\href@noop {} {\bibinfo  {journal} {Ph.D. thesis, University
  Claude Bernard Lyon-1, Lyon, France (1995)}\ }\BibitemShut {NoStop}%
\bibitem [{\citenamefont {Chabanat}\ \emph {et~al.}(1997)\citenamefont
  {Chabanat}, \citenamefont {Bonche}, \citenamefont {Haensel}, \citenamefont
  {Meyer},\ and\ \citenamefont {Schaeffer}}]{CHABANAT1997710}%
  \BibitemOpen
\bibfield  {journal} {  }\bibfield  {author} {\bibinfo {author} {\bibfnamefont
  {E.}~\bibnamefont {Chabanat}}, \bibinfo {author} {\bibfnamefont
  {P.}~\bibnamefont {Bonche}}, \bibinfo {author} {\bibfnamefont
  {P.}~\bibnamefont {Haensel}}, \bibinfo {author} {\bibfnamefont
  {J.}~\bibnamefont {Meyer}}, \ and\ \bibinfo {author} {\bibfnamefont
  {R.}~\bibnamefont {Schaeffer}},\ }\href {\doibase
  https://doi.org/10.1016/S0375-9474(97)00596-4} {\bibfield  {journal}
  {\bibinfo  {journal} {Nuclear Physics A}\ }\textbf {\bibinfo {volume}
  {627}},\ \bibinfo {pages} {710} (\bibinfo {year} {1997})}\BibitemShut
  {NoStop}%
\bibitem [{\citenamefont {Chabanat}\ \emph {et~al.}(1998)\citenamefont
  {Chabanat}, \citenamefont {Bonche}, \citenamefont {Haensel}, \citenamefont
  {Meyer},\ and\ \citenamefont {Schaeffer}}]{CHABANAT1998231}%
  \BibitemOpen
  \bibfield  {author} {\bibinfo {author} {\bibfnamefont {E.}~\bibnamefont
  {Chabanat}}, \bibinfo {author} {\bibfnamefont {P.}~\bibnamefont {Bonche}},
  \bibinfo {author} {\bibfnamefont {P.}~\bibnamefont {Haensel}}, \bibinfo
  {author} {\bibfnamefont {J.}~\bibnamefont {Meyer}}, \ and\ \bibinfo {author}
  {\bibfnamefont {R.}~\bibnamefont {Schaeffer}},\ }\href {\doibase
  https://doi.org/10.1016/S0375-9474(98)00180-8} {\bibfield  {journal}
  {\bibinfo  {journal} {Nuclear Physics A}\ }\textbf {\bibinfo {volume}
  {635}},\ \bibinfo {pages} {231} (\bibinfo {year} {1998})}\BibitemShut
  {NoStop}%
\bibitem [{\citenamefont {Bennour}\ \emph {et~al.}(1989)\citenamefont
  {Bennour}, \citenamefont {Heenen}, \citenamefont {Bonche}, \citenamefont
  {Dobaczewski},\ and\ \citenamefont {Flocard}}]{PhysRevC.40.2834}%
  \BibitemOpen
  \bibfield  {author} {\bibinfo {author} {\bibfnamefont {L.}~\bibnamefont
  {Bennour}}, \bibinfo {author} {\bibfnamefont {P.-H.}\ \bibnamefont {Heenen}},
  \bibinfo {author} {\bibfnamefont {P.}~\bibnamefont {Bonche}}, \bibinfo
  {author} {\bibfnamefont {J.}~\bibnamefont {Dobaczewski}}, \ and\ \bibinfo
  {author} {\bibfnamefont {H.}~\bibnamefont {Flocard}},\ }\href {\doibase
  10.1103/PhysRevC.40.2834} {\bibfield  {journal} {\bibinfo  {journal} {Phys.
  Rev. C}\ }\textbf {\bibinfo {volume} {40}},\ \bibinfo {pages} {2834}
  (\bibinfo {year} {1989})}\BibitemShut {NoStop}%
\bibitem [{\citenamefont {Reinhard}(1999)}]{REINHARD1999305}%
  \BibitemOpen
  \bibfield  {author} {\bibinfo {author} {\bibfnamefont {P.-G.}\ \bibnamefont
  {Reinhard}},\ }\href {\doibase https://doi.org/10.1016/S0375-9474(99)00076-7}
  {\bibfield  {journal} {\bibinfo  {journal} {Nuclear Physics A}\ }\textbf
  {\bibinfo {volume} {649}},\ \bibinfo {pages} {305} (\bibinfo {year}
  {1999})},\ \bibinfo {note} {giant Resonances}\BibitemShut {NoStop}%
\bibitem [{\citenamefont {Agrawal}\ \emph {et~al.}(2005)\citenamefont
  {Agrawal}, \citenamefont {Shlomo},\ and\ \citenamefont
  {Au}}]{PhysRevC.72.014310}%
  \BibitemOpen
  \bibfield  {author} {\bibinfo {author} {\bibfnamefont {B.~K.}\ \bibnamefont
  {Agrawal}}, \bibinfo {author} {\bibfnamefont {S.}~\bibnamefont {Shlomo}}, \
  and\ \bibinfo {author} {\bibfnamefont {V.~K.}\ \bibnamefont {Au}},\ }\href
  {\doibase 10.1103/PhysRevC.72.014310} {\bibfield  {journal} {\bibinfo
  {journal} {Phys. Rev. C}\ }\textbf {\bibinfo {volume} {72}},\ \bibinfo
  {pages} {014310} (\bibinfo {year} {2005})}\BibitemShut {NoStop}%
\bibitem [{\citenamefont {Agrawal}\ \emph {et~al.}(2003)\citenamefont
  {Agrawal}, \citenamefont {Shlomo},\ and\ \citenamefont
  {Sanzhur}}]{PhysRevC.67.034314}%
  \BibitemOpen
  \bibfield  {author} {\bibinfo {author} {\bibfnamefont {B.~K.}\ \bibnamefont
  {Agrawal}}, \bibinfo {author} {\bibfnamefont {S.}~\bibnamefont {Shlomo}}, \
  and\ \bibinfo {author} {\bibfnamefont {A.~I.}\ \bibnamefont {Sanzhur}},\
  }\href {\doibase 10.1103/PhysRevC.67.034314} {\bibfield  {journal} {\bibinfo
  {journal} {Phys. Rev. C}\ }\textbf {\bibinfo {volume} {67}},\ \bibinfo
  {pages} {034314} (\bibinfo {year} {2003})}\BibitemShut {NoStop}%
\bibitem [{\citenamefont {Friedrich}\ and\ \citenamefont
  {Reinhard}(1986)}]{PhysRevC.33.335}%
  \BibitemOpen
  \bibfield  {author} {\bibinfo {author} {\bibfnamefont {J.}~\bibnamefont
  {Friedrich}}\ and\ \bibinfo {author} {\bibfnamefont {P.-G.}\ \bibnamefont
  {Reinhard}},\ }\href {\doibase 10.1103/PhysRevC.33.335} {\bibfield  {journal}
  {\bibinfo  {journal} {Phys. Rev. C}\ }\textbf {\bibinfo {volume} {33}},\
  \bibinfo {pages} {335} (\bibinfo {year} {1986})}\BibitemShut {NoStop}%
\bibitem [{\citenamefont {Goriely}\ \emph {et~al.}(2010)\citenamefont
  {Goriely}, \citenamefont {Chamel},\ and\ \citenamefont
  {Pearson}}]{PhysRevC.82.035804}%
  \BibitemOpen
  \bibfield  {author} {\bibinfo {author} {\bibfnamefont {S.}~\bibnamefont
  {Goriely}}, \bibinfo {author} {\bibfnamefont {N.}~\bibnamefont {Chamel}}, \
  and\ \bibinfo {author} {\bibfnamefont {J.~M.}\ \bibnamefont {Pearson}},\
  }\href {\doibase 10.1103/PhysRevC.82.035804} {\bibfield  {journal} {\bibinfo
  {journal} {Phys. Rev. C}\ }\textbf {\bibinfo {volume} {82}},\ \bibinfo
  {pages} {035804} (\bibinfo {year} {2010})}\BibitemShut {NoStop}%
\bibitem [{\citenamefont {Goriely}\ \emph {et~al.}(2013)\citenamefont
  {Goriely}, \citenamefont {Chamel},\ and\ \citenamefont
  {Pearson}}]{PhysRevC.88.024308}%
  \BibitemOpen
  \bibfield  {author} {\bibinfo {author} {\bibfnamefont {S.}~\bibnamefont
  {Goriely}}, \bibinfo {author} {\bibfnamefont {N.}~\bibnamefont {Chamel}}, \
  and\ \bibinfo {author} {\bibfnamefont {J.~M.}\ \bibnamefont {Pearson}},\
  }\href {\doibase 10.1103/PhysRevC.88.024308} {\bibfield  {journal} {\bibinfo
  {journal} {Phys. Rev. C}\ }\textbf {\bibinfo {volume} {88}},\ \bibinfo
  {pages} {024308} (\bibinfo {year} {2013})}\BibitemShut {NoStop}%
\bibitem [{\citenamefont {{Baym}}\ \emph {et~al.}(1971)\citenamefont {{Baym}},
  \citenamefont {{Pethick}},\ and\ \citenamefont {{Sutherland}}}]{BPS}%
  \BibitemOpen
  \bibfield  {author} {\bibinfo {author} {\bibfnamefont {G.}~\bibnamefont
  {{Baym}}}, \bibinfo {author} {\bibfnamefont {C.}~\bibnamefont {{Pethick}}}, \
  and\ \bibinfo {author} {\bibfnamefont {P.}~\bibnamefont {{Sutherland}}},\
  }\href {\doibase 10.1086/151216} {\bibfield  {journal} {\bibinfo  {journal}
  {\apj}\ }\textbf {\bibinfo {volume} {170}},\ \bibinfo {pages} {299} (\bibinfo
  {year} {1971})}\BibitemShut {NoStop}%
\bibitem [{\citenamefont {Gulminelli}\ and\ \citenamefont
  {Raduta}(2015)}]{Gul2015}%
  \BibitemOpen
  \bibfield  {author} {\bibinfo {author} {\bibfnamefont {F.}~\bibnamefont
  {Gulminelli}}\ and\ \bibinfo {author} {\bibfnamefont {A.~R.}\ \bibnamefont
  {Raduta}},\ }\href {\doibase 10.1103/PhysRevC.92.055803} {\bibfield
  {journal} {\bibinfo  {journal} {Phys. Rev. C}\ }\textbf {\bibinfo {volume}
  {92}},\ \bibinfo {pages} {055803} (\bibinfo {year} {2015})}\BibitemShut
  {NoStop}%
\bibitem [{\citenamefont {{Fortin}}\ \emph {et~al.}(2016)\citenamefont
  {{Fortin}} \emph {et~al.}}]{Fortin}%
  \BibitemOpen
  \bibfield  {author} {\bibinfo {author} {\bibfnamefont {M.}~\bibnamefont
  {{Fortin}}} \emph {et~al.},\ }\href {\doibase 10.1103/PhysRevC.94.035804}
  {\bibfield  {journal} {\bibinfo  {journal} {\prc}\ }\textbf {\bibinfo
  {volume} {94}},\ \bibinfo {eid} {035804} (\bibinfo {year} {2016})},\ \Eprint
  {http://arxiv.org/abs/1604.01944} {arXiv:1604.01944 [astro-ph.SR]}
  \BibitemShut {NoStop}%
\bibitem [{\citenamefont {{Malik}}\ \emph {et~al.}(2018)\citenamefont {{Malik}}
  \emph {et~al.}}]{Tuhin}%
  \BibitemOpen
  \bibfield  {author} {\bibinfo {author} {\bibfnamefont {T.}~\bibnamefont
  {{Malik}}} \emph {et~al.},\ }\href {\doibase 10.1103/PhysRevC.98.035804}
  {\bibfield  {journal} {\bibinfo  {journal} {PhRvC}\ }\textbf {\bibinfo
  {volume} {98}},\ \bibinfo {pages} {035804} (\bibinfo {year} {2018})},\
  \Eprint {http://arxiv.org/abs/1805.11963} {arXiv:1805.11963 [nucl-th]}
  \BibitemShut {NoStop}%
\bibitem [{\citenamefont {Antoniadis}\ \emph {et~al.}(2013)\citenamefont
  {Antoniadis}, \citenamefont {Freire}, \citenamefont {Wex}, \citenamefont
  {Tauris}, \citenamefont {Lynch}, \citenamefont {van Kerkwijk}, \citenamefont
  {Kramer}, \citenamefont {Bassa}, \citenamefont {Dhillon}, \citenamefont
  {Driebe}, \citenamefont {Hessels}, \citenamefont {Kaspi}, \citenamefont
  {Kondratiev}, \citenamefont {Langer}, \citenamefont {Marsh}, \citenamefont
  {McLaughlin}, \citenamefont {Pennucci}, \citenamefont {Ransom}, \citenamefont
  {Stairs}, \citenamefont {van Leeuwen}, \citenamefont {Verbiest},\ and\
  \citenamefont {Whelan}}]{anto}%
  \BibitemOpen
  \bibfield  {author} {\bibinfo {author} {\bibfnamefont {J.}~\bibnamefont
  {Antoniadis}}, \bibinfo {author} {\bibfnamefont {P.~C.~C.}\ \bibnamefont
  {Freire}}, \bibinfo {author} {\bibfnamefont {N.}~\bibnamefont {Wex}},
  \bibinfo {author} {\bibfnamefont {T.~M.}\ \bibnamefont {Tauris}}, \bibinfo
  {author} {\bibfnamefont {R.~S.}\ \bibnamefont {Lynch}}, \bibinfo {author}
  {\bibfnamefont {M.~H.}\ \bibnamefont {van Kerkwijk}}, \bibinfo {author}
  {\bibfnamefont {M.}~\bibnamefont {Kramer}}, \bibinfo {author} {\bibfnamefont
  {C.}~\bibnamefont {Bassa}}, \bibinfo {author} {\bibfnamefont {V.~S.}\
  \bibnamefont {Dhillon}}, \bibinfo {author} {\bibfnamefont {T.}~\bibnamefont
  {Driebe}}, \bibinfo {author} {\bibfnamefont {J.~W.~T.}\ \bibnamefont
  {Hessels}}, \bibinfo {author} {\bibfnamefont {V.~M.}\ \bibnamefont {Kaspi}},
  \bibinfo {author} {\bibfnamefont {V.~I.}\ \bibnamefont {Kondratiev}},
  \bibinfo {author} {\bibfnamefont {N.}~\bibnamefont {Langer}}, \bibinfo
  {author} {\bibfnamefont {T.~R.}\ \bibnamefont {Marsh}}, \bibinfo {author}
  {\bibfnamefont {M.~A.}\ \bibnamefont {McLaughlin}}, \bibinfo {author}
  {\bibfnamefont {T.~T.}\ \bibnamefont {Pennucci}}, \bibinfo {author}
  {\bibfnamefont {S.~M.}\ \bibnamefont {Ransom}}, \bibinfo {author}
  {\bibfnamefont {I.~H.}\ \bibnamefont {Stairs}}, \bibinfo {author}
  {\bibfnamefont {J.}~\bibnamefont {van Leeuwen}}, \bibinfo {author}
  {\bibfnamefont {J.~P.~W.}\ \bibnamefont {Verbiest}}, \ and\ \bibinfo {author}
  {\bibfnamefont {D.~G.}\ \bibnamefont {Whelan}},\ }\href {\doibase
  10.1126/science.1233232} {\bibfield  {journal} {\bibinfo  {journal}
  {Science}\ }\textbf {\bibinfo {volume} {340}},\ \bibinfo {pages} {1233232}
  (\bibinfo {year} {2013})},\ \Eprint
  {http://arxiv.org/abs/https://www.science.org/doi/pdf/10.1126/science.1233232}
  {https://www.science.org/doi/pdf/10.1126/science.1233232} \BibitemShut
  {NoStop}%
\bibitem [{\citenamefont {Miller}\ \emph {et~al.}(2019)\citenamefont {Miller},
  \citenamefont {Lamb}, \citenamefont {Dittmann}, \citenamefont {Bogdanov},
  \citenamefont {Arzoumanian}, \citenamefont {Gendreau}, \citenamefont
  {Guillot}, \citenamefont {Harding}, \citenamefont {Ho}, \citenamefont
  {Lattimer}, \citenamefont {Ludlam}, \citenamefont {Mahmoodifar},
  \citenamefont {Morsink}, \citenamefont {Ray}, \citenamefont {Strohmayer},
  \citenamefont {Wood}, \citenamefont {Enoto}, \citenamefont {Foster},
  \citenamefont {Okajima}, \citenamefont {Prigozhin},\ and\ \citenamefont
  {Soong}}]{Miller_2019}%
  \BibitemOpen
  \bibfield  {author} {\bibinfo {author} {\bibfnamefont {M.~C.}\ \bibnamefont
  {Miller}}, \bibinfo {author} {\bibfnamefont {F.~K.}\ \bibnamefont {Lamb}},
  \bibinfo {author} {\bibfnamefont {A.~J.}\ \bibnamefont {Dittmann}}, \bibinfo
  {author} {\bibfnamefont {S.}~\bibnamefont {Bogdanov}}, \bibinfo {author}
  {\bibfnamefont {Z.}~\bibnamefont {Arzoumanian}}, \bibinfo {author}
  {\bibfnamefont {K.~C.}\ \bibnamefont {Gendreau}}, \bibinfo {author}
  {\bibfnamefont {S.}~\bibnamefont {Guillot}}, \bibinfo {author} {\bibfnamefont
  {A.~K.}\ \bibnamefont {Harding}}, \bibinfo {author} {\bibfnamefont
  {W.~C.~G.}\ \bibnamefont {Ho}}, \bibinfo {author} {\bibfnamefont {J.~M.}\
  \bibnamefont {Lattimer}}, \bibinfo {author} {\bibfnamefont {R.~M.}\
  \bibnamefont {Ludlam}}, \bibinfo {author} {\bibfnamefont {S.}~\bibnamefont
  {Mahmoodifar}}, \bibinfo {author} {\bibfnamefont {S.~M.}\ \bibnamefont
  {Morsink}}, \bibinfo {author} {\bibfnamefont {P.~S.}\ \bibnamefont {Ray}},
  \bibinfo {author} {\bibfnamefont {T.~E.}\ \bibnamefont {Strohmayer}},
  \bibinfo {author} {\bibfnamefont {K.~S.}\ \bibnamefont {Wood}}, \bibinfo
  {author} {\bibfnamefont {T.}~\bibnamefont {Enoto}}, \bibinfo {author}
  {\bibfnamefont {R.}~\bibnamefont {Foster}}, \bibinfo {author} {\bibfnamefont
  {T.}~\bibnamefont {Okajima}}, \bibinfo {author} {\bibfnamefont
  {G.}~\bibnamefont {Prigozhin}}, \ and\ \bibinfo {author} {\bibfnamefont
  {Y.}~\bibnamefont {Soong}},\ }\href {\doibase 10.3847/2041-8213/ab50c5}
  {\bibfield  {journal} {\bibinfo  {journal} {The Astrophysical Journal}\
  }\textbf {\bibinfo {volume} {887}},\ \bibinfo {pages} {L24} (\bibinfo {year}
  {2019})}\BibitemShut {NoStop}%
\bibitem [{\citenamefont {Riley}\ \emph {et~al.}(2019)\citenamefont {Riley},
  \citenamefont {Watts}, \citenamefont {Bogdanov}, \citenamefont {Ray},
  \citenamefont {Ludlam}, \citenamefont {Guillot}, \citenamefont {Arzoumanian},
  \citenamefont {Baker}, \citenamefont {Bilous}, \citenamefont {Chakrabarty},
  \citenamefont {Gendreau}, \citenamefont {Harding}, \citenamefont {Ho},
  \citenamefont {Lattimer}, \citenamefont {Morsink},\ and\ \citenamefont
  {Strohmayer}}]{Riley_2019}%
  \BibitemOpen
  \bibfield  {author} {\bibinfo {author} {\bibfnamefont {T.~E.}\ \bibnamefont
  {Riley}}, \bibinfo {author} {\bibfnamefont {A.~L.}\ \bibnamefont {Watts}},
  \bibinfo {author} {\bibfnamefont {S.}~\bibnamefont {Bogdanov}}, \bibinfo
  {author} {\bibfnamefont {P.~S.}\ \bibnamefont {Ray}}, \bibinfo {author}
  {\bibfnamefont {R.~M.}\ \bibnamefont {Ludlam}}, \bibinfo {author}
  {\bibfnamefont {S.}~\bibnamefont {Guillot}}, \bibinfo {author} {\bibfnamefont
  {Z.}~\bibnamefont {Arzoumanian}}, \bibinfo {author} {\bibfnamefont {C.~L.}\
  \bibnamefont {Baker}}, \bibinfo {author} {\bibfnamefont {A.~V.}\ \bibnamefont
  {Bilous}}, \bibinfo {author} {\bibfnamefont {D.}~\bibnamefont {Chakrabarty}},
  \bibinfo {author} {\bibfnamefont {K.~C.}\ \bibnamefont {Gendreau}}, \bibinfo
  {author} {\bibfnamefont {A.~K.}\ \bibnamefont {Harding}}, \bibinfo {author}
  {\bibfnamefont {W.~C.~G.}\ \bibnamefont {Ho}}, \bibinfo {author}
  {\bibfnamefont {J.~M.}\ \bibnamefont {Lattimer}}, \bibinfo {author}
  {\bibfnamefont {S.~M.}\ \bibnamefont {Morsink}}, \ and\ \bibinfo {author}
  {\bibfnamefont {T.~E.}\ \bibnamefont {Strohmayer}},\ }\href {\doibase
  10.3847/2041-8213/ab481c} {\bibfield  {journal} {\bibinfo  {journal} {The
  Astrophysical Journal}\ }\textbf {\bibinfo {volume} {887}},\ \bibinfo {pages}
  {L21} (\bibinfo {year} {2019})}\BibitemShut {NoStop}%
\bibitem [{\citenamefont {Fonseca}\ \emph {et~al.}(2021)\citenamefont
  {Fonseca}, \citenamefont {Cromartie}, \citenamefont {Pennucci}, \citenamefont
  {Ray}, \citenamefont {Kirichenko}, \citenamefont {Ransom}, \citenamefont
  {Demorest}, \citenamefont {Stairs}, \citenamefont {Arzoumanian},
  \citenamefont {Guillemot}, \citenamefont {Parthasarathy}, \citenamefont
  {Kerr}, \citenamefont {Cognard}, \citenamefont {Baker}, \citenamefont
  {Blumer}, \citenamefont {Brook}, \citenamefont {DeCesar}, \citenamefont
  {Dolch}, \citenamefont {Dong}, \citenamefont {Ferrara}, \citenamefont
  {Fiore}, \citenamefont {Garver-Daniels}, \citenamefont {Good}, \citenamefont
  {Jennings}, \citenamefont {Jones}, \citenamefont {Kaspi}, \citenamefont
  {Lam}, \citenamefont {Lorimer}, \citenamefont {Luo}, \citenamefont {McEwen},
  \citenamefont {McKee}, \citenamefont {McLaughlin}, \citenamefont {McMann},
  \citenamefont {Meyers}, \citenamefont {Naidu}, \citenamefont {Ng},
  \citenamefont {Nice}, \citenamefont {Pol}, \citenamefont {Radovan},
  \citenamefont {Shapiro-Albert}, \citenamefont {Tan}, \citenamefont
  {Tendulkar}, \citenamefont {Swiggum}, \citenamefont {Wahl},\ and\
  \citenamefont {Zhu}}]{Fonseca_2021}%
  \BibitemOpen
  \bibfield  {author} {\bibinfo {author} {\bibfnamefont {E.}~\bibnamefont
  {Fonseca}}, \bibinfo {author} {\bibfnamefont {H.~T.}\ \bibnamefont
  {Cromartie}}, \bibinfo {author} {\bibfnamefont {T.~T.}\ \bibnamefont
  {Pennucci}}, \bibinfo {author} {\bibfnamefont {P.~S.}\ \bibnamefont {Ray}},
  \bibinfo {author} {\bibfnamefont {A.~Y.}\ \bibnamefont {Kirichenko}},
  \bibinfo {author} {\bibfnamefont {S.~M.}\ \bibnamefont {Ransom}}, \bibinfo
  {author} {\bibfnamefont {P.~B.}\ \bibnamefont {Demorest}}, \bibinfo {author}
  {\bibfnamefont {I.~H.}\ \bibnamefont {Stairs}}, \bibinfo {author}
  {\bibfnamefont {Z.}~\bibnamefont {Arzoumanian}}, \bibinfo {author}
  {\bibfnamefont {L.}~\bibnamefont {Guillemot}}, \bibinfo {author}
  {\bibfnamefont {A.}~\bibnamefont {Parthasarathy}}, \bibinfo {author}
  {\bibfnamefont {M.}~\bibnamefont {Kerr}}, \bibinfo {author} {\bibfnamefont
  {I.}~\bibnamefont {Cognard}}, \bibinfo {author} {\bibfnamefont {P.~T.}\
  \bibnamefont {Baker}}, \bibinfo {author} {\bibfnamefont {H.}~\bibnamefont
  {Blumer}}, \bibinfo {author} {\bibfnamefont {P.~R.}\ \bibnamefont {Brook}},
  \bibinfo {author} {\bibfnamefont {M.}~\bibnamefont {DeCesar}}, \bibinfo
  {author} {\bibfnamefont {T.}~\bibnamefont {Dolch}}, \bibinfo {author}
  {\bibfnamefont {F.~A.}\ \bibnamefont {Dong}}, \bibinfo {author}
  {\bibfnamefont {E.~C.}\ \bibnamefont {Ferrara}}, \bibinfo {author}
  {\bibfnamefont {W.}~\bibnamefont {Fiore}}, \bibinfo {author} {\bibfnamefont
  {N.}~\bibnamefont {Garver-Daniels}}, \bibinfo {author} {\bibfnamefont
  {D.~C.}\ \bibnamefont {Good}}, \bibinfo {author} {\bibfnamefont
  {R.}~\bibnamefont {Jennings}}, \bibinfo {author} {\bibfnamefont {M.~L.}\
  \bibnamefont {Jones}}, \bibinfo {author} {\bibfnamefont {V.~M.}\ \bibnamefont
  {Kaspi}}, \bibinfo {author} {\bibfnamefont {M.~T.}\ \bibnamefont {Lam}},
  \bibinfo {author} {\bibfnamefont {D.~R.}\ \bibnamefont {Lorimer}}, \bibinfo
  {author} {\bibfnamefont {J.}~\bibnamefont {Luo}}, \bibinfo {author}
  {\bibfnamefont {A.}~\bibnamefont {McEwen}}, \bibinfo {author} {\bibfnamefont
  {J.~W.}\ \bibnamefont {McKee}}, \bibinfo {author} {\bibfnamefont {M.~A.}\
  \bibnamefont {McLaughlin}}, \bibinfo {author} {\bibfnamefont
  {N.}~\bibnamefont {McMann}}, \bibinfo {author} {\bibfnamefont {B.~W.}\
  \bibnamefont {Meyers}}, \bibinfo {author} {\bibfnamefont {A.}~\bibnamefont
  {Naidu}}, \bibinfo {author} {\bibfnamefont {C.}~\bibnamefont {Ng}}, \bibinfo
  {author} {\bibfnamefont {D.~J.}\ \bibnamefont {Nice}}, \bibinfo {author}
  {\bibfnamefont {N.}~\bibnamefont {Pol}}, \bibinfo {author} {\bibfnamefont
  {H.~A.}\ \bibnamefont {Radovan}}, \bibinfo {author} {\bibfnamefont
  {B.}~\bibnamefont {Shapiro-Albert}}, \bibinfo {author} {\bibfnamefont
  {C.~M.}\ \bibnamefont {Tan}}, \bibinfo {author} {\bibfnamefont {S.~P.}\
  \bibnamefont {Tendulkar}}, \bibinfo {author} {\bibfnamefont {J.~K.}\
  \bibnamefont {Swiggum}}, \bibinfo {author} {\bibfnamefont {H.~M.}\
  \bibnamefont {Wahl}}, \ and\ \bibinfo {author} {\bibfnamefont {W.~W.}\
  \bibnamefont {Zhu}},\ }\href {\doibase 10.3847/2041-8213/ac03b8} {\bibfield
  {journal} {\bibinfo  {journal} {The Astrophysical Journal Letters}\ }\textbf
  {\bibinfo {volume} {915}},\ \bibinfo {pages} {L12} (\bibinfo {year}
  {2021})}\BibitemShut {NoStop}%
\bibitem [{\citenamefont {Miller}\ \emph {et~al.}(2021)\citenamefont {Miller},
  \citenamefont {Lamb}, \citenamefont {Dittmann}, \citenamefont {Bogdanov},
  \citenamefont {Arzoumanian}, \citenamefont {Gendreau}, \citenamefont
  {Guillot}, \citenamefont {Ho}, \citenamefont {Lattimer}, \citenamefont
  {Loewenstein}, \citenamefont {Morsink}, \citenamefont {Ray}, \citenamefont
  {Wolff}, \citenamefont {Baker}, \citenamefont {Cazeau}, \citenamefont
  {Manthripragada}, \citenamefont {Markwardt}, \citenamefont {Okajima},
  \citenamefont {Pollard}, \citenamefont {Cognard}, \citenamefont {Cromartie},
  \citenamefont {Fonseca}, \citenamefont {Guillemot}, \citenamefont {Kerr},
  \citenamefont {Parthasarathy}, \citenamefont {Pennucci}, \citenamefont
  {Ransom},\ and\ \citenamefont {Stairs}}]{Miller_2021}%
  \BibitemOpen
  \bibfield  {author} {\bibinfo {author} {\bibfnamefont {M.~C.}\ \bibnamefont
  {Miller}}, \bibinfo {author} {\bibfnamefont {F.~K.}\ \bibnamefont {Lamb}},
  \bibinfo {author} {\bibfnamefont {A.~J.}\ \bibnamefont {Dittmann}}, \bibinfo
  {author} {\bibfnamefont {S.}~\bibnamefont {Bogdanov}}, \bibinfo {author}
  {\bibfnamefont {Z.}~\bibnamefont {Arzoumanian}}, \bibinfo {author}
  {\bibfnamefont {K.~C.}\ \bibnamefont {Gendreau}}, \bibinfo {author}
  {\bibfnamefont {S.}~\bibnamefont {Guillot}}, \bibinfo {author} {\bibfnamefont
  {W.~C.~G.}\ \bibnamefont {Ho}}, \bibinfo {author} {\bibfnamefont {J.~M.}\
  \bibnamefont {Lattimer}}, \bibinfo {author} {\bibfnamefont {M.}~\bibnamefont
  {Loewenstein}}, \bibinfo {author} {\bibfnamefont {S.~M.}\ \bibnamefont
  {Morsink}}, \bibinfo {author} {\bibfnamefont {P.~S.}\ \bibnamefont {Ray}},
  \bibinfo {author} {\bibfnamefont {M.~T.}\ \bibnamefont {Wolff}}, \bibinfo
  {author} {\bibfnamefont {C.~L.}\ \bibnamefont {Baker}}, \bibinfo {author}
  {\bibfnamefont {T.}~\bibnamefont {Cazeau}}, \bibinfo {author} {\bibfnamefont
  {S.}~\bibnamefont {Manthripragada}}, \bibinfo {author} {\bibfnamefont
  {C.~B.}\ \bibnamefont {Markwardt}}, \bibinfo {author} {\bibfnamefont
  {T.}~\bibnamefont {Okajima}}, \bibinfo {author} {\bibfnamefont
  {S.}~\bibnamefont {Pollard}}, \bibinfo {author} {\bibfnamefont
  {I.}~\bibnamefont {Cognard}}, \bibinfo {author} {\bibfnamefont {H.~T.}\
  \bibnamefont {Cromartie}}, \bibinfo {author} {\bibfnamefont {E.}~\bibnamefont
  {Fonseca}}, \bibinfo {author} {\bibfnamefont {L.}~\bibnamefont {Guillemot}},
  \bibinfo {author} {\bibfnamefont {M.}~\bibnamefont {Kerr}}, \bibinfo {author}
  {\bibfnamefont {A.}~\bibnamefont {Parthasarathy}}, \bibinfo {author}
  {\bibfnamefont {T.~T.}\ \bibnamefont {Pennucci}}, \bibinfo {author}
  {\bibfnamefont {S.}~\bibnamefont {Ransom}}, \ and\ \bibinfo {author}
  {\bibfnamefont {I.}~\bibnamefont {Stairs}},\ }\href {\doibase
  10.3847/2041-8213/ac089b} {\bibfield  {journal} {\bibinfo  {journal} {The
  Astrophysical Journal Letters}\ }\textbf {\bibinfo {volume} {918}},\ \bibinfo
  {pages} {L28} (\bibinfo {year} {2021})}\BibitemShut {NoStop}%
\bibitem [{\citenamefont {Tolman}(1939)}]{PhysRev.55.364}%
  \BibitemOpen
  \bibfield  {author} {\bibinfo {author} {\bibfnamefont {R.~C.}\ \bibnamefont
  {Tolman}},\ }\href {\doibase 10.1103/PhysRev.55.364} {\bibfield  {journal}
  {\bibinfo  {journal} {Phys. Rev.}\ }\textbf {\bibinfo {volume} {55}},\
  \bibinfo {pages} {364} (\bibinfo {year} {1939})}\BibitemShut {NoStop}%
\bibitem [{\citenamefont {Oppenheimer}\ and\ \citenamefont
  {Volkoff}(1939)}]{PhysRev.55.374}%
  \BibitemOpen
  \bibfield  {author} {\bibinfo {author} {\bibfnamefont {J.~R.}\ \bibnamefont
  {Oppenheimer}}\ and\ \bibinfo {author} {\bibfnamefont {G.~M.}\ \bibnamefont
  {Volkoff}},\ }\href {\doibase 10.1103/PhysRev.55.374} {\bibfield  {journal}
  {\bibinfo  {journal} {Phys. Rev.}\ }\textbf {\bibinfo {volume} {55}},\
  \bibinfo {pages} {374} (\bibinfo {year} {1939})}\BibitemShut {NoStop}%
\bibitem [{\citenamefont {Cromartie}\ \emph {et~al.}(2020)\citenamefont
  {Cromartie}, \citenamefont {Fonseca}, \citenamefont {Ransom}, \citenamefont
  {Demorest}, \citenamefont {Arzoumanian}, \citenamefont {Blumer},
  \citenamefont {Brook}, \citenamefont {DeCesar}, \citenamefont {Dolch},
  \citenamefont {Ellis} \emph {et~al.}}]{cromartie2020relativistic}%
  \BibitemOpen
  \bibfield  {author} {\bibinfo {author} {\bibfnamefont {H.~T.}\ \bibnamefont
  {Cromartie}}, \bibinfo {author} {\bibfnamefont {E.}~\bibnamefont {Fonseca}},
  \bibinfo {author} {\bibfnamefont {S.~M.}\ \bibnamefont {Ransom}}, \bibinfo
  {author} {\bibfnamefont {P.~B.}\ \bibnamefont {Demorest}}, \bibinfo {author}
  {\bibfnamefont {Z.}~\bibnamefont {Arzoumanian}}, \bibinfo {author}
  {\bibfnamefont {H.}~\bibnamefont {Blumer}}, \bibinfo {author} {\bibfnamefont
  {P.~R.}\ \bibnamefont {Brook}}, \bibinfo {author} {\bibfnamefont {M.~E.}\
  \bibnamefont {DeCesar}}, \bibinfo {author} {\bibfnamefont {T.}~\bibnamefont
  {Dolch}}, \bibinfo {author} {\bibfnamefont {J.~A.}\ \bibnamefont {Ellis}},
  \emph {et~al.},\ }\href@noop {} {\bibfield  {journal} {\bibinfo  {journal}
  {Nature Astronomy}\ }\textbf {\bibinfo {volume} {4}},\ \bibinfo {pages} {72}
  (\bibinfo {year} {2020})}\BibitemShut {NoStop}%
\bibitem [{\citenamefont {Riley}\ \emph {et~al.}(2021)\citenamefont {Riley},
  \citenamefont {Watts}, \citenamefont {Ray}, \citenamefont {Bogdanov},
  \citenamefont {Guillot}, \citenamefont {Morsink}, \citenamefont {Bilous},
  \citenamefont {Arzoumanian}, \citenamefont {Choudhury}, \citenamefont
  {Deneva}, \citenamefont {Gendreau}, \citenamefont {Harding}, \citenamefont
  {Ho}, \citenamefont {Lattimer}, \citenamefont {Loewenstein}, \citenamefont
  {Ludlam}, \citenamefont {Markwardt}, \citenamefont {Okajima}, \citenamefont
  {Prescod-Weinstein}, \citenamefont {Remillard}, \citenamefont {Wolff},
  \citenamefont {Fonseca}, \citenamefont {Cromartie}, \citenamefont {Kerr},
  \citenamefont {Pennucci}, \citenamefont {Parthasarathy}, \citenamefont
  {Ransom}, \citenamefont {Stairs}, \citenamefont {Guillemot},\ and\
  \citenamefont {Cognard}}]{Riley_2021}%
  \BibitemOpen
  \bibfield  {author} {\bibinfo {author} {\bibfnamefont {T.~E.}\ \bibnamefont
  {Riley}}, \bibinfo {author} {\bibfnamefont {A.~L.}\ \bibnamefont {Watts}},
  \bibinfo {author} {\bibfnamefont {P.~S.}\ \bibnamefont {Ray}}, \bibinfo
  {author} {\bibfnamefont {S.}~\bibnamefont {Bogdanov}}, \bibinfo {author}
  {\bibfnamefont {S.}~\bibnamefont {Guillot}}, \bibinfo {author} {\bibfnamefont
  {S.~M.}\ \bibnamefont {Morsink}}, \bibinfo {author} {\bibfnamefont {A.~V.}\
  \bibnamefont {Bilous}}, \bibinfo {author} {\bibfnamefont {Z.}~\bibnamefont
  {Arzoumanian}}, \bibinfo {author} {\bibfnamefont {D.}~\bibnamefont
  {Choudhury}}, \bibinfo {author} {\bibfnamefont {J.~S.}\ \bibnamefont
  {Deneva}}, \bibinfo {author} {\bibfnamefont {K.~C.}\ \bibnamefont
  {Gendreau}}, \bibinfo {author} {\bibfnamefont {A.~K.}\ \bibnamefont
  {Harding}}, \bibinfo {author} {\bibfnamefont {W.~C.~G.}\ \bibnamefont {Ho}},
  \bibinfo {author} {\bibfnamefont {J.~M.}\ \bibnamefont {Lattimer}}, \bibinfo
  {author} {\bibfnamefont {M.}~\bibnamefont {Loewenstein}}, \bibinfo {author}
  {\bibfnamefont {R.~M.}\ \bibnamefont {Ludlam}}, \bibinfo {author}
  {\bibfnamefont {C.~B.}\ \bibnamefont {Markwardt}}, \bibinfo {author}
  {\bibfnamefont {T.}~\bibnamefont {Okajima}}, \bibinfo {author} {\bibfnamefont
  {C.}~\bibnamefont {Prescod-Weinstein}}, \bibinfo {author} {\bibfnamefont
  {R.~A.}\ \bibnamefont {Remillard}}, \bibinfo {author} {\bibfnamefont {M.~T.}\
  \bibnamefont {Wolff}}, \bibinfo {author} {\bibfnamefont {E.}~\bibnamefont
  {Fonseca}}, \bibinfo {author} {\bibfnamefont {H.~T.}\ \bibnamefont
  {Cromartie}}, \bibinfo {author} {\bibfnamefont {M.}~\bibnamefont {Kerr}},
  \bibinfo {author} {\bibfnamefont {T.~T.}\ \bibnamefont {Pennucci}}, \bibinfo
  {author} {\bibfnamefont {A.}~\bibnamefont {Parthasarathy}}, \bibinfo {author}
  {\bibfnamefont {S.}~\bibnamefont {Ransom}}, \bibinfo {author} {\bibfnamefont
  {I.}~\bibnamefont {Stairs}}, \bibinfo {author} {\bibfnamefont
  {L.}~\bibnamefont {Guillemot}}, \ and\ \bibinfo {author} {\bibfnamefont
  {I.}~\bibnamefont {Cognard}},\ }\href {\doibase 10.3847/2041-8213/ac0a81}
  {\bibfield  {journal} {\bibinfo  {journal} {The Astrophysical Journal
  Letters}\ }\textbf {\bibinfo {volume} {918}},\ \bibinfo {pages} {L27}
  (\bibinfo {year} {2021})}\BibitemShut {NoStop}%
\bibitem [{\citenamefont {Hinderer}(2009)}]{Hinderer_2009}%
  \BibitemOpen
  \bibfield  {author} {\bibinfo {author} {\bibfnamefont {T.}~\bibnamefont
  {Hinderer}},\ }\href {\doibase 10.1088/0004-637x/697/1/964} {\bibfield
  {journal} {\bibinfo  {journal} {The Astrophysical Journal}\ }\textbf
  {\bibinfo {volume} {697}},\ \bibinfo {pages} {964} (\bibinfo {year}
  {2009})}\BibitemShut {NoStop}%
\bibitem [{\citenamefont {Hinderer}\ \emph {et~al.}(2010)\citenamefont
  {Hinderer}, \citenamefont {Lackey}, \citenamefont {Lang},\ and\ \citenamefont
  {Read}}]{PhysRevD.81.123016}%
  \BibitemOpen
  \bibfield  {author} {\bibinfo {author} {\bibfnamefont {T.}~\bibnamefont
  {Hinderer}}, \bibinfo {author} {\bibfnamefont {B.~D.}\ \bibnamefont
  {Lackey}}, \bibinfo {author} {\bibfnamefont {R.~N.}\ \bibnamefont {Lang}}, \
  and\ \bibinfo {author} {\bibfnamefont {J.~S.}\ \bibnamefont {Read}},\ }\href
  {\doibase 10.1103/PhysRevD.81.123016} {\bibfield  {journal} {\bibinfo
  {journal} {Phys. Rev. D}\ }\textbf {\bibinfo {volume} {81}},\ \bibinfo
  {pages} {123016} (\bibinfo {year} {2010})}\BibitemShut {NoStop}%
\bibitem [{\citenamefont {Zhao}\ and\ \citenamefont
  {Lattimer}(2018)}]{zhao2018tidal}%
  \BibitemOpen
  \bibfield  {author} {\bibinfo {author} {\bibfnamefont {T.}~\bibnamefont
  {Zhao}}\ and\ \bibinfo {author} {\bibfnamefont {J.~M.}\ \bibnamefont
  {Lattimer}},\ }\href@noop {} {\bibfield  {journal} {\bibinfo  {journal}
  {Physical Review D}\ }\textbf {\bibinfo {volume} {98}},\ \bibinfo {pages}
  {063020} (\bibinfo {year} {2018})}\BibitemShut {NoStop}%
\bibitem [{\citenamefont {Abbott}\ \emph {et~al.}(2020)\citenamefont {Abbott}
  \emph {et~al.}}]{Abbott:2020uma}%
  \BibitemOpen
  \bibfield  {author} {\bibinfo {author} {\bibfnamefont {B.}~\bibnamefont
  {Abbott}} \emph {et~al.} (\bibinfo {collaboration} {LIGO Scientific,
  Virgo}),\ }\href {\doibase 10.3847/2041-8213/ab75f5} {\bibfield  {journal}
  {\bibinfo  {journal} {Astrophys. J. Lett.}\ }\textbf {\bibinfo {volume}
  {892}},\ \bibinfo {pages} {L3} (\bibinfo {year} {2020})},\ \Eprint
  {http://arxiv.org/abs/2001.01761} {arXiv:2001.01761} \BibitemShut {NoStop}%
\bibitem [{\citenamefont {Andersson}\ and\ \citenamefont
  {Kokkotas}(1998{\natexlab{b}})}]{10.1046/j.1365-8711.1998.01840.x}%
  \BibitemOpen
  \bibfield  {author} {\bibinfo {author} {\bibfnamefont {N.}~\bibnamefont
  {Andersson}}\ and\ \bibinfo {author} {\bibfnamefont {K.~D.}\ \bibnamefont
  {Kokkotas}},\ }\href {\doibase 10.1046/j.1365-8711.1998.01840.x} {\bibfield
  {journal} {\bibinfo  {journal} {Monthly Notices of the Royal Astronomical
  Society}\ }\textbf {\bibinfo {volume} {299}},\ \bibinfo {pages} {1059}
  (\bibinfo {year} {1998}{\natexlab{b}})},\ \Eprint
  {http://arxiv.org/abs/https://academic.oup.com/mnras/article-pdf/299/4/1059/3869494/299-4-1059.pdf}
  {https://academic.oup.com/mnras/article-pdf/299/4/1059/3869494/299-4-1059.pdf}
  \BibitemShut {NoStop}%
\bibitem [{\citenamefont {CHIRENTI}\ \emph {et~al.}(2012)\citenamefont
  {CHIRENTI}, \citenamefont {SILVEIRA},\ and\ \citenamefont
  {AGUIAR}}]{CHIRENTI_2012}%
  \BibitemOpen
  \bibfield  {author} {\bibinfo {author} {\bibfnamefont {C.}~\bibnamefont
  {CHIRENTI}}, \bibinfo {author} {\bibfnamefont {P.~R.}\ \bibnamefont
  {SILVEIRA}}, \ and\ \bibinfo {author} {\bibfnamefont {O.~D.}\ \bibnamefont
  {AGUIAR}},\ }\href {\doibase 10.1142/s2010194512008185} {\bibfield  {journal}
  {\bibinfo  {journal} {International Journal of Modern Physics: Conference
  Series}\ }\textbf {\bibinfo {volume} {18}},\ \bibinfo {pages} {48} (\bibinfo
  {year} {2012})}\BibitemShut {NoStop}%
\bibitem [{\citenamefont {Sotani}\ \emph
  {et~al.}(2011{\natexlab{b}})\citenamefont {Sotani}, \citenamefont {Yasutake},
  \citenamefont {Maruyama},\ and\ \citenamefont {Tatsumi}}]{sotani11}%
  \BibitemOpen
  \bibfield  {author} {\bibinfo {author} {\bibfnamefont {H.}~\bibnamefont
  {Sotani}}, \bibinfo {author} {\bibfnamefont {N.}~\bibnamefont {Yasutake}},
  \bibinfo {author} {\bibfnamefont {T.}~\bibnamefont {Maruyama}}, \ and\
  \bibinfo {author} {\bibfnamefont {T.}~\bibnamefont {Tatsumi}},\ }\href
  {\doibase 10.1103/PhysRevD.83.024014} {\bibfield  {journal} {\bibinfo
  {journal} {Phys. Rev. D}\ }\textbf {\bibinfo {volume} {83}},\ \bibinfo
  {pages} {024014} (\bibinfo {year} {2011}{\natexlab{b}})}\BibitemShut
  {NoStop}%
\bibitem [{\citenamefont {Benesty}\ \emph {et~al.}(2009)\citenamefont
  {Benesty}, \citenamefont {Chen}, \citenamefont {Huang},\ and\ \citenamefont
  {Cohen}}]{Benesty2009}%
  \BibitemOpen
  \bibfield  {author} {\bibinfo {author} {\bibfnamefont {J.}~\bibnamefont
  {Benesty}}, \bibinfo {author} {\bibfnamefont {J.}~\bibnamefont {Chen}},
  \bibinfo {author} {\bibfnamefont {Y.}~\bibnamefont {Huang}}, \ and\ \bibinfo
  {author} {\bibfnamefont {I.}~\bibnamefont {Cohen}},\ }\enquote {\bibinfo
  {title} {Pearson correlation coefficient},}\ in\ \href {\doibase
  10.1007/978-3-642-00296-0_5} {\emph {\bibinfo {booktitle} {Noise Reduction in
  Speech Processing}}}\ (\bibinfo  {publisher} {Springer Berlin Heidelberg},\
  \bibinfo {address} {Berlin, Heidelberg},\ \bibinfo {year} {2009})\ pp.\
  \bibinfo {pages} {1--4}\BibitemShut {NoStop}%
\bibitem [{\citenamefont {Brandt}()}]{Brandt97}%
  \BibitemOpen
  \bibfield  {author} {\bibinfo {author} {\bibfnamefont {S.}~\bibnamefont
  {Brandt}},\ }\href@noop {} {\bibinfo  {journal} {Statistical and
  Computational Methods in Data Analysis (Springer, New York, 3rd English
  edition, 1997)}\ }\BibitemShut {NoStop}%
\bibitem [{\citenamefont {Malik}\ \emph {et~al.}(2020)\citenamefont {Malik},
  \citenamefont {Agrawal}, \citenamefont {Provid\^encia},\ and\ \citenamefont
  {De}}]{Tuhin2020}%
  \BibitemOpen
\bibfield  {journal} {  }\bibfield  {author} {\bibinfo {author} {\bibfnamefont
  {T.}~\bibnamefont {Malik}}, \bibinfo {author} {\bibfnamefont {B.~K.}\
  \bibnamefont {Agrawal}}, \bibinfo {author} {\bibfnamefont {C.~m.~c.}\
  \bibnamefont {Provid\^encia}}, \ and\ \bibinfo {author} {\bibfnamefont
  {J.~N.}\ \bibnamefont {De}},\ }\href {\doibase 10.1103/PhysRevC.102.052801}
  {\bibfield  {journal} {\bibinfo  {journal} {Phys. Rev. C}\ }\textbf {\bibinfo
  {volume} {102}},\ \bibinfo {pages} {052801} (\bibinfo {year}
  {2020})}\BibitemShut {NoStop}%
\bibitem [{\citenamefont {Pradhan}\ \emph
  {et~al.}(2022{\natexlab{a}})\citenamefont {Pradhan}, \citenamefont
  {Chatterjee}, \citenamefont {Lanoye},\ and\ \citenamefont
  {Jaikumar}}]{Pradhan2022a}%
  \BibitemOpen
  \bibfield  {author} {\bibinfo {author} {\bibfnamefont {B.~K.}\ \bibnamefont
  {Pradhan}}, \bibinfo {author} {\bibfnamefont {D.}~\bibnamefont {Chatterjee}},
  \bibinfo {author} {\bibfnamefont {M.}~\bibnamefont {Lanoye}}, \ and\ \bibinfo
  {author} {\bibfnamefont {P.}~\bibnamefont {Jaikumar}},\ }\href@noop {}
  {\bibfield  {journal} {\bibinfo  {journal} {arXiv preprint
  10.48550/ARXIV.2203.03141}\ } (\bibinfo {year}
  {2022}{\natexlab{a}})}\BibitemShut {NoStop}%
\bibitem [{\citenamefont {Abbott}(2018)}]{GW170817_radii}%
  \BibitemOpen
  \bibfield  {author} {\bibinfo {author} {\bibfnamefont {B.~P.~{\it et. al.}.}\
  \bibnamefont {Abbott}} (\bibinfo {collaboration} {The LIGO Scientific
  Collaboration and the Virgo Collaboration}),\ }\href {\doibase
  10.1103/PhysRevLett.121.161101} {\bibfield  {journal} {\bibinfo  {journal}
  {Phys. Rev. Lett.}\ }\textbf {\bibinfo {volume} {121}},\ \bibinfo {pages}
  {161101} (\bibinfo {year} {2018})}\BibitemShut {NoStop}%
\bibitem [{\citenamefont {Carson}\ \emph {et~al.}(2019)\citenamefont {Carson},
  \citenamefont {Steiner},\ and\ \citenamefont {Yagi}}]{Carson2019}%
  \BibitemOpen
  \bibfield  {author} {\bibinfo {author} {\bibfnamefont {Z.}~\bibnamefont
  {Carson}}, \bibinfo {author} {\bibfnamefont {A.~W.}\ \bibnamefont {Steiner}},
  \ and\ \bibinfo {author} {\bibfnamefont {K.}~\bibnamefont {Yagi}},\ }\href
  {\doibase 10.1103/PhysRevD.99.043010} {\bibfield  {journal} {\bibinfo
  {journal} {Phys. Rev. D}\ }\textbf {\bibinfo {volume} {99}},\ \bibinfo
  {pages} {043010} (\bibinfo {year} {2019})}\BibitemShut {NoStop}%
\bibitem [{\citenamefont {Tsang}\ \emph {et~al.}(2020)\citenamefont {Tsang},
  \citenamefont {Tsang}, \citenamefont {Danielewicz}, \citenamefont {Lynch},\
  and\ \citenamefont {Fattoyev}}]{Tsang2020}%
  \BibitemOpen
  \bibfield  {author} {\bibinfo {author} {\bibfnamefont {C.~Y.}\ \bibnamefont
  {Tsang}}, \bibinfo {author} {\bibfnamefont {M.~B.}\ \bibnamefont {Tsang}},
  \bibinfo {author} {\bibfnamefont {P.}~\bibnamefont {Danielewicz}}, \bibinfo
  {author} {\bibfnamefont {W.~G.}\ \bibnamefont {Lynch}}, \ and\ \bibinfo
  {author} {\bibfnamefont {F.~J.}\ \bibnamefont {Fattoyev}},\ }\href {\doibase
  10.1103/PhysRevC.102.045808} {\bibfield  {journal} {\bibinfo  {journal}
  {Phys. Rev. C}\ }\textbf {\bibinfo {volume} {102}},\ \bibinfo {pages}
  {045808} (\bibinfo {year} {2020})}\BibitemShut {NoStop}%
\bibitem [{\citenamefont {Reed}\ \emph {et~al.}(2021)\citenamefont {Reed},
  \citenamefont {Fattoyev}, \citenamefont {Horowitz},\ and\ \citenamefont
  {Piekarewicz}}]{Reed2021}%
  \BibitemOpen
  \bibfield  {author} {\bibinfo {author} {\bibfnamefont {B.~T.}\ \bibnamefont
  {Reed}}, \bibinfo {author} {\bibfnamefont {F.~J.}\ \bibnamefont {Fattoyev}},
  \bibinfo {author} {\bibfnamefont {C.~J.}\ \bibnamefont {Horowitz}}, \ and\
  \bibinfo {author} {\bibfnamefont {J.}~\bibnamefont {Piekarewicz}},\ }\href
  {\doibase 10.1103/PhysRevLett.126.172503} {\bibfield  {journal} {\bibinfo
  {journal} {Phys. Rev. Lett.}\ }\textbf {\bibinfo {volume} {126}},\ \bibinfo
  {pages} {172503} (\bibinfo {year} {2021})},\ \Eprint
  {http://arxiv.org/abs/2101.03193} {arXiv:2101.03193} \BibitemShut {NoStop}%
\bibitem [{\citenamefont {Essick}\ \emph {et~al.}(2021)\citenamefont {Essick},
  \citenamefont {Tews}, \citenamefont {Landry},\ and\ \citenamefont
  {Schwenk}}]{Reed2021A}%
  \BibitemOpen
  \bibfield  {author} {\bibinfo {author} {\bibfnamefont {R.}~\bibnamefont
  {Essick}}, \bibinfo {author} {\bibfnamefont {I.}~\bibnamefont {Tews}},
  \bibinfo {author} {\bibfnamefont {P.}~\bibnamefont {Landry}}, \ and\ \bibinfo
  {author} {\bibfnamefont {A.}~\bibnamefont {Schwenk}},\ }\href {\doibase
  10.1103/PhysRevLett.127.192701} {\bibfield  {journal} {\bibinfo  {journal}
  {Phys. Rev. Lett.}\ }\textbf {\bibinfo {volume} {127}},\ \bibinfo {pages}
  {192701} (\bibinfo {year} {2021})}\BibitemShut {NoStop}%
\bibitem [{\citenamefont {Biswas}(2021)}]{Biswas:2021yge}%
  \BibitemOpen
  \bibfield  {author} {\bibinfo {author} {\bibfnamefont {B.}~\bibnamefont
  {Biswas}},\ }\href {\doibase 10.3847/1538-4357/ac1c72} {\bibfield  {journal}
  {\bibinfo  {journal} {Astrophys. J.}\ }\textbf {\bibinfo {volume} {921}},\
  \bibinfo {pages} {63} (\bibinfo {year} {2021})},\ \Eprint
  {http://arxiv.org/abs/2105.02886} {arXiv:2105.02886} \BibitemShut {NoStop}%
\bibitem [{\citenamefont {Margueron}\ \emph {et~al.}(2018)\citenamefont
  {Margueron}, \citenamefont {Hoffmann~Casali},\ and\ \citenamefont
  {Gulminelli}}]{Margueron2018}%
  \BibitemOpen
  \bibfield  {author} {\bibinfo {author} {\bibfnamefont {J.}~\bibnamefont
  {Margueron}}, \bibinfo {author} {\bibfnamefont {R.}~\bibnamefont
  {Hoffmann~Casali}}, \ and\ \bibinfo {author} {\bibfnamefont {F.}~\bibnamefont
  {Gulminelli}},\ }\href {\doibase 10.1103/PhysRevC.97.025805} {\bibfield
  {journal} {\bibinfo  {journal} {Phys. Rev. C}\ }\textbf {\bibinfo {volume}
  {97}},\ \bibinfo {pages} {025805} (\bibinfo {year} {2018})}\BibitemShut
  {NoStop}%
\bibitem [{\citenamefont {Malik}\ \emph {et~al.}(2018)\citenamefont {Malik},
  \citenamefont {Alam}, \citenamefont {Fortin}, \citenamefont {Provid\^encia},
  \citenamefont {Agrawal}, \citenamefont {Jha}, \citenamefont {Kumar},\ and\
  \citenamefont {Patra}}]{Malik2018}%
  \BibitemOpen
  \bibfield  {author} {\bibinfo {author} {\bibfnamefont {T.}~\bibnamefont
  {Malik}}, \bibinfo {author} {\bibfnamefont {N.}~\bibnamefont {Alam}},
  \bibinfo {author} {\bibfnamefont {M.}~\bibnamefont {Fortin}}, \bibinfo
  {author} {\bibfnamefont {C.}~\bibnamefont {Provid\^encia}}, \bibinfo {author}
  {\bibfnamefont {B.~K.}\ \bibnamefont {Agrawal}}, \bibinfo {author}
  {\bibfnamefont {T.~K.}\ \bibnamefont {Jha}}, \bibinfo {author} {\bibfnamefont
  {B.}~\bibnamefont {Kumar}}, \ and\ \bibinfo {author} {\bibfnamefont {S.~K.}\
  \bibnamefont {Patra}},\ }\href {\doibase 10.1103/PhysRevC.98.035804}
  {\bibfield  {journal} {\bibinfo  {journal} {Phys. Rev. C}\ }\textbf {\bibinfo
  {volume} {98}},\ \bibinfo {pages} {035804} (\bibinfo {year}
  {2018})}\BibitemShut {NoStop}%
\bibitem [{\citenamefont {De}\ \emph {et~al.}(2018)\citenamefont {De},
  \citenamefont {Finstad}, \citenamefont {Lattimer}, \citenamefont {Brown},
  \citenamefont {Berger},\ and\ \citenamefont {Biwer}}]{De18}%
  \BibitemOpen
  \bibfield  {author} {\bibinfo {author} {\bibfnamefont {S.}~\bibnamefont
  {De}}, \bibinfo {author} {\bibfnamefont {D.}~\bibnamefont {Finstad}},
  \bibinfo {author} {\bibfnamefont {J.~M.}\ \bibnamefont {Lattimer}}, \bibinfo
  {author} {\bibfnamefont {D.~A.}\ \bibnamefont {Brown}}, \bibinfo {author}
  {\bibfnamefont {E.}~\bibnamefont {Berger}}, \ and\ \bibinfo {author}
  {\bibfnamefont {C.~M.}\ \bibnamefont {Biwer}},\ }\href {\doibase
  10.1103/PhysRevLett.121.091102} {\bibfield  {journal} {\bibinfo  {journal}
  {Phys. Rev. Lett.}\ }\textbf {\bibinfo {volume} {121}},\ \bibinfo {pages}
  {091102} (\bibinfo {year} {2018})}\BibitemShut {NoStop}%
\bibitem [{\citenamefont {Agrawal}\ and\ \citenamefont
  {Malik}(2021)}]{AgrawalCRC}%
  \BibitemOpen
  \bibfield  {author} {\bibinfo {author} {\bibfnamefont {B.}~\bibnamefont
  {Agrawal}}\ and\ \bibinfo {author} {\bibfnamefont {T.}~\bibnamefont
  {Malik}},\ }\href@noop {} {\emph {\bibinfo {title} {Nuclear Structure
  Physics}}}\ (\bibinfo  {publisher} {CRC Press},\ \bibinfo {year}
  {2021})\BibitemShut {NoStop}%
\bibitem [{\citenamefont {Khan}\ \emph {et~al.}(2012)\citenamefont {Khan},
  \citenamefont {Margueron},\ and\ \citenamefont {Vida\~na}}]{Khan2012}%
  \BibitemOpen
  \bibfield  {author} {\bibinfo {author} {\bibfnamefont {E.}~\bibnamefont
  {Khan}}, \bibinfo {author} {\bibfnamefont {J.}~\bibnamefont {Margueron}}, \
  and\ \bibinfo {author} {\bibfnamefont {I.}~\bibnamefont {Vida\~na}},\ }\href
  {\doibase 10.1103/PhysRevLett.109.092501} {\bibfield  {journal} {\bibinfo
  {journal} {Phys. Rev. Lett.}\ }\textbf {\bibinfo {volume} {109}},\ \bibinfo
  {pages} {092501} (\bibinfo {year} {2012})}\BibitemShut {NoStop}%
\bibitem [{\citenamefont {Khan}\ and\ \citenamefont
  {Margueron}(2013)}]{Khan2013}%
  \BibitemOpen
  \bibfield  {author} {\bibinfo {author} {\bibfnamefont {E.}~\bibnamefont
  {Khan}}\ and\ \bibinfo {author} {\bibfnamefont {J.}~\bibnamefont
  {Margueron}},\ }\href {\doibase 10.1103/PhysRevC.88.034319} {\bibfield
  {journal} {\bibinfo  {journal} {Phys. Rev. C}\ }\textbf {\bibinfo {volume}
  {88}},\ \bibinfo {pages} {034319} (\bibinfo {year} {2013})}\BibitemShut
  {NoStop}%
\bibitem [{\citenamefont {Tews}\ \emph {et~al.}(2017)\citenamefont {Tews},
  \citenamefont {Lattimer}, \citenamefont {Ohnishi},\ and\ \citenamefont
  {Kolomeitsev}}]{Tews2017}%
  \BibitemOpen
  \bibfield  {author} {\bibinfo {author} {\bibfnamefont {I.}~\bibnamefont
  {Tews}}, \bibinfo {author} {\bibfnamefont {J.~M.}\ \bibnamefont {Lattimer}},
  \bibinfo {author} {\bibfnamefont {A.}~\bibnamefont {Ohnishi}}, \ and\
  \bibinfo {author} {\bibfnamefont {E.~E.}\ \bibnamefont {Kolomeitsev}},\
  }\href {\doibase 10.3847/1538-4357/aa8db9} {\bibfield  {journal} {\bibinfo
  {journal} {The Astrophysical Journal}\ }\textbf {\bibinfo {volume} {848}},\
  \bibinfo {pages} {105} (\bibinfo {year} {2017})}\BibitemShut {NoStop}%
\bibitem [{\citenamefont {Mondal}\ \emph {et~al.}(2017)\citenamefont {Mondal},
  \citenamefont {Agrawal}, \citenamefont {De}, \citenamefont {Samaddar},
  \citenamefont {Centelles},\ and\ \citenamefont {Vi\~nas}}]{Mondal2017}%
  \BibitemOpen
  \bibfield  {author} {\bibinfo {author} {\bibfnamefont {C.}~\bibnamefont
  {Mondal}}, \bibinfo {author} {\bibfnamefont {B.~K.}\ \bibnamefont {Agrawal}},
  \bibinfo {author} {\bibfnamefont {J.~N.}\ \bibnamefont {De}}, \bibinfo
  {author} {\bibfnamefont {S.~K.}\ \bibnamefont {Samaddar}}, \bibinfo {author}
  {\bibfnamefont {M.}~\bibnamefont {Centelles}}, \ and\ \bibinfo {author}
  {\bibfnamefont {X.}~\bibnamefont {Vi\~nas}},\ }\href {\doibase
  10.1103/PhysRevC.96.021302} {\bibfield  {journal} {\bibinfo  {journal} {Phys.
  Rev. C}\ }\textbf {\bibinfo {volume} {96}},\ \bibinfo {pages} {021302}
  (\bibinfo {year} {2017})}\BibitemShut {NoStop}%
\bibitem [{\citenamefont {Li}\ and\ \citenamefont {Magno}(2020)}]{Li2020}%
  \BibitemOpen
  \bibfield  {author} {\bibinfo {author} {\bibfnamefont {B.-A.}\ \bibnamefont
  {Li}}\ and\ \bibinfo {author} {\bibfnamefont {M.}~\bibnamefont {Magno}},\
  }\href {\doibase 10.1103/PhysRevC.102.045807} {\bibfield  {journal} {\bibinfo
   {journal} {Phys. Rev. C}\ }\textbf {\bibinfo {volume} {102}},\ \bibinfo
  {pages} {045807} (\bibinfo {year} {2020})}\BibitemShut {NoStop}%
\bibitem [{\citenamefont {Holt}\ and\ \citenamefont {Lim}(2018)}]{Holt2018}%
  \BibitemOpen
  \bibfield  {author} {\bibinfo {author} {\bibfnamefont {J.~W.}\ \bibnamefont
  {Holt}}\ and\ \bibinfo {author} {\bibfnamefont {Y.}~\bibnamefont {Lim}},\
  }\href {\doibase https://doi.org/10.1016/j.physletb.2018.07.038} {\bibfield
  {journal} {\bibinfo  {journal} {Physics Letters B}\ }\textbf {\bibinfo
  {volume} {784}},\ \bibinfo {pages} {77} (\bibinfo {year} {2018})}\BibitemShut
  {NoStop}%
\bibitem [{\citenamefont {{Lindblom}}\ and\ \citenamefont
  {{Detweiler}}(1983{\natexlab{b}})}]{Dscale}%
  \BibitemOpen
  \bibfield  {author} {\bibinfo {author} {\bibfnamefont {L.}~\bibnamefont
  {{Lindblom}}}\ and\ \bibinfo {author} {\bibfnamefont {S.~L.}\ \bibnamefont
  {{Detweiler}}},\ }\href {\doibase 10.1086/190884} {\bibfield  {journal}
  {\bibinfo  {journal} {\apjs}\ }\textbf {\bibinfo {volume} {53}},\ \bibinfo
  {pages} {73} (\bibinfo {year} {1983}{\natexlab{b}})}\BibitemShut {NoStop}%
\bibitem [{\citenamefont {Burgio}\ \emph {et~al.}(2011)\citenamefont {Burgio},
  \citenamefont {Ferrari}, \citenamefont {Gualtieri},\ and\ \citenamefont
  {Schulze}}]{Burgio11}%
  \BibitemOpen
  \bibfield  {author} {\bibinfo {author} {\bibfnamefont {G.~F.}\ \bibnamefont
  {Burgio}}, \bibinfo {author} {\bibfnamefont {V.}~\bibnamefont {Ferrari}},
  \bibinfo {author} {\bibfnamefont {L.}~\bibnamefont {Gualtieri}}, \ and\
  \bibinfo {author} {\bibfnamefont {H.-J.}\ \bibnamefont {Schulze}},\ }\href
  {\doibase 10.1103/PhysRevD.84.044017} {\bibfield  {journal} {\bibinfo
  {journal} {Phys. Rev. D}\ }\textbf {\bibinfo {volume} {84}},\ \bibinfo
  {pages} {044017} (\bibinfo {year} {2011})}\BibitemShut {NoStop}%
\bibitem [{\citenamefont {Chirenti}\ \emph
  {et~al.}(2015{\natexlab{b}})\citenamefont {Chirenti}, \citenamefont
  {de~Souza},\ and\ \citenamefont {Kastaun}}]{PhysRevD.91.044034}%
  \BibitemOpen
  \bibfield  {author} {\bibinfo {author} {\bibfnamefont {C.}~\bibnamefont
  {Chirenti}}, \bibinfo {author} {\bibfnamefont {G.~H.}\ \bibnamefont
  {de~Souza}}, \ and\ \bibinfo {author} {\bibfnamefont {W.}~\bibnamefont
  {Kastaun}},\ }\href {\doibase 10.1103/PhysRevD.91.044034} {\bibfield
  {journal} {\bibinfo  {journal} {Phys. Rev. D}\ }\textbf {\bibinfo {volume}
  {91}},\ \bibinfo {pages} {044034} (\bibinfo {year}
  {2015}{\natexlab{b}})}\BibitemShut {NoStop}%
\bibitem [{\citenamefont {{Yoshida}}\ and\ \citenamefont
  {{Eriguchi}}(1997)}]{1997ApJ...490..779Y}%
  \BibitemOpen
  \bibfield  {author} {\bibinfo {author} {\bibfnamefont {S.}~\bibnamefont
  {{Yoshida}}}\ and\ \bibinfo {author} {\bibfnamefont {Y.}~\bibnamefont
  {{Eriguchi}}},\ }\href {\doibase 10.1086/304918} {\bibfield  {journal}
  {\bibinfo  {journal} {\apj}\ }\textbf {\bibinfo {volume} {490}},\ \bibinfo
  {pages} {779} (\bibinfo {year} {1997})},\ \Eprint
  {http://arxiv.org/abs/astro-ph/9704111} {arXiv:astro-ph/9704111 [astro-ph]}
  \BibitemShut {NoStop}%
\bibitem [{\citenamefont {Lindblom}\ and\ \citenamefont
  {Splinter}(1990)}]{lindblom1990accuracy}%
  \BibitemOpen
  \bibfield  {author} {\bibinfo {author} {\bibfnamefont {L.}~\bibnamefont
  {Lindblom}}\ and\ \bibinfo {author} {\bibfnamefont {R.~J.}\ \bibnamefont
  {Splinter}},\ }\href@noop {} {\bibfield  {journal} {\bibinfo  {journal} {The
  Astrophysical Journal}\ }\textbf {\bibinfo {volume} {348}},\ \bibinfo {pages}
  {198} (\bibinfo {year} {1990})}\BibitemShut {NoStop}%
\bibitem [{\citenamefont {Rosofsky}\ \emph {et~al.}(2019)\citenamefont
  {Rosofsky}, \citenamefont {Gold}, \citenamefont {Chirenti}, \citenamefont
  {Huerta},\ and\ \citenamefont {Miller}}]{rosofsky2019probing}%
  \BibitemOpen
  \bibfield  {author} {\bibinfo {author} {\bibfnamefont {S.~G.}\ \bibnamefont
  {Rosofsky}}, \bibinfo {author} {\bibfnamefont {R.}~\bibnamefont {Gold}},
  \bibinfo {author} {\bibfnamefont {C.}~\bibnamefont {Chirenti}}, \bibinfo
  {author} {\bibfnamefont {E.}~\bibnamefont {Huerta}}, \ and\ \bibinfo {author}
  {\bibfnamefont {M.~C.}\ \bibnamefont {Miller}},\ }\href@noop {} {\bibfield
  {journal} {\bibinfo  {journal} {Physical Review D}\ }\textbf {\bibinfo
  {volume} {99}},\ \bibinfo {pages} {084024} (\bibinfo {year}
  {2019})}\BibitemShut {NoStop}%
\bibitem [{\citenamefont {Shashank}\ \emph {et~al.}(2021)\citenamefont
  {Shashank}, \citenamefont {Nouri},\ and\ \citenamefont
  {Gupta}}]{shashank2021f}%
  \BibitemOpen
  \bibfield  {author} {\bibinfo {author} {\bibfnamefont {S.}~\bibnamefont
  {Shashank}}, \bibinfo {author} {\bibfnamefont {F.~H.}\ \bibnamefont {Nouri}},
  \ and\ \bibinfo {author} {\bibfnamefont {A.}~\bibnamefont {Gupta}},\
  }\href@noop {} {\bibfield  {journal} {\bibinfo  {journal} {arXiv preprint
  arXiv:2108.04643}\ } (\bibinfo {year} {2021})}\BibitemShut {NoStop}%
\bibitem [{\citenamefont {Pradhan}\ \emph
  {et~al.}(2022{\natexlab{b}})\citenamefont {Pradhan}, \citenamefont
  {Chatterjee}, \citenamefont {Lanoye},\ and\ \citenamefont
  {Jaikumar}}]{Pradhan_2022}%
  \BibitemOpen
  \bibfield  {author} {\bibinfo {author} {\bibfnamefont {B.~K.}\ \bibnamefont
  {Pradhan}}, \bibinfo {author} {\bibfnamefont {D.}~\bibnamefont {Chatterjee}},
  \bibinfo {author} {\bibfnamefont {M.}~\bibnamefont {Lanoye}}, \ and\ \bibinfo
  {author} {\bibfnamefont {P.}~\bibnamefont {Jaikumar}},\ }\href {\doibase
  10.1103/physrevc.106.015805} {\bibfield  {journal} {\bibinfo  {journal}
  {Physical Review C}\ }\textbf {\bibinfo {volume} {106}} (\bibinfo {year}
  {2022}{\natexlab{b}}),\ 10.1103/physrevc.106.015805}\BibitemShut {NoStop}%
\end{thebibliography}%
\end{document}